\documentclass[]{aastex631}
\usepackage{bm} 
\usepackage{color}
\usepackage{graphicx}

\newcommand{\msun}{{M}_\odot}
\newcommand{\msunyr}{{M}_\odot\,{\rm yr}^{-1}}

\newcommand{\cc}{{\rm cm^{-3}}}

\newcommand{\vect}[1]{\mbox{\boldmath$#1$}}

\shorttitle{Twisted Pseudodisk II}
\shortauthors{Machida, Hirano and Basu}

\begin{document}
\title{Twisted Pseudodisk and Asymmetric Mass Accretion on the Circumstellar Disk II: Parameter Survey}

\correspondingauthor{Masahiro N. Machida}
\email{machida.masahiro.018@m.kyushu-u.ac.jp}

\author[0000-0002-0963-0872]{Masahiro N. Machida}
\affiliation{Department of Earth and Planetary Sciences, Faculty of Science, Kyushu University, Fukuoka 819-0395, Japan}
\affiliation{Department of Physics and Astronomy, University of Western Ontario, London, ON N6A 3K7, Canada}

\author[0000-0002-4317-767X]{Shingo Hirano}
\affiliation{Department of Applied Physics, Faculty of Engineering, Kanagawa University, Kanagawa 221-0802, Japan}
\affiliation{Department of Astronomy, School of Science, University of Tokyo, Tokyo 113-0033, Japan}

\author[0000-0003-0855-350X]{Shantanu Basu}
\affiliation{Canadian Institute for Theoretical Astrophysics, University of Toronto, 60 St. George Street, Toronto, ON M5S 3H8, Canada}
\affiliation{Department of Physics and Astronomy, University of Western Ontario, London, ON N6A 3K7, Canada}

\begin{abstract}
Using three-dimensional magnetohydrodynamic simulations accounting for ohmic dissipation, we investigate the long-term evolution of the infalling envelope for $\sim 100,000$\,yr after protostar formation. 
The primary control parameter is the initial misalignment angle $\theta_0$ between the rotation axis and the magnetic field, varied from $0^\circ$ to $90^\circ$. 
In all misaligned models ($\theta_0 \ne 0^\circ$), the pseudodisk (or infalling envelope) surrounding the circumstellar disk develops a highly distorted and complex morphology.
Gas accretion from the envelope proceeds through discrete, channel-like flows, producing asymmetric streams rather than axisymmetric infall.
Protostellar outflows are launched even in these complex environments, and inflow and outflow regions are strongly interwoven.
Outflows transport angular momentum anisotropically, leading to time-dependent changes in disk orientation and outflow direction, which further enhance envelope asymmetry.
We examine the relative orientation between gas flows and magnetic fields and find that, although statistical tendencies exist, accretion streams cannot be uniquely characterized as either parallel or perpendicular to the magnetic field, particularly in strongly misaligned cases.
We also show that most infalling gas accretes at nearly the free-fall velocity, with only a minor fraction significantly slowed within the pseudodisk.
Asymmetric accretion and stream-like structures persist beyond 10,000\,au, demonstrating that large-scale asymmetry can arise intrinsically.
Our results show that complex streams and asymmetric envelopes naturally emerge from magnetized core physics, even in the absence of initial turbulence or external gas accretion.
\end{abstract}


\section{Introduction}
\label{sec:intro}
Recent high-angular-resolution observations, particularly with the Atacama Large Millimeter/submillimeter Array (ALMA), have revealed that gas accretion onto circumstellar disks during star formation is often highly non-axisymmetric \citep{pineda23}.
In addition to disk structures, several systems exhibit filamentary gas flows directly connected to disks, commonly referred to as streamers \citep{Yen2019,Pineda2020,Cacciapuoti2024,Kido2025}.
These structures span a wide range of spatial scales, from disk scales ($\sim 10$--100 au) to flows extending beyond 1,000 au.
In some cases, streamers are found in the vicinity of protostellar outflows \citep[e.g.,][]{Flores2023,Kido2023,Aso2023,Tanious2024}, while in other systems the streamer-like features are not isolated, but are instead embedded within a highly complex infalling envelope \citep{tokuda14,tokuda16,tokuda17,tokuda18,tokuda23,tokuda24,riaz24}.
Despite these growing observational discoveries, the physical origin of streamers and asymmetric accretion remains poorly understood.
 
Several mechanisms have been proposed to explain streamer formation and asymmetric accretion, including anisotropic accretion driven by turbulence within molecular cloud cores, late-time inflow of gas or dense clumps from outside the core, and interactions between infalling gas and protostellar outflows \citep{pineda23}.
Numerical simulations have also demonstrated that complex and asymmetric accretion can develop in turbulent environments. 
Turbulent core-collapse simulations have shown that gas accretes onto protostellar disks through narrow, anisotropic channels \citep{Seifried2013,Seifried2015}. 
In turbulent magnetized cores, \citet{Tu2024} also found highly distorted ``gravo-magneto sheetlets,'' which they interpreted as turbulence-warped versions of classical pseudodisks.
Complex circumstellar structures can also develop in purely hydrodynamical simulations of turbulent star-forming environments \citep[e.g.,][]{Bate2018}.
Multi-scale simulations starting from turbulent molecular-cloud or clump-scale environments have also produced complex accretion flows and streamer-like structures around forming protostellar systems \citep{Kuffmeier2017,Kuffmeier2018,Kuffmeier2019,Lebreuilly2021,Lebreuilly2024,Lebreuilly2025}. 
More recently, \citet{Yang2025} followed disk formation using initial conditions extracted from a molecular-cloud simulation, retaining the complex structures inherited from the larger-scale environment. 
Taken together, these studies indicate that turbulence and the surrounding star-forming environment can contribute to asymmetric accretion and streamer-like structures.

In contrast, observations have also identified flattened, disk-like envelope structures --- often interpreted as pseudodisks --- surrounding circumstellar disks \citep{Lee14,Lee2019}. 
These findings indicate that the infalling envelope (or pseudodisk) plays a crucial role in regulating how gas is supplied to disks and protostars. 
To interpret such observations, it is therefore also important to determine whether asymmetric accretion can develop within an individual collapsing core, even in the absence of turbulence and continuous gas supply from larger scales.

From a theoretical perspective, stars form through the gravitational collapse of molecular cloud cores. 
As collapse proceeds, the central density increases, and the characteristic free-fall timescale decreases as $t_{\rm ff} \propto \rho^{-1/2}$. 
When the central number density reaches $n \sim 10^{10}$--$10^{11}\,\cc$, the gas becomes optically thick, and a hydrostatic object known as the first adiabatic core forms. 
As the density within the first core gradually increases, molecular hydrogen begins to dissociate at $n \simeq 10^{16}\,\cc$, triggering the second collapse. 
Once the central density exceeds $\sim 10^{22}\,\cc$, the gas again becomes adiabatic, halting further collapse and leading to the formation of a protostar with an initial mass of order $10^{-3}\,M_\odot$ \citep{larson69,masunaga00}.
After protostar formation, the surrounding lower-density gas does not evolve as rapidly as the central region but continues to fall inward, thereby forming an infalling envelope \citep{machida10,machida11,machida13,machida24,Machida2026}.
Observations further indicate that prestellar cores possess magnetic energies comparable to their gravitational energies, implying that magnetic fields play a significant role in collapse and accretion processes \citep{crutcher99,crutcher10}.
As magnetized cloud core collapses, the Lorentz force produces a flattened, disk-like structure known as a pseudodisk \citep{Galli1993}.
After protostar formation, the pseudodisk surrounds the circumstellar disk and continues to supply gas inward.
Although the pseudodisk is a disk-like structure shaped by magnetic fields, the magnetic field does not completely suppress gravitational collapse, and gas within the pseudodisk continues to accrete inward \citep[e.g.,][]{machida05a,Machida2006ApJ}.
Thus, in the vicinity of the disk, the pseudodisk and the infalling envelope 
represent partially overlapping descriptions of the accreting gas \citep{Hirano2019}.
The pseudodisk can be viewed as a magnetically shaped component of the infalling envelope, although the infalling envelope itself is not limited to the pseudodisk.

Understanding the evolution of the infalling envelope is therefore essential for clarifying how protostars and disks grow and how asymmetric accretion structures such as streamers arise.
Many numerical studies of star formation have adopted idealized initial conditions in which the magnetic field and angular momentum vectors are aligned \citep[see review by][]{tsukamoto23b}.
In simple aligned-core models, the disk, pseudodisk, and outflow tend to share a common axis, and strongly asymmetric accretion does not readily emerge.
In reality, both observations and theoretical studies suggest that magnetic fields and angular momentum vectors can be misaligned in star-forming cores \citep{Hull2013,Misugi2024}.
Previous numerical studies have investigated the effects of such misalignment on the collapse of magnetized cores, demonstrating that it can significantly affect magnetic braking, disk formation, and the relative orientations of the magnetic field, disk, and angular momentum vectors \citep{Matsumoto2004,Joos2012,Li2013,Tsukamoto2018}.

Although previous studies have investigated the effects of misalignment, relatively few studies have followed the long-term evolution after protostar formation with sufficient spatial resolution to resolve circumstellar disks \citep{machida14, Machida2020d, Matsumoto17}.
In particular, the morphology and time evolution of infalling envelopes or pseudodisks in misaligned configurations remain largely unexplored. 
This motivates a systematic examination of envelope evolution in well-resolved, isolated core models.

\cite{Machida2026} demonstrated that asymmetric accretion and streamer-like flows can arise from misalignment between the large-scale magnetic field and the angular momentum vector, even in the absence of strong turbulence or external inflow.
However, that study focused on a single misalignment angle of $30^\circ$ and adopted a relatively large prestellar core.
Recently, \citet{Ahmad2026} performed simulations with a similar setup to ours.
In their study, they adopted an initial misalignment between the magnetic field and the rotation axis (with a misalignment angle of $10^\circ$) and performed simulations without imposing turbulence. 
The spatial resolution and the simulation time after protostellar formation were comparable to 
those in our simulations. 
They showed that this misalignment leads to asymmetric accretion that drives a turbulent state in the central region.
In particular, they demonstrated that gas inflow from above the disk is non-negligible and can produce an eccentric disk.
They mainly investigated the impact of asymmetric accretion on disk formation and evolution.
In contrast, \citet{Machida2026} focused on the evolution of the envelope and pseudodisk outside the disk.
However, in both studies, the explored parameter space was limited to a single model.

In this study, we systematically vary the initial misalignment angle between the magnetic field and the angular momentum vector and consider prestellar cores of different sizes.
Our goal is to determine how common the deformation of pseudodisks and infalling envelopes is, and how robust asymmetric accretion and the formation of streamer-like structures are across parameter space.
To facilitate comparisons with future observations, we also investigate the orientations of gas streams relative to the magnetic field and the velocities of infalling gas.
Ultimately, we show that even in an isolated core without initial turbulence or continuous gas supply from larger scales, complex infalling envelopes and streamer-like accretion flows can arise from the misalignment between the magnetic field and rotation axis.

The structure of this paper is as follows: 
Models and numerical settings are described in Section 2, and simulation results are presented in Section 3. 
We discuss diagnostics relevant to observations and the possible spatial scale of asymmetric accretion in Section 4. 
A summary is presented in Section 5.


\section{Models and Numerical Settings}
The numerical settings and initial conditions used in this study are nearly identical to those described in \citet{Machida2026}.
Therefore, we provide only a brief summary here.

While \citet{Machida2026} presented results only for a single model (T30L), the present paper includes a total of 12 models, including an unmagnetized model (B00), to investigate the dependence on the initial misalignment angle between the magnetic field and the rotation axis.
We consider two types of prestellar cores with different sizes as the initial conditions. 
Each prestellar core follows a critical Bonnor-Ebert (B.E.) density profile, characterized by the central density $n_{\rm c}$ and the isothermal temperature $T_{\rm iso}$. 
We fix the isothermal temperature at $T_{\rm iso} = 10$\,K and adopt different central number densities of $n_{\rm c} = 10^4\,\cc$ and $10^6\,\cc$. 

For $n_c = 10^6\,\cc$, the critical B.E. radius is $R_{\rm cl} = 4,800$\,au, while for $n_{\rm c} = 10^4\,\cc$, it is $R_{\rm cl}=48,000$\,au.  
In all models, the density of the entire core is enhanced by a factor of $f_{\rm BE}=1.68$ to promote gravitational collapse \citep{machida13}. 
The mass of the prestellar core (or the B.E. sphere) is $M_{\rm cl}=0.81\,\msun$ for $n_{\rm c} = 10^6\,\cc$ and $M_{\rm cl}=8.1 \msun$ for $n_{\rm c} = 10^4\,\cc$.
Hereafter, we refer to the model with $n_{c} = 10^6\,\cc$ as the \textit{small-core model}, and the one with $n_{\rm c} = 10^4\,\cc$ as the \textit{large-core model}.

A uniform magnetic field $B_0$ is imposed over the entire computational domain, and each prestellar core rotates rigidly with an angular velocity $\Omega_0$. 
To reduce the number of free parameters, we adopt the dimensionless parameters $\alpha_0$, $\beta_0$, and $\gamma_0$ for both the small and large-core models \citep{Matsushita17,matsushita2018}. 
These correspond to the initial ratios of thermal, rotational, and magnetic energy to gravitational energy, respectively. 
In all models, we adopt $\alpha_0 = 0.5$, $\beta_0 = 0.02$, and $\gamma_0 = 0.1$.
The initial mass-to-flux ratio $\mu_0$, normalized by the critical value $(2 \pi G^{1/2})^{-1}$, is also fixed to $\mu_0 = 3.0$ in all models. 
Because the energy ratios are fixed but the core masses differ between the small and large-core models, 
$B_0$ and $\Omega_0$ are uniquely determined for each core size.
For the small-core models, $B_0=55\,\mu{\rm G}$ and $\Omega_0=2.6\times10^{-13}\,{\rm s}^{-1}$, whereas for the large-core models $B_0=5.5\,\mu{\rm G}$ and $\Omega_0=2.6\times10^{-14}\,{\rm s}^{-1}$.

In addition to the central number density $n_{\rm c}$, we treat the angle between the magnetic field and the rotation axis $\theta_0$ as a model parameter \citep{Machida2006ApJ}.  
All simulations are performed in Cartesian coordinates. 
We align the magnetic field with the $z$-axis and tilt the rotation axis from the $z$-axis toward the $x$-axis by $\theta_0$.
We consider $\theta_0 = 0^\circ, 5^\circ, 10^\circ, 30^\circ, 45^\circ, 60^\circ, 80^\circ$, and $90^\circ$.
Table~\ref{table:1} summarizes the model parameters ($\theta_0$, $B_0$, $\Omega_0$, $M_{\rm cl}$, and $R_{\rm cl}$) and the naming convention: the two-digit number following ``T'' indicates $\theta_0$ in degrees, and the suffix ``L'' denotes the large-core models. 
Models without the suffix ``L'' correspond to the small-core models.
We include the unmagnetized model B00, which has $B_0 = 0$, and therefore $\theta_0$ is not defined.
In addition, we include the large-core model T30L, which was used in \citet{Machida2026}, as a reference model.

We solve the resistive MHD equations with self-gravity using our nested-grid code \citep{machida04,machida05a,machida06,machida10,Machida2014ApJL,machida16,Machida2020d,machida24,Basu24}.
We include only Ohmic dissipation as a non-ideal MHD effect and do not include ambipolar diffusion or the Hall effect.
The possible effects of ambipolar diffusion and the Hall effect are discussed in \S~\ref{sec:nonidealMHD}.
The basic equations solved are the same as equations [1]--[5] in \citet{Basu24}.
Each grid is composed of $(i, j, k) = (64, 64, 64)$ cells at each level. 
We denote the box size and cell width of the $l$-th grid level by $L(l)$ and $h(l)$, respectively.
The small-core models use 13 levels ($l_{\rm max} = 13$), while the large-core models use 16 levels ($l_{\rm max} = 16$), where $l_{\rm max}$ is the maximum grid level.
In both models, the prestellar core is fully contained within the $l=5$ grid. 
Therefore, the box size of the $l=5$ grid, denoted by $L(5)$, corresponds to twice the critical B.E. radius.
For the large-core models, $L(5) = 9.5 \times 10^4$\,au and $h(5) = 1500$\,au. 
The coarsest grid ($l=1$) has $L(1) = 1.5 \times 10^6$\,au and $h(1) = 2.4 \times 10^4$\,au, while the finest grid ($l=16$) has $L(16) = 46$\,au and $h(16) = 0.74$\,au.
For the small-core models, $L(5) = 9.5 \times 10^3$\,au and $h(5) = 150$\,au. 
The largest grid has $L(1) = 1.5 \times 10^5$\,au and $h(1) = 2.3 \times 10^3$\,au, while the smallest grid reaches $L(13) = 37$\,au and $h(13) = 0.58$\,au.
Thus, the highest spatial resolution is slightly finer in the small-core models than in the large-core models.
We initialize the grid hierarchy up to level $l=5$ at $t=0$, and we subsequently introduce finer levels so that the local Jeans length is always resolved by at least 16 cells (Truelove criterion).

As in our previous studies \citep{machida12}, the density outside the B.E. radius ($r > r_{\rm cl}$) is set to be uniform, and gas inflow from this region is prohibited.
However, outflow from the central region to the outer region ($r > r_{\rm cl}$) is allowed \citep{Nozaki2025}. 
The ambient medium outside the critical B.E. sphere ($r>r_{\rm cl}$) mimics the interstellar medium \citep{Nozaki2023}.
To prevent artificial reflection of Alfv\'en waves generated near the center, a large surrounding region is included beyond the prestellar core \citep{machida13}.

When the central density exceeds $n_{\rm sink} = 10^{13}\,\cc$, a sink cell is introduced \citep{machida10,machida20}. 
After sink formation, we remove 1\% of the gas with $n > n_{\rm sink}$ within $r_{\rm sink} = 1$\,au and add it to the central point mass that acts as a gravitational source, following \citet{Machida2026}.
Using these setups, we performed simulations for each model listed in Table~\ref{table:1} up to approximately $10^5$\,yr after protostar formation.

\begin{table}[htbp]
\begin{center}
\begin{tabular}{ccccccc} 
    \hline
    Model & $\theta_0$ [$^\circ$] & $n_{\rm c}$ [$\cc$] & $B_{\rm 0}$ [$\mu\,{\rm G}$] & $\Omega_{\rm 0}$ [$10^{-14}\,{\rm s}^{-1}$]  & $M_{\rm cl}$ [$\msun$] & $R_{\rm cl}$ [au] \\ 
    \hline   
    T00  & 0    &$10^6$ & 55  & 25  & 0.81 & 4,800  \\
    T05  & 5    &$10^6$ & 55  & 25  & 0.81 & 4,800  \\ 
    T10  & 10   &$10^6$ & 55  & 25  & 0.81 & 4,800  \\
    T30  & 30   &$10^6$ & 55  & 25  & 0.81 & 4,800  \\
    T45  & 45   &$10^6$ & 55  & 25  & 0.81 & 4,800  \\
    T60  & 60   &$10^6$ & 55  & 25  & 0.81 & 4,800  \\
    T80  & 80   &$10^6$ & 55  & 25  & 0.81 & 4,800  \\
    T90  & 90   &$10^6$ & 55  & 25  & 0.81 & 4,800  \\
    T30L & 30   &$10^4$ & 5.5 & 2.5 & 8.1  & 48,000 \\
    T60L & 60   &$10^4$ & 5.5 & 2.5 & 8.1  & 48,000 \\
    T80L & 80   &$10^4$ & 5.5 & 2.5 & 8.1  & 48,000 \\
    B00  & (30) &$10^6$ & 0   & 25  & 0.81 & 4,800  \\
    \hline
\end{tabular}
\caption{
Model name and parameters.
Columns list the initial misalignment angle $\theta_0$, central number density $n_{\rm c}$, magnetic field strength $B_0$, angular velocity $\Omega_0$, cloud mass $M_{\rm cl}$, and core radius $R_{\rm cl}$.
For model B00, the value of $\theta_0$ in parentheses only specifies the orientation of the rotation axis tilted from the $z$-axis toward the $x$-axis, since the model has no magnetic field.
}
\label{table:1}
\end{center}
\end{table}

\section{Results}
\label{sec:results}
\begin{figure*}
\centering
\includegraphics[width=1.0\columnwidth]{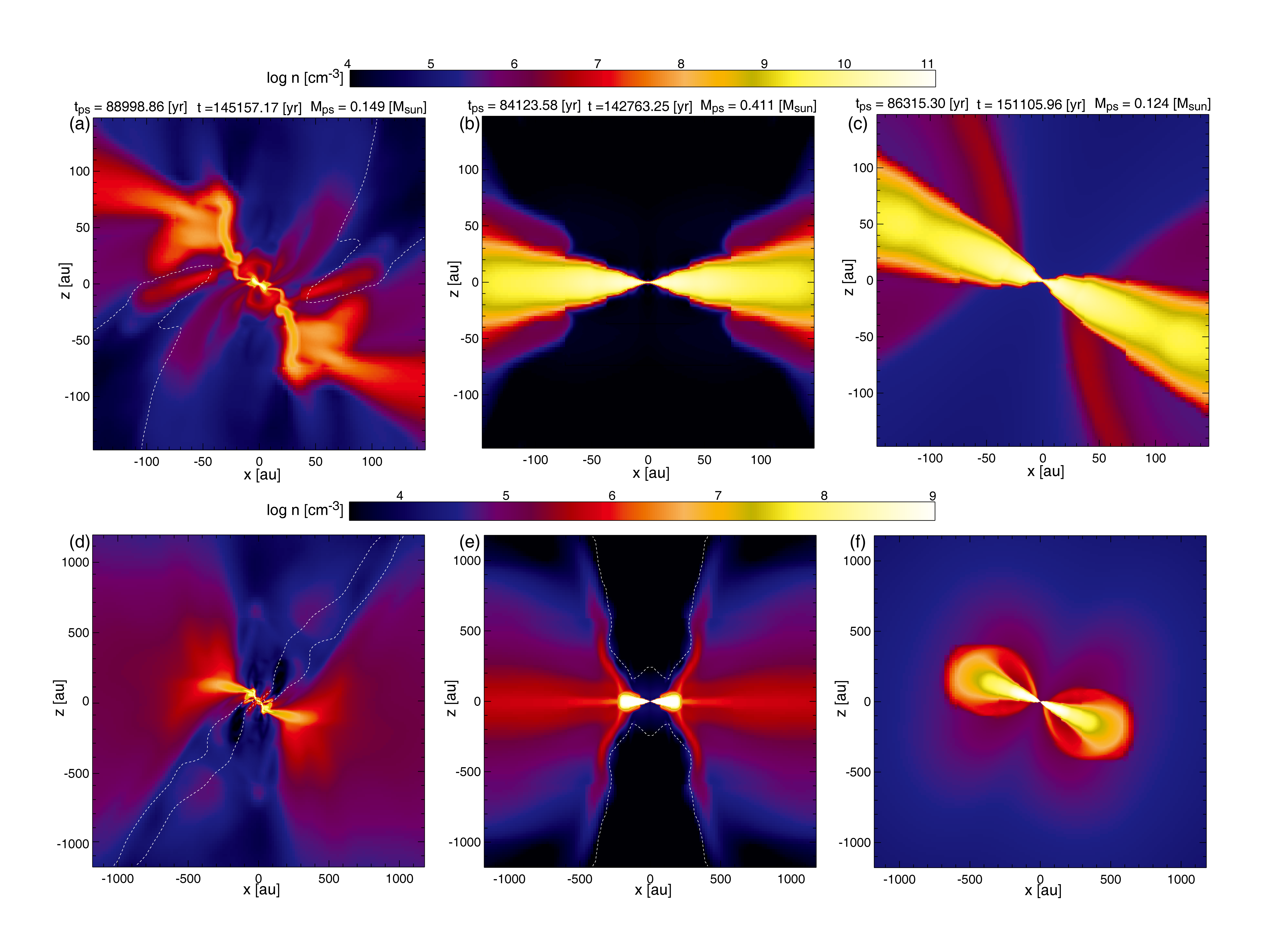}
\caption{
Density slices on the $y=0$ plane at two scales, 300\,au (top) and 2,000\,au (bottom), for models T60 (left), T00 (center), and B00 (right) at $t_{\rm ps}\simeq8\times10^{4}$\,yr.
Each panel lists elapsed time after protostar formation $t_{\rm ps}$, elapsed time after the prestellar cloud begins to collapse $t$, and protostellar mass $M_{\rm ps}$.
The white dotted curve in panels (a), (d), and (e) indicates the boundary between the inflowing ($v_r <0 $) and outflowing $(v_r > 0)$ regions.
}
\label{fig:1}
\end{figure*}

As shown in our previous studies \citep{Machida2006ApJ,Hirano2019,hirano20,Machida2020d,machida21,Machida2026}, the misalignment between the global magnetic field and the rotation axis of the prestellar core causes the structure of the infalling envelope or pseudodisk to undergo significant deformation over time.
Figure~\ref{fig:1} shows the density distribution on the $y=0$ plane at two different scales, 300\,au (top panels) and 2,000\,au (bottom panels), $\sim 8 \times 10^4$\,yr after protostar formation, for three representative models (T60, T00, and B00).

In the initial state, magnetic fields are aligned with the rotation axis for model T00 (the middle panels of Figure~\ref{fig:1}). 
In this model, the gas entrained by outflow and an outflow cavity can be seen along the $z$-axis around $x \approx \pm 500$\,au (Figure~\ref{fig:1}e). 
Apart from this cavity, the overall structure is disk-like, as seen in Figures~\ref{fig:1}(b) and (e).
The high-density region near the center in Figure~\ref{fig:1}(b) corresponds to the rotationally supported disk, while the outer high-density region at $|x| \gtrsim 300$\,au in Figure~\ref{fig:1}(e) corresponds to the pseudodisk. 
In both regions, the disk planes are flat and undistorted, with the same normal direction (along the $z$-axis) at all radii.
Figures~\ref{fig:1}(c) and (f) show the unmagnetized model B00. 
In this model, the rotation axis is inclined by $30^\circ$ from the $z$-axis toward the $x$-axis in Cartesian coordinates. 
As a result, the disk formed in this coordinate frame is also tilted. 
However, the disk remains flat and undistorted, with a uniform normal direction across all radii.
In this model, no outflow is driven from either the disk or the high-density region, while the relatively dense shell-like structure above the disk is supported by gas pressure and rotation.

In contrast, the misaligned model T60, shown in Figures~\ref{fig:1}(a) and (d), shows a drastic radial change in the disk orientation. 
A disk-like structure is visible in the central region of Figure~\ref{fig:1}(a), but the surrounding region does not show a clear rotationally supported disk or pseudodisk structure, as seen in Figures~\ref{fig:1}(b) and (e).  Instead, it exhibits a highly complex morphology.
The central region in Figure~\ref{fig:1}(a) does form a thin disk, but its orientation (i.e., normal direction) varies significantly with spatial scale \citep[see also ][]{Hirano2019, hirano20, Machida2020d}. 
The outflow, traced by the white dotted line in Figures~\ref{fig:1}(a) and (d), shows a significantly more complicated propagation path compared to the aligned case in Figure~\ref{fig:1}(e).

The disk and envelope structures seen in model T00 (aligned model) and B00 (unmagnetized model) are frequently observed in idealized simulations \citep[see review by][]{tsukamoto23b}, but they may not be realistic. 
In reality, magnetic fields should be considered, and perfect alignment between the magnetic field and rotation axis is unlikely. 
Therefore, misaligned models may better represent actual star-forming environments.

In this section, we first show the spatial and temporal variation of the disk and its surrounding structure for the typical misaligned small-core model T60. 
We then examine the time evolution of the envelope morphology and physical quantities for small-core models with different misalignment parameters, including T00. 
Finally, we present the results of the large-core models.

\subsection{Typical Model}
\begin{figure*}
\centering
\includegraphics[width=0.90\columnwidth]{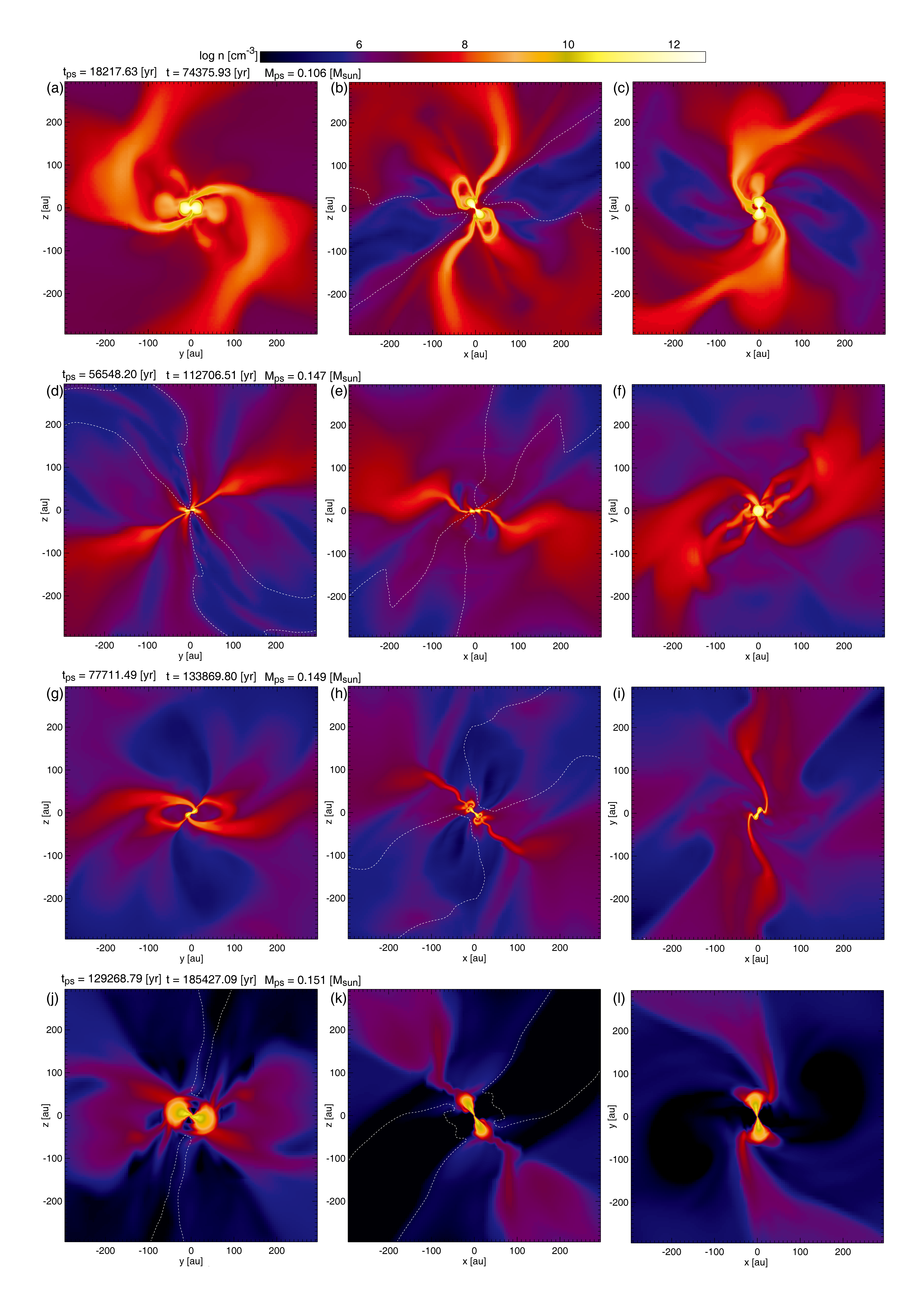}
\caption{
Time sequence of density slices for model T60 on the $x=0$ (left), $y=0$ (center), and $z=0$ (right) planes.
Each left panel lists $t_{\rm ps}$, $t$, and $M_{\rm ps}$.
The white dotted curve in panels (b), (d), (e), (h), (j), and (k) indicates the boundary between the inflowing ($v_r <0$) and outflowing $(v_r > 0)$ regions.
(
Animation: Time evolution of the density and velocity distributions in the $x=0$, $y=0$, and $z=0$ planes for model T60. 
The elapsed time after protostar formation, $t_{\rm ps}$, the elapsed time since the beginning of the simulation, $t$, and the protostellar mass $M_{\rm ps}$ are indicated. 
The animation duration is 3\,s. 
The animation shows that the orientation of the pseudodisk normal and the outflow direction change with time.
)
}
\label{fig:2}
\end{figure*}

\begin{figure*}
\centering
\includegraphics[width=0.90\columnwidth]{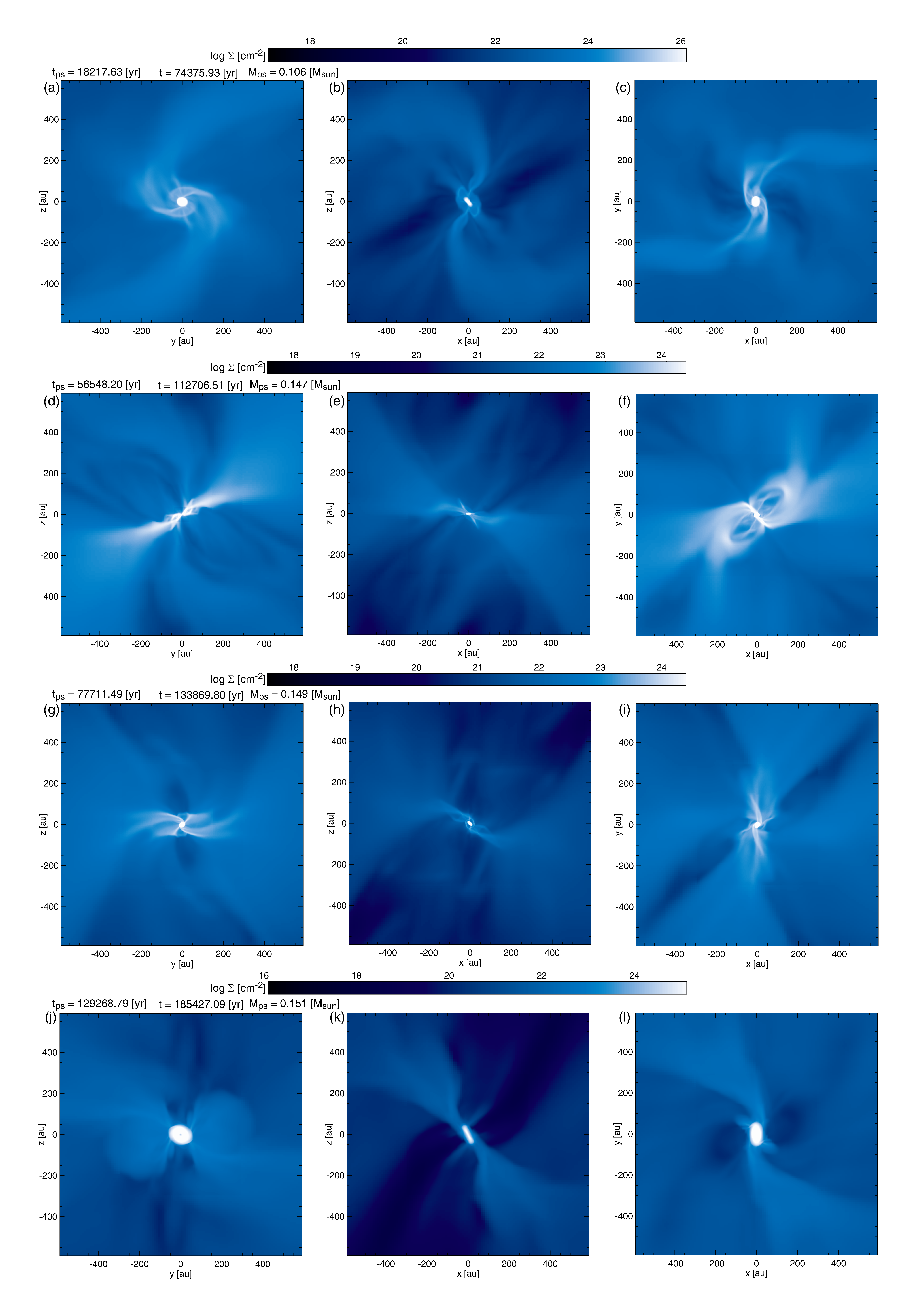}
\caption{
Same as Figure~\ref{fig:2}, but showing the surface density integrated along the line of sight. 
(Animation: Time evolution of the surface density distributions in the $x=0$, $y=0$, and $z=0$ planes for model T60. 
The elapsed time after protostar formation, $t_{\rm ps}$, the elapsed time since the beginning of the simulation, $t$, and the protostellar mass, $M_{\rm ps}$, are indicated. 
The animation duration is 3\,s. 
The animation shows that the orientation of the pseudodisk changes and the surface density of the envelope decreases with time.
)
}
\label{fig:3}
\end{figure*}

Figure~\ref{fig:2} shows the time evolution of the density distribution for model T60 on the $x=0$ (left), $y=0$ (center), and $z=0$ (right) cutting planes. 
The spatial scale is approximately 600\,au.  
Figure~\ref{fig:3} presents the time evolution of the integrated density (i.e., surface density) along the $x$- (left), $y$- (center), and $z$-axes (right). 
The spatial scale of the panels in Figure~\ref{fig:3} is approximately 1,200\,au.  
In each figure, time progresses from top to bottom, and the time epochs in Figures~\ref{fig:2} and \ref{fig:3} are identical.

In Figures~\ref{fig:2}(a)--(c), the yellow regions correspond roughly to the rotationally supported disk. 
At this epoch ($t_{\rm ps}=18217.63$\,yr), the pseudodisk is already distorted by rotation, forming spiral-like structures around the circumstellar disk.  
In Figures~\ref{fig:3}(a)--(c), the central white regions correspond to the circumstellar disk. 
These panels indicate that, while the disk itself maintains an almost circular shape, the surrounding structures are complex and the gas is distributed asymmetrically.  
The spiral structures seen in Figures~\ref{fig:2}(a)--(c) can also be identified beyond 1,000\,au as seen in Figures~\ref{fig:3}(a)--(c).

In Figures~\ref{fig:2}(d) and \ref{fig:3}(d), clear cavity-like structures produced by outflows can be seen toward the upper left and lower right relative to the center.
Figures~\ref{fig:2}(e) and \ref{fig:3}(e) show that the pseudodisk or infalling envelope surrounding the central rotationally supported disk is highly warped or undulating. 
The outflow propagates toward the upper right and lower left in these panels. 
In Figure~\ref{fig:3}(e), the slightly denser gas located above and below the center corresponds to outflowing gas ejected from the central region.
In Figures~\ref{fig:2}(f) and \ref{fig:3}(f), hole-like features appear toward the upper right and lower left from the center.
These are likely formed by the twisting of a very thin disk-like pseudodisk \citep[see also][]{Ahmad2026}. 
Since the formation process of each structure is beyond the scope of this paper, we do not discuss it in detail.  
Figures~\ref{fig:2}(g)--(i) show that gas flows into the disk from particular directions; accretion onto the disk does not necessarily occur solely from the side or directly from above. 
As seen in Figure~\ref{fig:2}(h), vertical accretion from above and below the disk may be partially suppressed by the outflow, whereas in Figure~\ref{fig:2}(g), gas accretion onto the outer edge of the disk occurs from above. 
Figures~\ref{fig:3}(g)--(i) also demonstrate that the gas distribution at this stage is non-uniform, which is partly due to the presence of low-density regions along the outflow direction (Figure~\ref{fig:3}(h)).

Figures~\ref{fig:2}(j)--(l) and \ref{fig:3}(j)--(l) show the structures at the end of the simulation. 
Compared to Figures~\ref{fig:3}(a)--(c), the circumstellar disk in Figures~\ref{fig:3}(j)--(l) has grown significantly.  
Figures~\ref{fig:2}(j) and (k) reveal that, because the outflow is active, there is almost no accretion from above the disk. 
At this stage, accretion occurs mainly from the lateral (midplane) direction of the disk. 
More precisely, at the disk edge, gas also accretes vertically from above (Figure~\ref{fig:2}(l)). 
Figure~\ref{fig:2}(k) also shows that the gas density is low along the outflow direction, while a (distorted) disk-like envelope exists in the direction perpendicular to the outflow.

Both Figures~\ref{fig:2} and \ref{fig:3} illustrate the time evolution of the envelope or pseudodisk surrounding the circumstellar disk over the first $\sim 10^5$\,yr after protostar formation. 
These figures show that the envelope morphology changes significantly over time, and the accretion pattern onto the disk is also expected to vary substantially with time.

\begin{figure*}
\centering
\includegraphics[width=1.0\columnwidth]{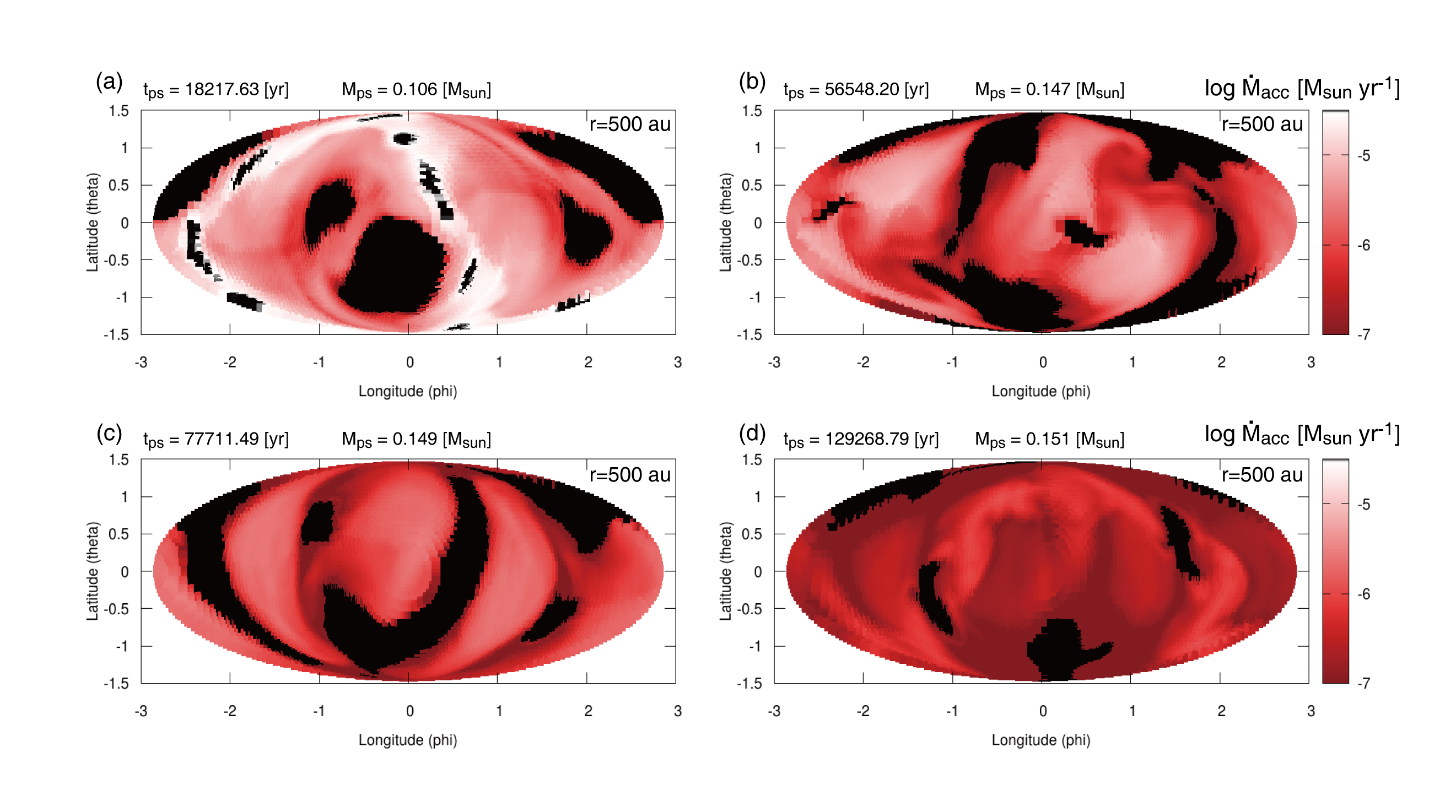}
\caption{
Mass accretion rate map on a spherical surface of radius 500\,au for model T60 at the same epochs as Figure~\ref{fig:1}.
Black regions indicate outflow ($v_r > 0$).
The vertical and horizontal axes represent latitude ($-\pi/2 < \theta < \pi/2$) and longitude ($-\pi < \phi < \pi$) in radians, respectively.
}
\label{fig:4}
\end{figure*}

Figure~\ref{fig:4} shows the mass accretion rate on a spherical surface of radius 500\,au at each epoch corresponding to Figures~\ref{fig:2} and \ref{fig:3}. 
For clarity, each point in the accretion-rate map is converted to the total value integrated over the full sphere (for details, see \citealt{Machida2026}).  
In each panel, the black regions correspond to outflow regions with $v_r>0$, while the other colors represent inflow regions with $v_r<0$. 
At $\sim 2\times 10^4$\,yr after protostar formation (Figure~\ref{fig:4}(a)), bright regions (white or light red) dominate, indicating mass accretion rates exceeding $10^{-5}\,\msunyr$. 
In contrast, at the end of the simulation (Figure~\ref{fig:4}(d)), darker red regions dominate, corresponding to mass accretion rates of less than $10^{-6}$--$10^{-7}\,\msunyr$.  
Figure~\ref{fig:4} also shows that inflow and outflow regions are complexly interwoven
and that their directions change with time.
Furthermore, the outflow region in Figure~\ref{fig:4}(d) is narrower than in Figure~\ref{fig:4}(a), indicating that the outflow weakens on the 500\,au scale.

\begin{figure*}
\centering
\includegraphics[width=1.0\columnwidth]{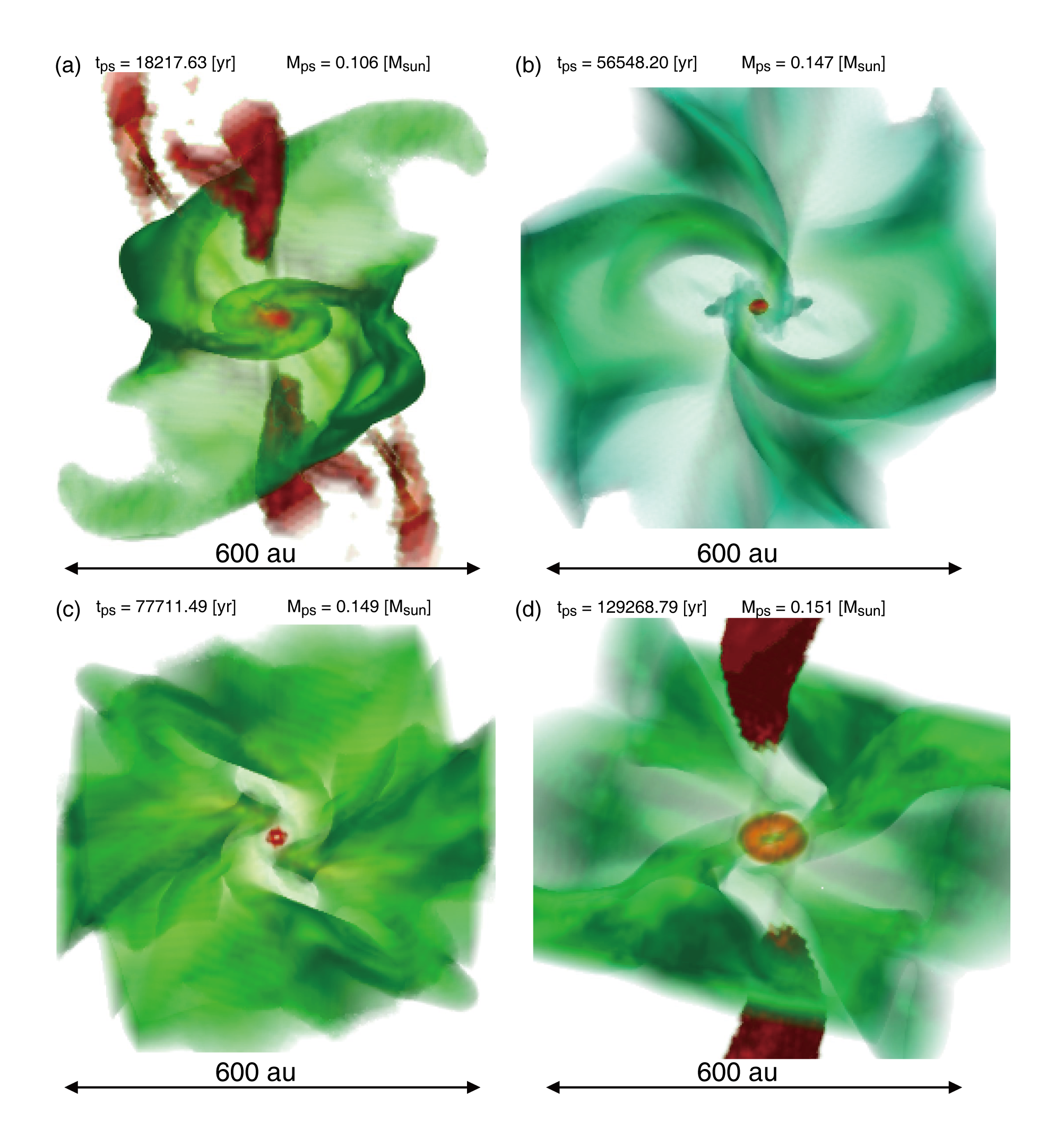}
\caption{
Volume rendering of the density distribution around the central region for model T60.
Colors indicate the envelope (green), disk (orange), and outflow (red; shown only in panels (a) and (d)).
Each panel lists $t_{\rm ps}$, $M_{\rm ps}$, and the spatial scale.
The viewing angle in each panel is adjusted to clearly show the central structure.
}
\label{fig:5}
\end{figure*}

Figure~\ref{fig:5} displays the density structures at each epoch in Figures~\ref{fig:2}--\ref{fig:4} as viewed via ray tracing from the center. 
The density decreases from red to green; the red regions above and below the disk in Figures~\ref{fig:5}(a) and (d) correspond to the outflows
\footnote{
In this volume rendering, the color scale transitions from red to green with decreasing density. 
The circumstellar disk and part of the outflow appear in red because they correspond to regions of high density ($>10^8\,\cc$). 
No explicit density threshold or additional selection criteria are imposed to identify the disk or outflow.
}. 
In Figures~\ref{fig:5}(b) and (c), the outflow is not visible because its density was temporarily comparable to or lower than that of the surrounding gas, implying that the outflow is driven in a non-steady manner (for details, see \S~\ref{sec:parameters}).  
Figure~\ref{fig:5}(a) also shows that gas flows from the envelope to the disk along spiral-like streams. 
In Figures~\ref{fig:5}(b) and (d), the inflowing gas toward the center is channeled through multiple streams. 
In Figure~\ref{fig:5}(c), the overall structure is complex, and the high-density streams are not directly connected to the disk, suggesting that accretion from the envelope to the disk is temporarily reduced at this epoch.

\begin{figure*}
\centering
\includegraphics[width=0.8\columnwidth]{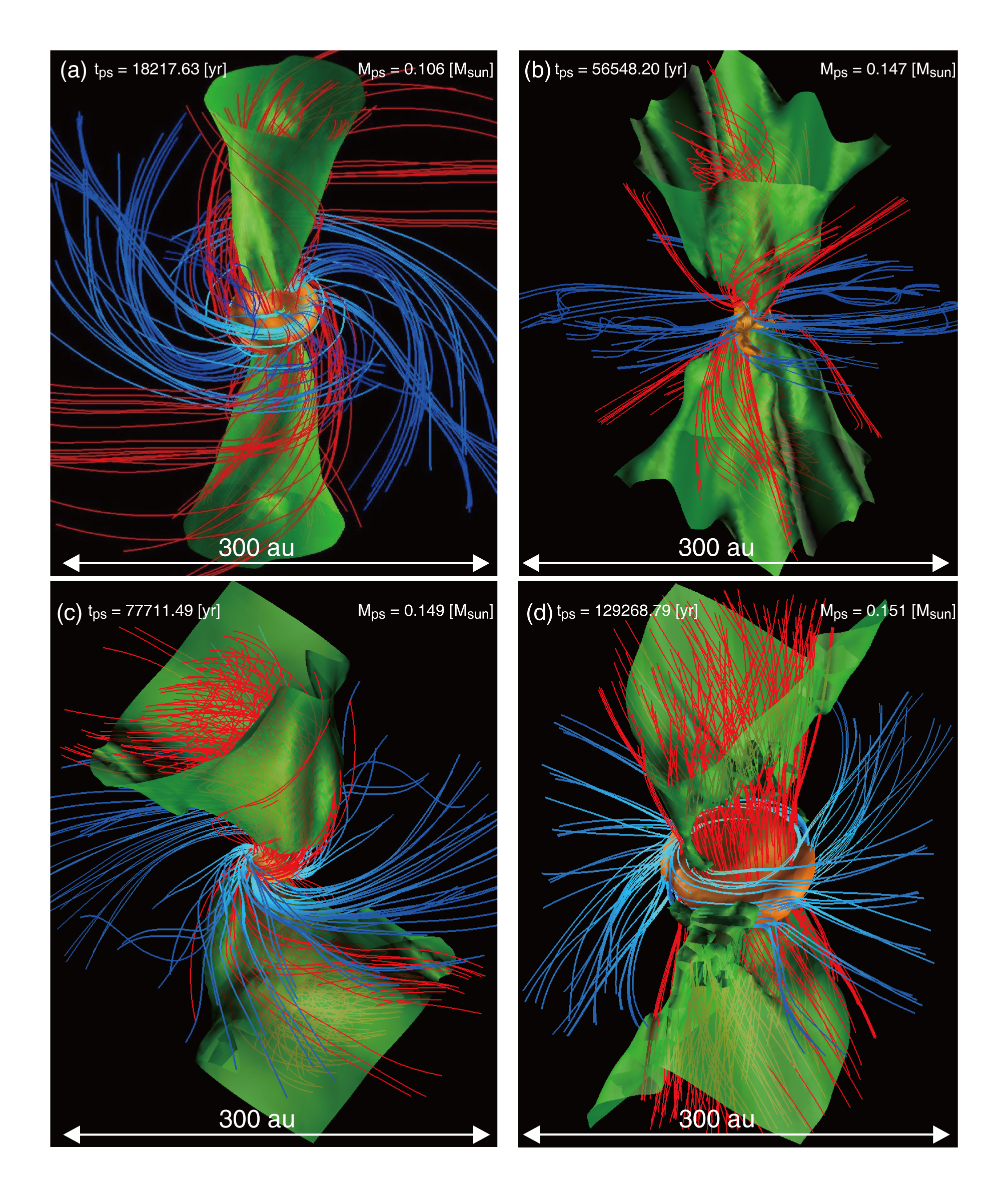}
\caption{
Three-dimensional structures for model T60 at the same epochs as Figure~\ref{fig:2}: disk (orange surface), outflow (green surface), inflow (blue lines), and magnetic field (red lines).
Each panel lists $t_{\rm ps}$, $M_{\rm ps}$, and the spatial scale.
The viewing angle in each panel is adjusted to clearly show the central structure.
For simplicity, the outflow is defined as the region with a radial velocity exceeding the sound speed ($v_r > c_{\rm s}$), while the disk is defined as the non-outflow region with densities greater than $10^9\,\cc$ (for details, see \S\ref{sec:parameters} and \S\ref{sec:A1}).
}
\label{fig:6}
\end{figure*}

Figure~\ref{fig:6} plots the three-dimensional structure of the high-density regions, outflows, magnetic field lines, and accretion streams for each epoch. 
The magnetic field lines are integrated from the high-density regions ($>10^8\,\cc$), showing only areas with strong fields. 
The accretion streams are traced from the regions on each surface of the $l=10$ box, for which the box size is $L(10)=300$\,au, where the accretion rate is high, following the method of \citet{Machida2026}.
Thus, only the streams with high accretion rates are shown.  
Figure~\ref{fig:6} shows that the high-density region forms a disk-like structure, with outflows launched roughly along the vertical direction from the disk plane. 
Gas flows toward the central region primarily from directions nearly perpendicular to the outflow axis, but through several distinct accretion bands.  
Although it is difficult to investigate the relative orientations of the magnetic field lines and streams from this figure, the magnetic field is typically neither perfectly parallel nor perfectly perpendicular to the accretion stream.
The relation between the inflow direction and the magnetic field orientation is discussed in \S~\ref{sec:angleVB}.

\subsection{Models with Different Parameters}
\label{sec:parameters}
\begin{figure*}
\centering
\includegraphics[width=0.9\columnwidth]{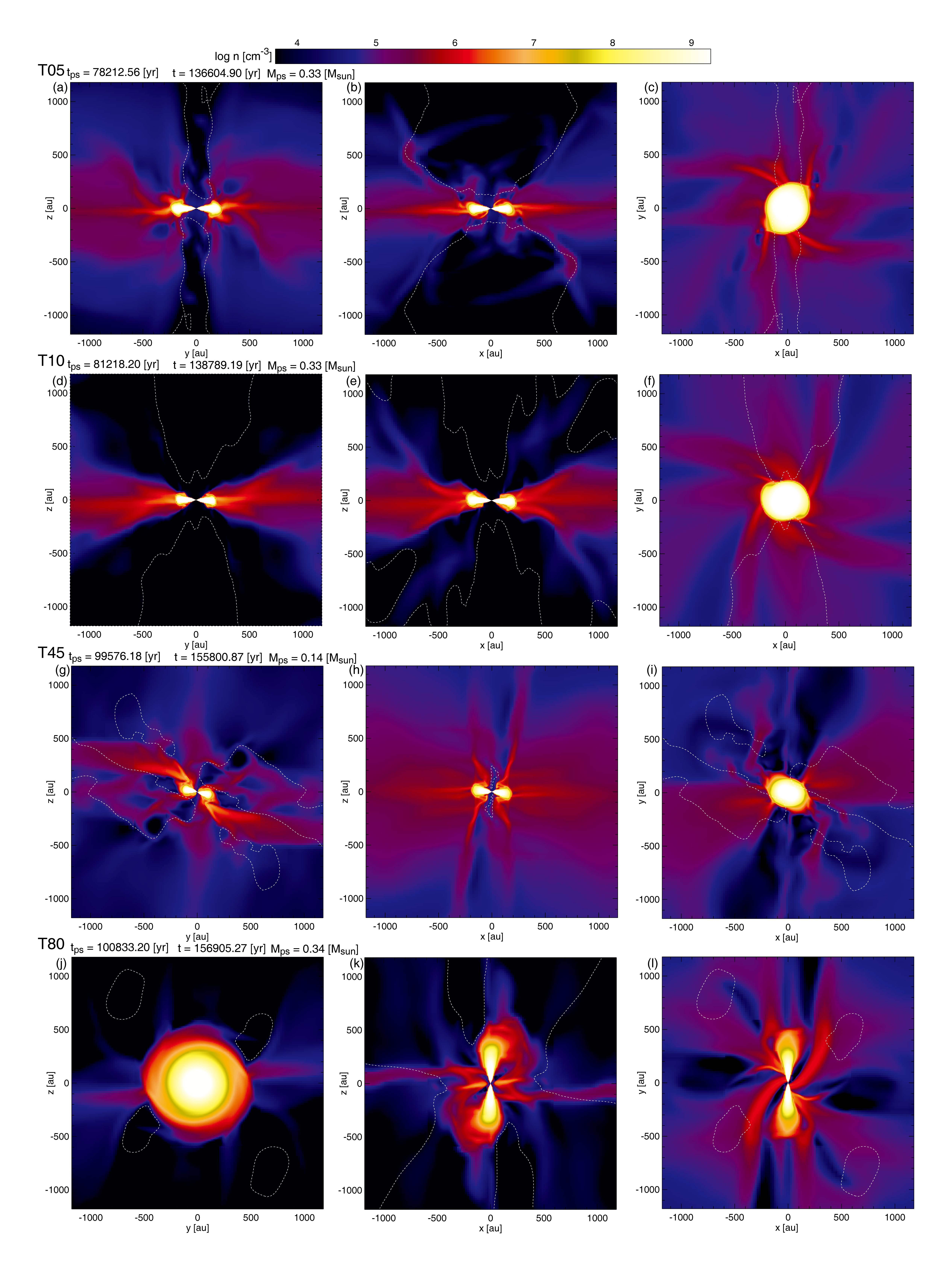}
\caption{
Density slices on the $x=0$ (left), $y=0$ (center), and $z=0$ (right) planes for models T05 (a--c), T10 (d--f), T45 (g--i), and T80 (j--l).
Each leftmost panel lists the model name, $t_{\rm ps}$, $t$, and $M_{\rm ps}$.
The white dotted curve marks the boundary between inflow ($v_r<0$) and outflow ($v_r>0$).
}
\label{fig:7}
\end{figure*}

In this subsection, we investigate the dependence of the development of asymmetries and the structure of the accretion streams on the initial parameter $\theta_0$.
We describe the results for small-core models T00, T05, T10, T30, T45, T60, T80, T90, and B00.
Figure~\ref{fig:7} plots the density distributions on the $x=0$ (left), $y=0$ (center), and $z=0$ (right) planes at the end of the simulation for models T05, T10, T45, and T80. 
In each panel, the yellow--white regions indicate high-density areas that roughly correspond to the circumstellar (rotationally supported) disk. 
The white dotted curve marks the boundary between inflowing ($v_r<0$) and outflowing ($v_r>0$) regions,
and the enclosed area corresponds to the outflow region.
Model T05 (Figures~\ref{fig:7}(a)--(c)) is similar to the aligned model T00 (Figures~\ref{fig:1}(b) and (e)).
The disk and the pseudodisk (disk-like infalling envelope) are roughly aligned, and the outflow is launched vertically from the disk.
However, Figure~\ref{fig:7}(a) shows that for $|y| > 500$\,au, the disk is significantly distorted, forming a spur-like structure.
In addition, near the equatorial plane, gas accretes onto the disk from multiple, well-separated azimuthal directions (Figure~\ref{fig:7}c). 
These results indicate that even a small difference of $\theta_0 = 5^\circ$ between the rotation axis and magnetic field can substantially change the structure of the infalling envelope or the circumstellar environment compared with the aligned case ($\theta_0=0^\circ$, model T00).

Model T10, shown in Figures~\ref{fig:7}(d)--(f), also exhibits a significantly distorted pseudodisk or infalling envelope, similar to model T05. 
As in model T05, gas accretes onto the disk in the equatorial plane from several specific azimuthal directions (Figure~\ref{fig:7}(f)).
In model T45, shown in Figures~\ref{fig:7}(g)--(i), the infalling envelope (or pseudodisk) has a highly complex morphology. 
At this stage, the outflow is temporarily weakened, but a clear outflow cavity above and below the disk can still be seen in Figure~\ref{fig:7}(h).

Model T80, shown in Figures~\ref{fig:7}(j)--(l), initially has its rotation axis roughly along the $x$-axis, resulting in a disk that forms nearly perpendicular to the initial magnetic field (Figures~\ref{fig:7}(k) and (l)).
The overall gas accretion pattern is highly complex, but Figure~\ref{fig:7}(l) shows gas directly falling onto the inner edge of the disk or into the immediate vicinity of the protostar (sink). 
As shown in Figure~\ref{fig:7}(k), the (bipolar) outflow is launched in both the vertical and parallel directions relative to the disk’s short axis. Such outflows have also been observed recently, and the component parallel to the disk plane is referred to as a ``secondary outflow'' \citep{Okoda2021,sato23,Sai2024}.

\begin{figure*}
\centering
\includegraphics[width=1.0\columnwidth]{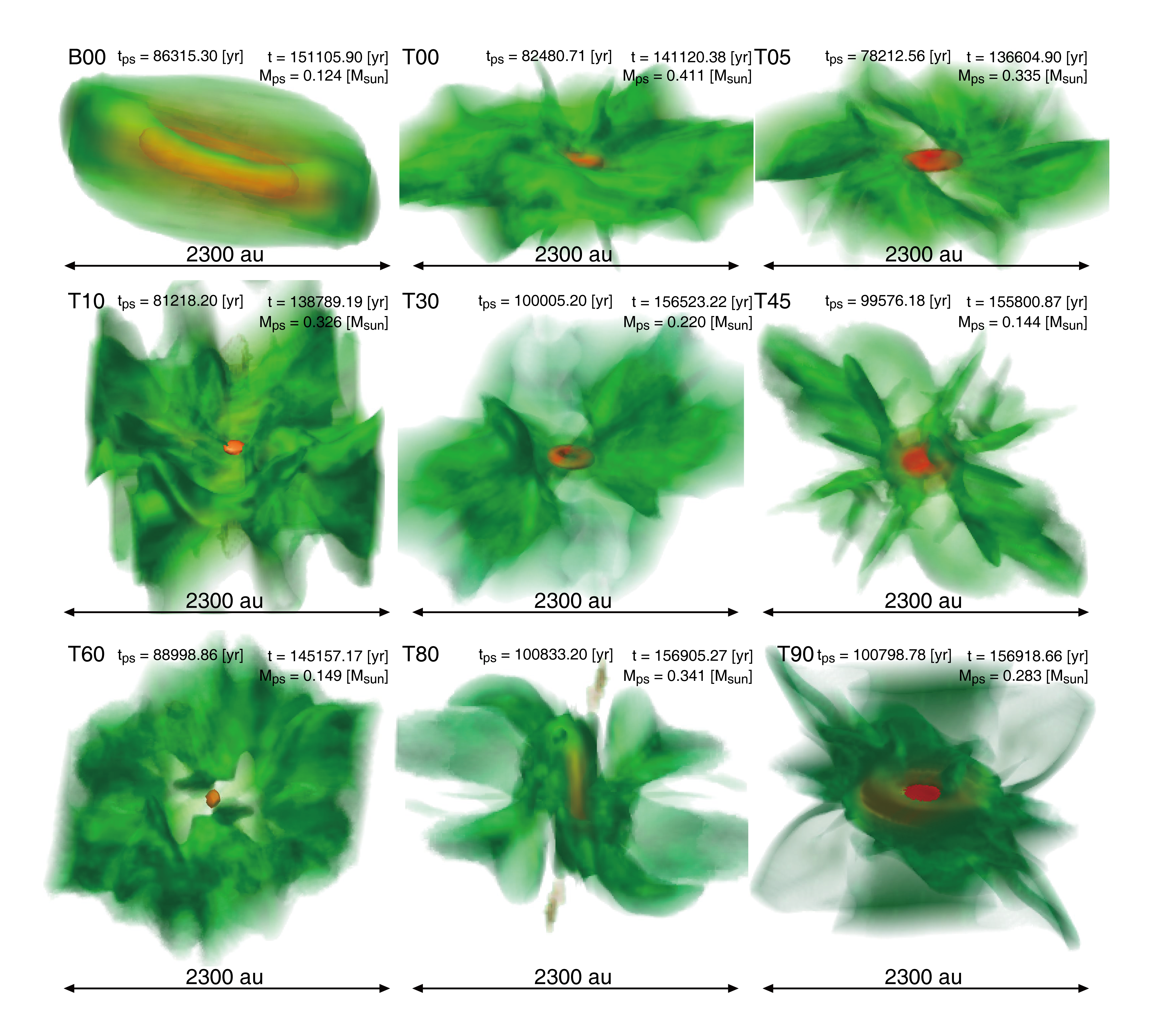}
\caption{
Volume renderings of the density distribution around the central region at $\sim10^5$ yr after protostar formation for models B00, T00, T05, T10, T30, T45, T60, T80, and T90: envelope (green) and disk (red).
Each panel lists $t_{\rm ps}$, $t$, $M_{\rm ps}$, and the spatial scale.
The viewing angle in each panel is adjusted to clearly show the central structure.
}
\label{fig:8}
\end{figure*}

Figure~\ref{fig:8} shows the density structures of all the small-core models, visualized via volume rendering in the same manner as in Figure~\ref{fig:5}.
In the unmagnetized model B00, a well-defined disk enclosed by the disk-like envelope is evident.   
In model T00, the spike-like structures extending above and below the pseudodisk (or disk) correspond to the outflow cavities. 
Because the outflow sweeps along the surface of the pseudodisk \citep{Basu24}, the pseudodisk surface is not smooth but exhibits irregularities, although the overall structure remains disk-like.
The envelope (pseudodisk) in model T05 retains an overall thin disk-like structure, but spiral features can also be identified.

In model T10, the central circumstellar disk (red) is surrounded by finger-like protrusions. 
Some of these correspond to outflow cavities, while others are twisted and distorted portions of the pseudodisk.  
The envelopes (or pseudodisks) in models T30 and T45 exhibit complex morphologies in addition to the finger-like protrusions.  
In model T45, the spike-like features extending from the central region are particularly prominent, and gas accretes asymmetrically onto the circumstellar disk through several channels.
Models T60, T80, and T90 also have infalling envelopes (or pseudodisks) with highly complex structures, indicating that the gas surrounding the disk deviates greatly from a uniform distribution.

\begin{figure*}
\centering
\includegraphics[width=1.0\columnwidth]{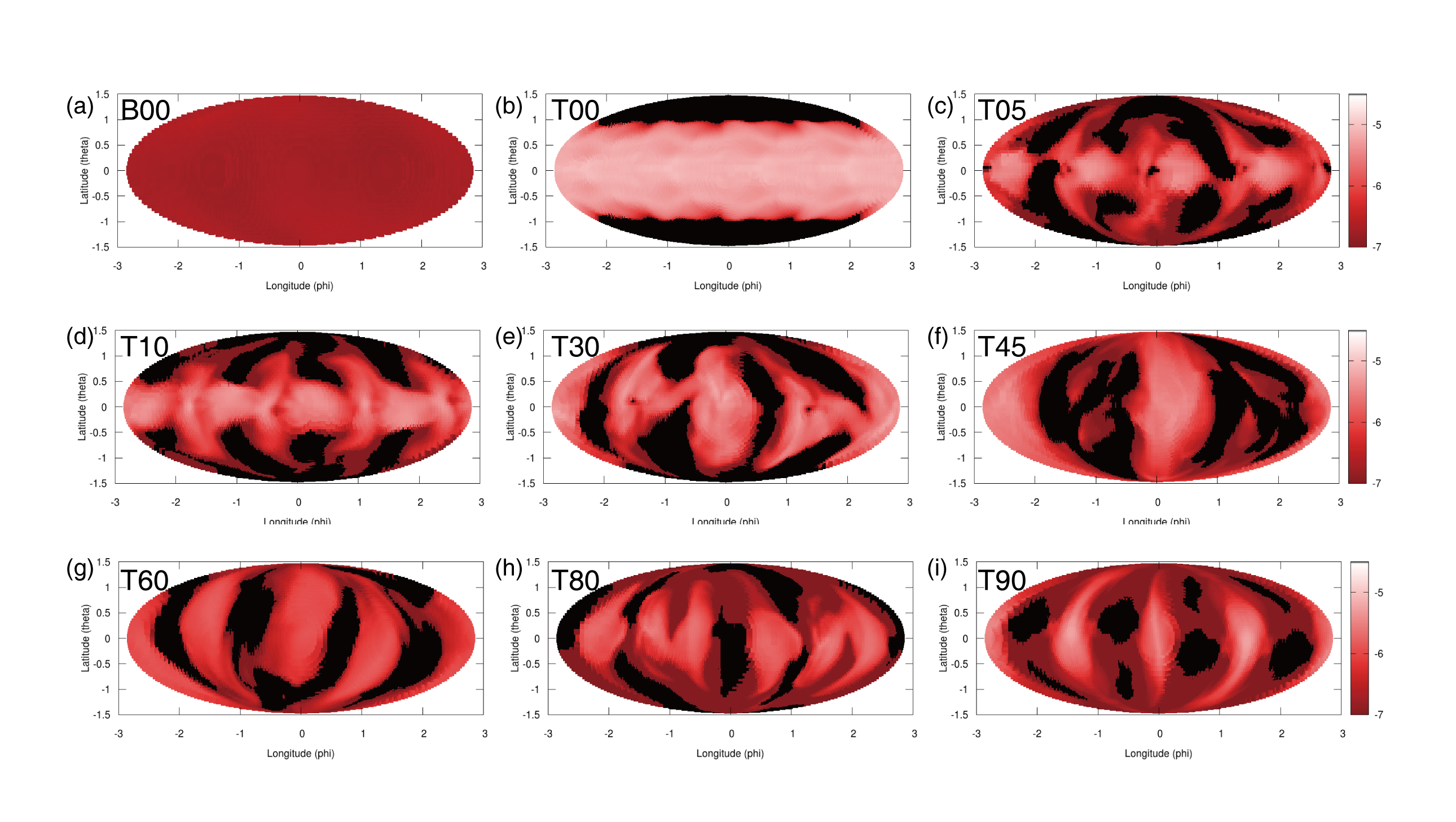}
\caption{
Mass accretion rate maps on a spherical surface of radius 1,000\,au for models B00, T00, T05, T10, T30, T45, T60, T80, and T90 at the same epochs as Figure~\ref{fig:8}.
Black regions indicate outflow ($v_r > 0$).
The vertical and horizontal axes represent latitude ($-\pi/2 < \theta < \pi/2$) and longitude ($-\pi < \phi < \pi$) in radians, respectively.
}
\label{fig:9}
\end{figure*}

Figure~\ref{fig:9} shows the mass accretion rate on a spherical surface of radius 1,000\,au at the end of the simulation. 
The spatial scale is nearly the same as in Figure~\ref{fig:8}. 
As in Figure~\ref{fig:4}, the black regions indicate outflow ($v_r>0$), while the colored regions indicate inflow ($v_r<0$).

In the unmagnetized model B00, the accretion pattern at 1,000\,au is nearly isotropic, and the accretion rate at the final epoch is $\lesssim 10^{-6}\,\msunyr$.  
In the aligned model T00, accretion occurs from the lateral (equatorial) direction, with outflows appearing along the vertical (polar) direction.
In contrast, the other misalignment models (T05, T10, T30, T45, T60, T80, T90) show inflow and outflow regions that are highly intermixed.
In all models, the initial magnetic field was oriented vertically (north--south). 
As a result, in models T05, T10, and T30, the outflow regions tend to be located near the polar regions.  
On the other hand, in models T45, T60, and T80, the outflows are not concentrated in the polar regions. 
In model T90, no outflow is driven in the polar regions.

Furthermore, in models T05, T45, T80, and T90, there is a clear separation between regions of high and low accretion rates. 
For example, in model T05, specific regions on the equatorial plane have high accretion rates of $\sim 10^{-5}\msunyr$, whereas the inflow regions near the poles have accretion rates as low as $\sim 10^{-7}\,\msunyr$.
Therefore, compared with models B00 and T00, the misalignment models exhibit a distinct contrast between regions of high and low accretion rates, with differences exceeding two orders of magnitude. 
In summary, compared with B00 and T00, the misaligned models clearly exhibit asymmetric accretion and asymmetric inflow.
From Figure~\ref{fig:9}, we can also expect that the inflow rate or accretion rate is likely to be non-steady.

\begin{figure*}
\centering
\includegraphics[width=1.0\columnwidth]{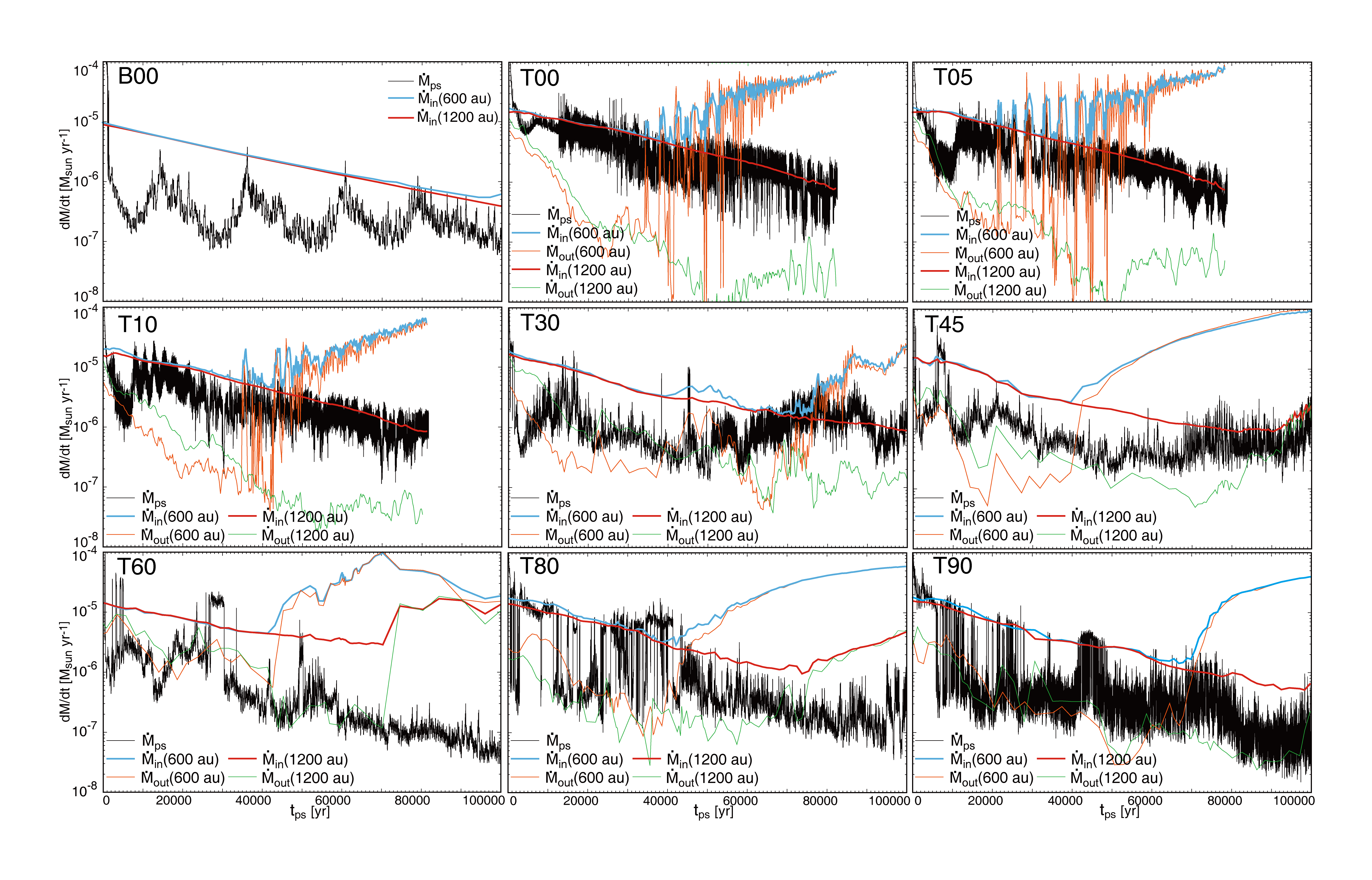}
\caption{
Time evolution of the mass accretion rate onto the protostar (black) and the inflow (cyan and red) and outflow (purple and orange) rates measured on spherical surfaces at $r=600$ and 1,200\,au for models B00, T00, T05, T10, T30, T45, T60, T80, and T90.
}
\label{fig:10}
\end{figure*}

Figure~\ref{fig:10} plots the mass accretion rate onto the central protostar (sink), as well as the inflow and outflow rates at 600\,au and 1,200\,au estimated on the corresponding grid surfaces, for all small-core models.
\footnote{
The inflow and outflow rates at any scale or grid surface are derived in equations~(7) and (8) of \citet{machida24}.
}.
In model B00, the inflow rate decreases monotonically.
Note that the inflow rate at 600\,au shows a slight increase around $t_{\rm ps} \sim 70,000$--$100,000$\,yr. 
This is because the outer edge of the circumstellar (rotationally supported) disk reaches close to 600\,au (Figure~\ref{fig:1}f).
The accretion rate from the disk onto the central protostar is non-steady, gradually declining while oscillating. 
This is due to the development of spiral structures caused by gravitational instability in the circumstellar disk \citep{machida10}.

Models T00, T05, and T10 show similar trends in the accretion, inflow, and outflow rates.
Although the mass accretion onto the central protostar (sink) is highly variable, the accretion rate decreases overall with time.
In these models, the inflow rate at 600\,au increases about 20,000--30,000\,yr after protostar formation.
Around the same period, the outflow rate also increases to values comparable to the inflow rate.
This is because the bulk of the outflow reaches 600\,au around this period, lifting circumstellar gas, some of which subsequently falls back. The fallback gas contributes to the enhanced inflow rate.
Despite the strong mass ejection from the central region by the outflow, the mass accretion rate onto the protostar does not change significantly.
This implies that the net inflow rate (inflow rate $-$ outflow rate) remains nearly constant.

The inflow and outflow rates in model T30 are similar to those in models T00, T05, and T10.
However, model T30 shows a temporary increase in the mass accretion rate onto the protostar at $t_{\rm ps} \sim 55,000$--$90,000$\,yr. 
This occurs because the infalling envelope (or pseudodisk) is substantially deformed, leading to corresponding changes in both the inflow and accretion rates.

Models T00, T05, and T10 were performed for a shorter duration ($t_{\rm ps} \lesssim 80,000$\,yr) than the other models ($t_{\rm ps} \gtrsim 100,000$\,yr), owing to the shorter timesteps resulting from the high Alfv\'en and outflow speeds. 
As a result, the bulk of the outflow did not reach 1,200\,au within these runs, and the sharp rise in the inflow and outflow rates seen at 600\,au is not observed at 1,200\,au.

In models T45 and T60, both the outflow and inflow rates at 600\,au increase from $t_{\rm ps} \sim 40,000$\,yr. 
At $t_{\rm ps} \gtrsim 70,000$--$80,000$\,yr, the outflow and inflow rates at 1,200\,au also increase. 
In these models, the inflow rate decreases monotonically until the outflow reaches 600 or 1,200\,au. 
After the outflow reaches these radii, the inflow rate temporarily rises, but the oscillations are not large. 
This indicates that the envelope evolves gradually in these cases and does not trigger rapid, large-amplitude changes in the inflow rate.

Models T80 and T90 show trends similar to those of models T45 and T60. 
However, the mass accretion rates onto the protostar in models T80 and T90 are slightly higher than in models T45 and T60. 
This is likely because magnetic braking is more effective in the envelope near the disk, allowing gas to fall efficiently onto the protostar \citep{hirano20}. 
In addition, model T90 exhibits a delayed onset of strong outflow activity reaching 600\,au (later than $t_{\rm ps} = 70,000$\,yr) compared to the other models. 
This is also considered to be due to a magnetic field configuration in the central region and to magnetic braking that transports angular momentum away from the disk.

\begin{figure*}
\centering
\includegraphics[width=1.0\columnwidth]{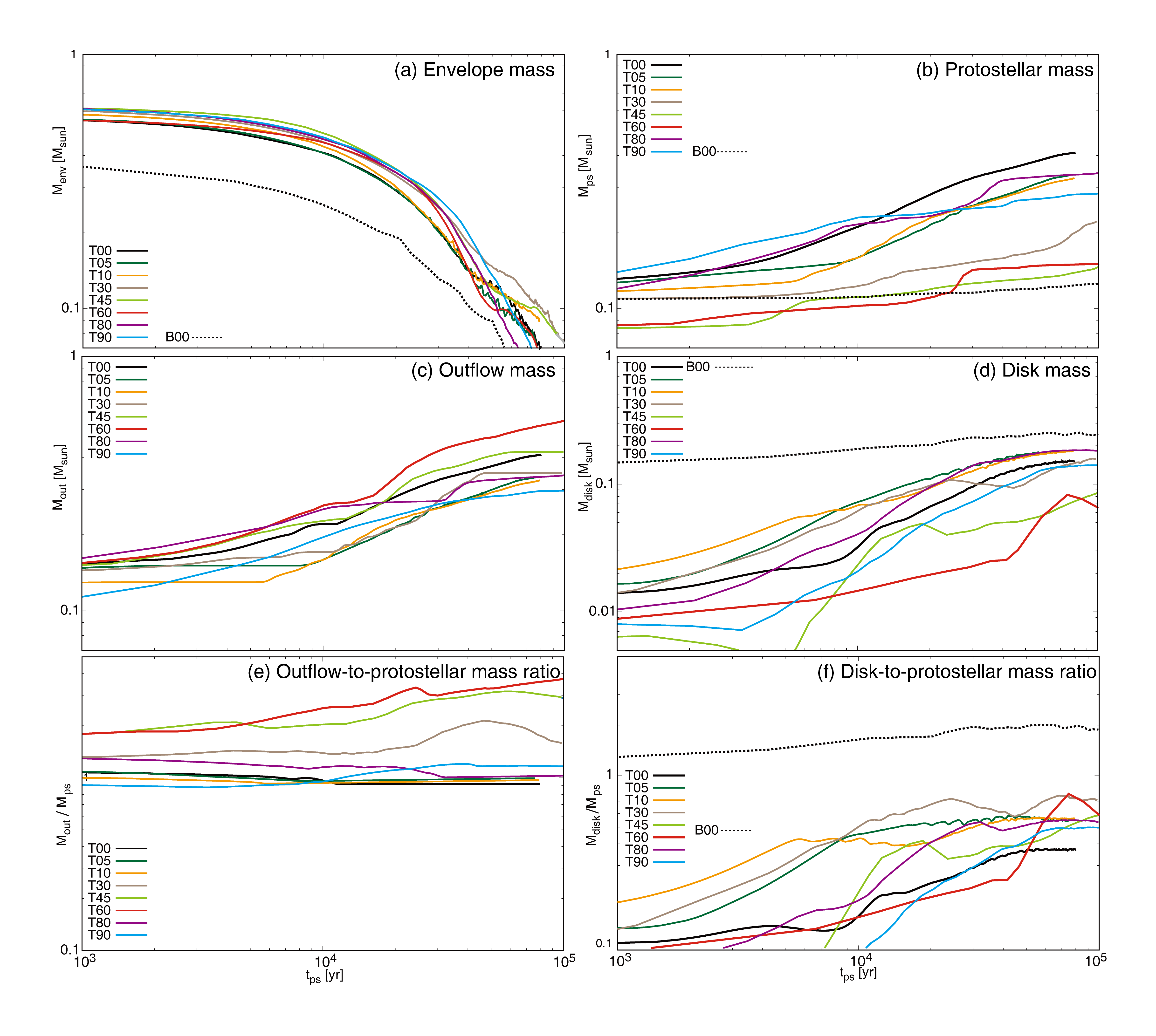}
\caption{
Time evolution of masses and mass ratios for models T00, T05, T10, T30, T45, T60, T80, T90, and B00: (a) envelope mass $M_{\rm env}$, (b) protostellar mass $M_{\rm ps}$, (c) outflow mass $M_{\rm out}$, (d) disk mass $M_{\rm disk}$, (e) outflow-to-protostellar mass ratio, and (f) disk-to-protostellar mass ratio.
}
\label{fig:11}
\end{figure*}

Figure~\ref{fig:11} plots, as functions of the time after protostar formation ($t_{\rm ps}$), the envelope mass (panel a), protostellar mass (panel b), outflow mass (panel c), disk mass (panel d), outflow-to-protostellar mass ratio (panel e), and disk-to-protostellar mass ratio (panel f) for the small-core models. 
The outflow mass $M_{\rm out}$ is calculated by integrating the mass in regions with $v_r > c_{\rm s}$. 
The disk mass $M_{\rm disk}$ is, for simplicity, calculated by integrating regions with $n > 10^9\,\cc$ that are not part of the outflow region ($v_r < c_{\rm s}$). 
This density-based criterion is less rigorous than disk identification methods used in other studies that explicitly examine the gas kinematics. 
The dependence of the outflow and disk masses on the adopted criteria is examined in \S\ref{sec:A1}.
The protostellar mass $M_{\rm ps}$ corresponds to the sink mass. 
The envelope mass $M_{\rm env}$ is estimated as \citep{machida13,machida24} 
\begin{equation}
M_{\rm env} = M_{\rm env,0} - M_{\rm ps} - M_{\rm disk} - M_{\rm out},
\end{equation}
where $M_{\rm env,0}$ is the initial molecular cloud core mass ($M_{\rm env,0}=0.81\,\msun$ for all small-core models). 
The total mass $M_{\rm env}+M_{\rm ps}+M_{\rm disk}+M_{\rm out}$ is conserved to within $0.1\,\%$ of $M_{\rm env,0}$ throughout the simulations.
By the end of the simulations, the envelope mass decreases to $\lesssim0.08\,\msun$ in all models (Figure~\ref{fig:11}(a)), implying that more than 90\% of the initial core mass has either accreted onto the central region or been expelled by the outflow.

Although the envelope evolves similarly among all magnetized models, the envelope is depleted most rapidly in the unmagnetized model B00, reflecting the absence of magnetic support.
From Figure~\ref{fig:11}(b), it is evident that, despite identical initial prestellar cloud masses, the final stellar masses differ between models. 
The unmagnetized model B00 produces the smallest protostellar mass but the largest disk mass (Figures~\ref{fig:11}(b) and (d)). 
Among the magnetized models, those with relatively small or large initial misalignment angles (T00, T05, T10, T80, T90) yield relatively large final protostellar masses, whereas models T45 and T60 produce smaller protostellar masses \citep[see also][]{hirano20}.

In contrast, as shown in Figure~\ref{fig:11}(c), the outflow masses in models T45 and T60 are larger than in the other models \cite[see also][]{Hennebelle2009}. 
The disk masses in models T45 and T60 are smaller than in the other models, although the differences among the magnetized models remain within a factor of $\sim 3$ (Figure~\ref{fig:11}(d)). 
Because models T45 and T60 drive strong outflows from the central region, they efficiently remove mass from the central region, leading to lower stellar and disk masses.
Their outflow-to-protostar mass ratios reach $M_{\rm out}/M_{\rm ps}\gtrsim3$ at late times (Figure~\ref{fig:11}e), implying that $\gtrsim70\%$ of the initial mass is expelled.
Other magnetized models also have $M_{\rm out}/M_{\rm ps} \simeq 1$, implying that roughly half of the prestellar core mass is removed by the protostellar outflow.

The disk-to-protostar mass ratio at the end of the simulations is $M_{\rm disk}/M_{\rm ps}\sim0.3$--0.7 for the magnetized models (Figure~\ref{fig:11}f).
Because these ratios vary over time, the disk mass likely changes significantly due to variations in gravitational instability and magnetic braking efficiency \citep{Machida2026}. 
As this study focuses on the large-scale structure, we do not further analyze the internal disk structure here.

\section{Discussion}
\subsection{Angle Between Stream Velocity and Magnetic Field}
\label{sec:angleVB}
\begin{figure*}
\centering
\includegraphics[width=1.0\columnwidth]{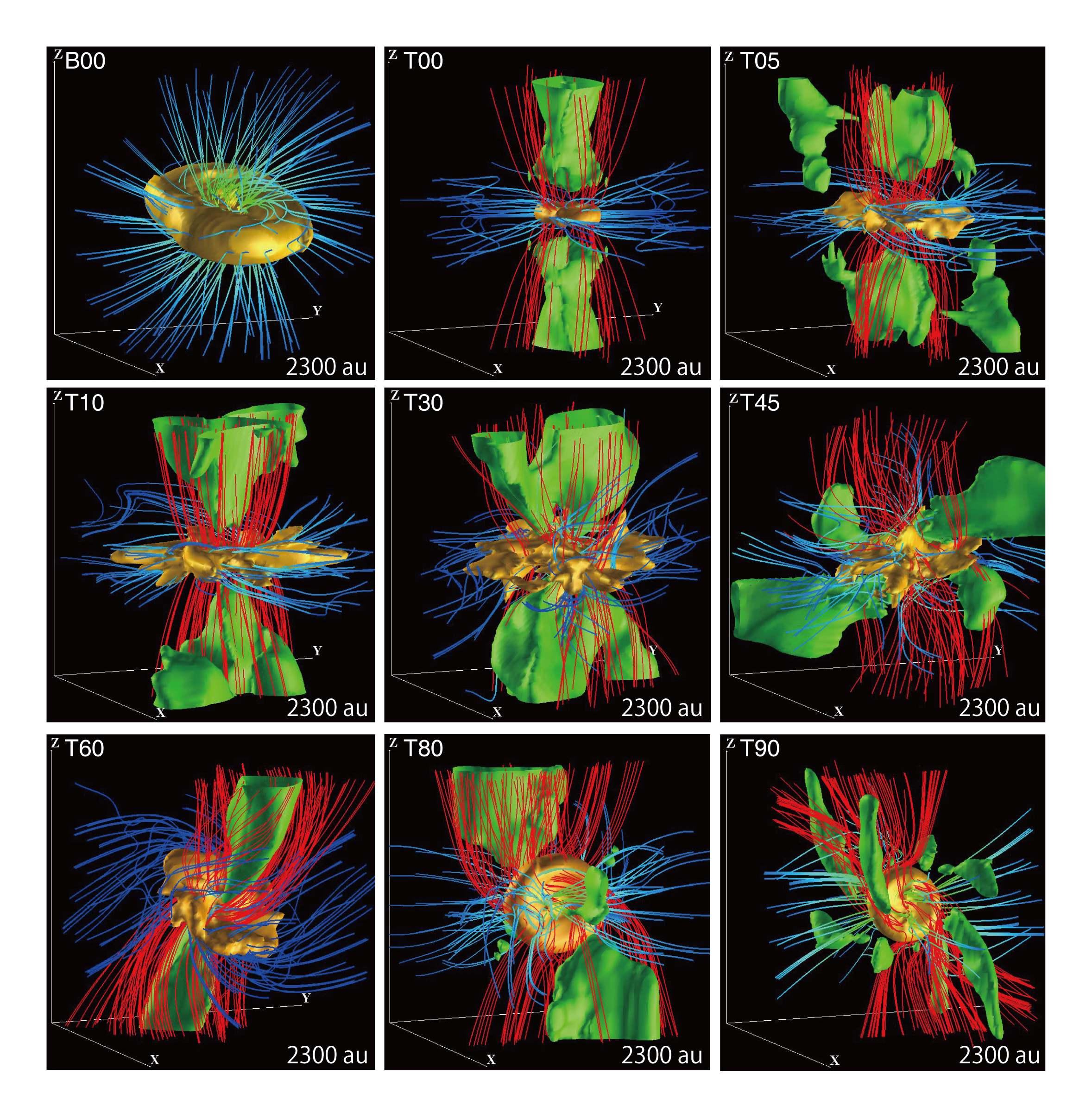}
\caption{
Three-dimensional structures for models B00, T00, T05, T10, T30, T45, T60, T80, and T90 at the same epoch as Figure~\ref{fig:8}: disk-like envelope (orange surface), outflow (green surface), inflow (blue lines), and magnetic field (red lines). 
The model name and spatial scale are indicated in each panel.
}
\label{fig:12}
\end{figure*}

In this subsection, we discuss the angle between the velocity vector of the infalling gas and the magnetic field vector in the magnetized small-core models.  
Figure~\ref{fig:12} summarizes the three-dimensional morphology at the end of the simulations for the small-core models, showing the disk-like envelope and outflow structures, inflow streamlines, and magnetic field lines. 
The visualization is constructed in the same manner as Figure~\ref{fig:6}. 
To better visualize the envelope structure, an iso-density surface at $n = 10^7\,\cc$ is adopted in Figure~\ref{fig:12}, whereas Figure~\ref{fig:6} uses an iso-density surface at $n = 10^9\,\cc$ to highlight the disk structure.
In the unmagnetized model B00, gas flows into the disk both from above and below, as well as from the sides.
In models T00, T05, T10, and T30, the magnetic field lines are oriented from the lower to the upper side, exhibiting a nearly hourglass morphology, and the outflows also propagate vertically from the central region.  
In these models, the disk normal is roughly aligned with both the magnetic field and the outflow axis, and gas predominantly accretes from lateral directions.  
However, the high-density envelope region in model T30 shows a somewhat more complex structure, and gas inflow is not restricted to lateral directions only.

The outflow in model T45 exhibits a quadrupolar morphology, and multiple outflow components are also observed in models T80 and T90 \citep[see also][]{Hirano2026}.  
Model T60 shows a structure similar to a bipolar outflow.
However,  the high-density infalling envelope has a complex morphology, and gas accretes toward the center from various directions.  
In models T80 and T90, the disk is inclined sideways, with its normal pointing in the $x$-direction.
As a result, the net magnetic field direction is perpendicular to the disk normal.  
Moreover, in the vicinity of the center, the magnetic field lines are strongly wound due to the rotation of the disk-like envelope.

Comparing the magnetic field lines and streamlines in Figure~\ref{fig:12}, models with small initial misalignment angles tend to show streamlines nearly perpendicular to the magnetic field.
In contrast, for large misalignment angles, the geometry becomes more complex, and regions where streamlines run nearly parallel to the magnetic field are also present.

\begin{figure*}
\centering
\includegraphics[width=1.0\columnwidth]{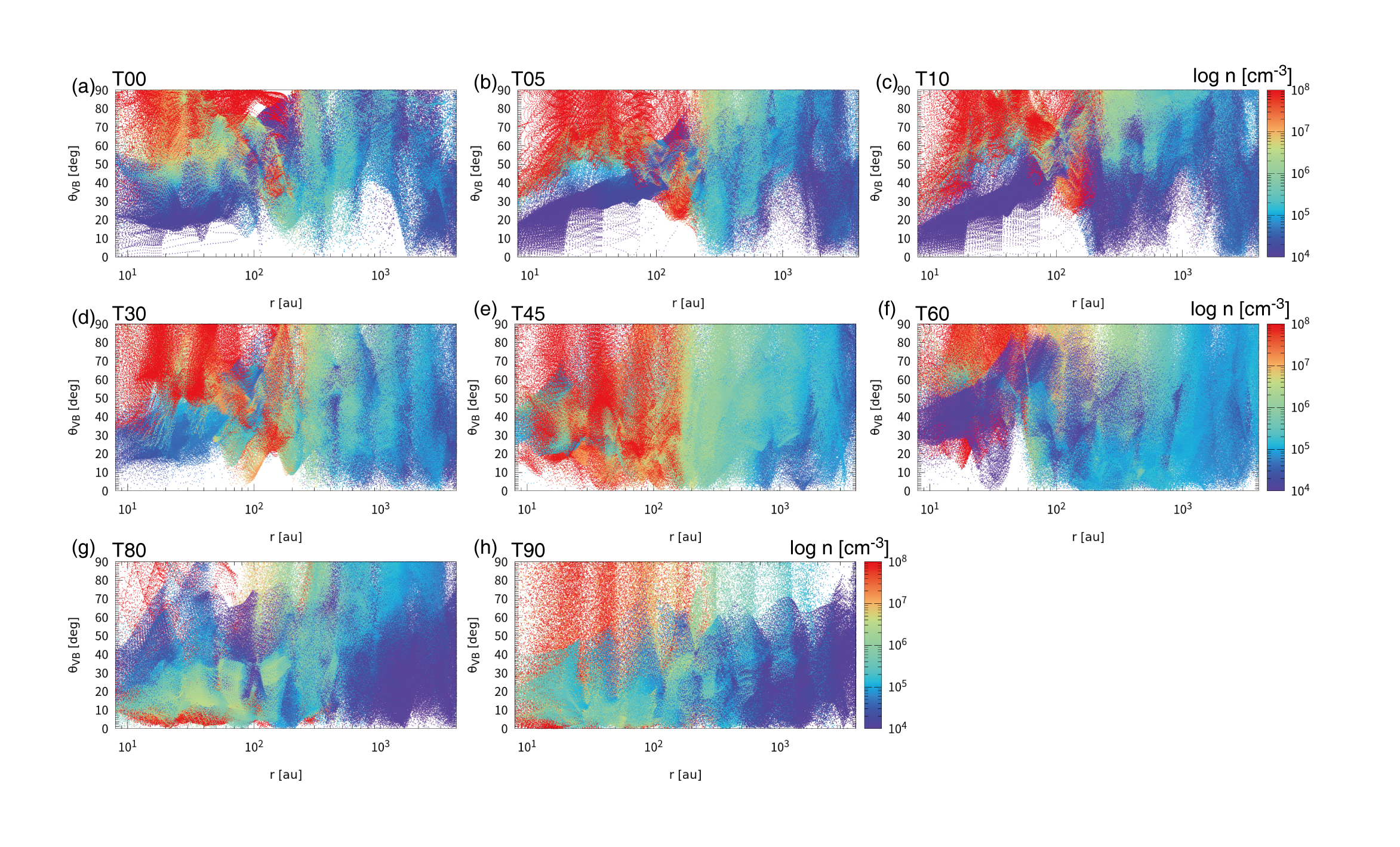}
\caption{
Distribution of infalling gas ($v_r<0$) in the $r$–$\theta_{\rm VB}$ plane at the same epoch as Figure~\ref{fig:8} for the small-core models. 
Here, $r$ is the radial distance measured from the sink position, and $\theta_{\rm VB}$ is the acute angle ($0^\circ \le \theta_{\rm VB} \le 90^\circ$) between the infall velocity vector $\vect{V}$ and the magnetic field vector $\vect{B}$. 
Colors indicate the gas number density in each cell.
} 
 \label{fig:13}
\end{figure*}

To examine the relative orientation, we calculate the angle between the magnetic field vector $\vect{B}$ and the velocity vector $\vect{V}$ in every computational cell.
Figure~\ref{fig:13} plots, for all small-core models, the angle $\theta_{\rm VB}$ between $\vect{V}$ and $\vect{B}$ for infalling gas ($v_r < 0$) as a function of radius from the protostar.   
Here, we adopt the acute angle, and the gas density is indicated by color.\footnote{  
The striped patterns visible in each panel arise from the use of nested grids, where the resolution (number of cells) changes from level to level.  
Because the resolution is higher and the number of cells increases closer to the center, the distributions appear wavy. 
However, this is purely a display artifact.
}

In all models, the angle between $\vect{V}$ and $\vect{B}$ for infalling gas is broadly distributed over the range $0^\circ < \theta_{\rm VB} < 90^\circ$.  
Focusing on the region within 100--1,000\,au from the central protostar, in models T00, T05, T10, and T30, the high-density gas exhibits large $\theta_{\rm VB}$ values, clustering near $90^\circ$.  
This trend reflects the structure of the pseudodisk in the aligned case: the pseudodisk is a high-density structure threaded vertically by magnetic field lines, and gas within it accretes predominantly perpendicular to the field.  
In contrast, low-density gas tends to fall along the magnetic field lines from above the pseudodisk, resulting in smaller $\theta_{\rm VB}$ values.

In model T45, the high-density gas is widely distributed over $0^\circ < \theta_{\rm VB} < 90^\circ$.  
The low-density gas angles are more frequently found in the range $20^\circ < \theta_{\rm VB} < 60^\circ$, but no well-defined preferred angle is apparent.
In model T60, most high-density gas has $\theta_{\rm VB} > 50^\circ$, whereas low-density gas is predominantly distributed over $20^\circ < \theta_{\rm VB} < 70^\circ$.  
Model T90 exhibits trends similar to model T60.  
Model T80, in contrast, shows a tendency toward smaller $\theta_{\rm VB}$ values compared with the other models.  

These trends are expected to change over time.  
Therefore, it is not possible to state definitively that the streams falling onto the central disk are either parallel or perpendicular to the magnetic field.  
Even if polarization observations reveal the magnetic field configuration near a streamer, the field geometry alone may not uniquely determine the gas-flow direction.


To quantify the relative orientation between the magnetic field and the infalling gas velocity, we calculated the mass fractions of the infalling gas for three ranges of the angle between the magnetic field and velocity vectors.
The resulting radial distributions are presented in Appendix~\ref{sec:A2}.
The relative orientation varies substantially with both radius and the initial misalignment angle, supporting the above interpretation.

\subsection{Infall Velocity}
\begin{figure*}
\centering
\includegraphics[width=1.0\columnwidth]{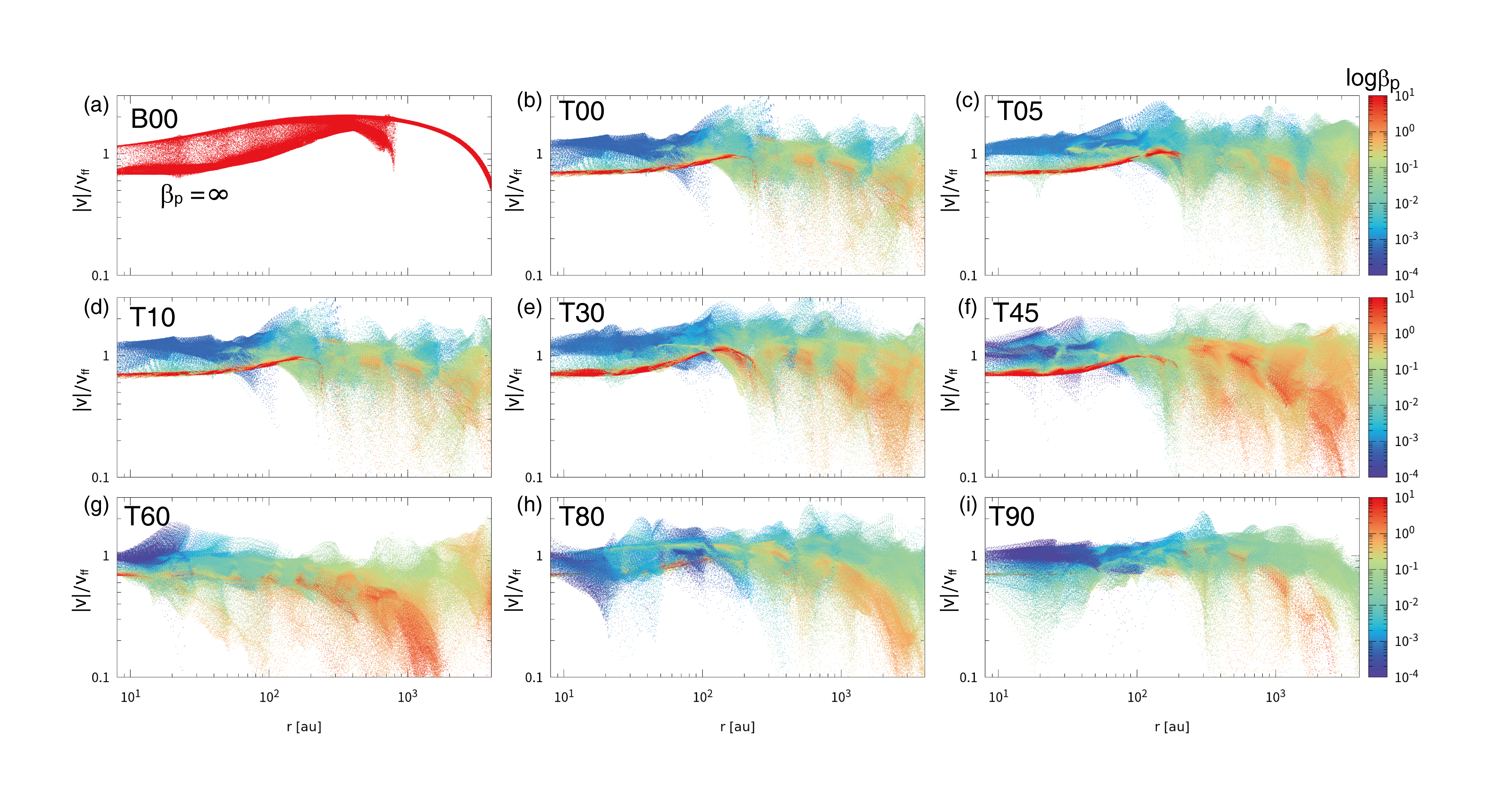}
\caption{
Distribution of infalling gas ($v_r<0$) in the $r$–$\vert \vect{v} \vert/v_{\rm ff}$ plane at the same epochs as Figure~\ref{fig:8} for models B00, T00, T05, T10, T30, T45, T60, T80, and T90. 
Here, $r$ is the radial distance from the central sink particle, and the vertical axis shows the normalized radial velocity $\vert \vect{v} \vert/v_{\rm ff}$. 
Colors indicate the plasma beta $\beta_{\rm p}$ of each cell.
}
 \label{fig:14}
\end{figure*}

The velocity of the infalling gas is a crucial diagnostic for measuring the mass of the central protostar and for understanding the origin of streamers.
Observations have shown that the infalling gas sometimes has velocities smaller than the free-fall velocity (the so-called `factor of 3 problem'; \citealt{Takakuwa2013,ohashi14,Chou2014,aso15,Takakuwa2020,Sai2022}).  
It is also considered that observed streamers exhibit either the free-fall velocity or a velocity gradient consistent with free-fall motion.  

To compare the velocity of the infalling gas with the free-fall velocity, Figure~\ref{fig:14} plots the ratio $\vert \vect{v} \vert/v_{\rm ff}$ for all infalling cells with $v_r<0$ in the computational domain, where the free-fall velocity is defined as
\begin{equation}
v_{\rm ff} \equiv \sqrt{\frac{2GM_{\rm ps}}{r}}.
\label{eq:freefall}
\end{equation}
In Figure~\ref{fig:14}, the color represents the plasma beta $\beta_{\rm p}$ in each cell.

In model B00, the normalized (infall) velocity is distributed in the range $0.7 < \vert \vect{v} \vert/v_{\rm ff} < 1.1$.  
Since the free-fall velocity is defined as in equation~(\ref{eq:freefall}), it differs from the Keplerian velocity $\sqrt{GM_{\rm ps}/r}$ by a factor of $\sqrt{2} \simeq 1.4$.  
Accordingly, the inner region with $\vert \vect{v} \vert /v_{\rm ff} \simeq 0.7$ corresponds to the Keplerian disk.  
In model B00, a Keplerian disk with a radius of 500--600\,au forms.  
At larger radii, the disk mass becomes non-negligible, leading to an increase in the rotation velocity, with $\vert \vect{v} \vert/v_{\rm ff}$ approaching unity.  
In the infalling envelope, the normalized velocity slightly exceeds unity, $\vert \vect{v} \vert/v_{\rm ff} \gtrsim 1$.  
\citet{Hirano2025} pointed out that during the accretion phase, the contribution from the disk and envelope masses makes the actual infall velocity slightly larger than the value derived solely from the central stellar mass $M_{\rm ps}$.  
However, the distribution of $\vert \vect{v} \vert/v_{\rm ff}$ around unity in model B00 indicates that the gas accreting onto the disk has nearly free-fall velocities.  
Therefore, in the absence of magnetic fields, the infalling gas can be regarded as essentially falling at the free-fall velocity.

We then focus on the magnetized models.  
In all such models, thin red curves with $\beta_{\rm p} \gtrsim 10$ are seen, extending inward to radii of roughly 100\,au. 
They trace the Keplerian or rotationally supported disk as well as the inner edge of the pseudodisk.
In rotationally supported disks, magnetic fields are dissipated and the plasma beta becomes high.  
As a result, magnetic braking becomes inefficient \citep{Machida2011PASJ}, and the disk undergoes nearly Keplerian rotation while slowly accreting toward the center.
Because the gas velocity is normalized by the free-fall velocity rather than by the Keplerian velocity, $\vert \vect{v} \vert/v_{\rm ff} \sim 0.7$ corresponds to Keplerian rotation.  
These red curves indicate that the gas in these regions rotates nearly at the Keplerian speed.  

Focusing on the inner $\sim$100\,au, models with small misalignment angles (models T00, T05, T10, T30, T45) show that the infalling gas velocity is typically in the range $0.7 \lesssim \vert \vect{v} \vert/v_{\rm ff} \lesssim 1.1$, i.e., close to free-fall velocity.  
In contrast, models T60, T80, and T90 contain a non-negligible fraction of gas with velocities in the range $0.1 < \vert \vect{v} \vert /v_{\rm ff} < 0.7$.  
These models possess more complex magnetic field structures, indicating that part of the gas is substantially decelerated by magnetic forces.  

On larger scales ($100 \,{\rm au} < r < 4,000 \,{\rm au}$), the ratios $\vert \vect{v} \vert/v_{\rm ff}$ are more broadly distributed in the range $0.1 < \vert \vect{v} \vert/v_{\rm ff} < 1.1$ for all the small-core models.  
Nevertheless, most of the gas falls toward the center with $\vert \vect{v} \vert/v_{\rm ff} \sim 1$, indicating that the bulk of the envelope accretes nearly at the free-fall velocity.  
The velocity distribution is also regulated by plasma beta.
In magnetically dominated regions (low $\beta_{\rm p}$), the gas tends to move closer to $v_{\rm ff}$, while in high-$\beta_{\rm p}$ regions the gas can fall substantially below $v_{\rm ff}$.
In the envelope, low-$\beta_{\rm p}$ regions generally correspond to low-density gas, which is less affected by pressure gradients and thus falls directly inward.  
Conversely, high-$\beta_{\rm p}$ regions correspond to denser gas, typically located in the pseudodisk, where the gas is in a quasi-static state with gas pressure and magnetic tension nearly balancing gravity.  
As a result, the infall velocity is reduced in such regions.  


Although most of the infalling gas has velocities close to the free-fall velocity, a small fraction has $|v|/v_{\rm ff}\sim0.3$, comparable to the slow infall associated with the factor-of-three problem.
\citet{Hirano2025} also found localized regions with $v_{\rm infall}/v_{\rm ff}\sim0.3$ near the outer edge of the pseudodisk, where the infalling gas is decelerated by the magnetic force.
Thus, the small fraction of slowly infalling gas found in our simulations is qualitatively consistent with the localized slow-infall regions found in \citet{Hirano2025}.
It is difficult to directly compare the fraction of slowly infalling gas in our simulations with the occurrence of slow infall in observations.
To identify the factor-of-three problem observationally, both the infall velocity and the protostellar mass, which is required to estimate the free-fall velocity, need to be determined.
The protostellar mass is generally estimated from Keplerian rotation of the circumstellar disk.
Since such measurements are difficult for young embedded protostars, the number of sources for which the factor-of-three problem can be examined is still limited \citep[see][and references therein]{Hirano2025}.
In addition, observed infall velocities are inferred from the line-of-sight velocity and can therefore be affected by projection effects.
In addition, optical-depth effects may cause molecular-line observations to preferentially trace only part of the infalling gas along the line of sight.
Therefore, a direct quantitative comparison between the fraction of slowly infalling gas in our simulations and observations is not straightforward. 

\begin{figure*}
\centering
\includegraphics[width=0.9\columnwidth]{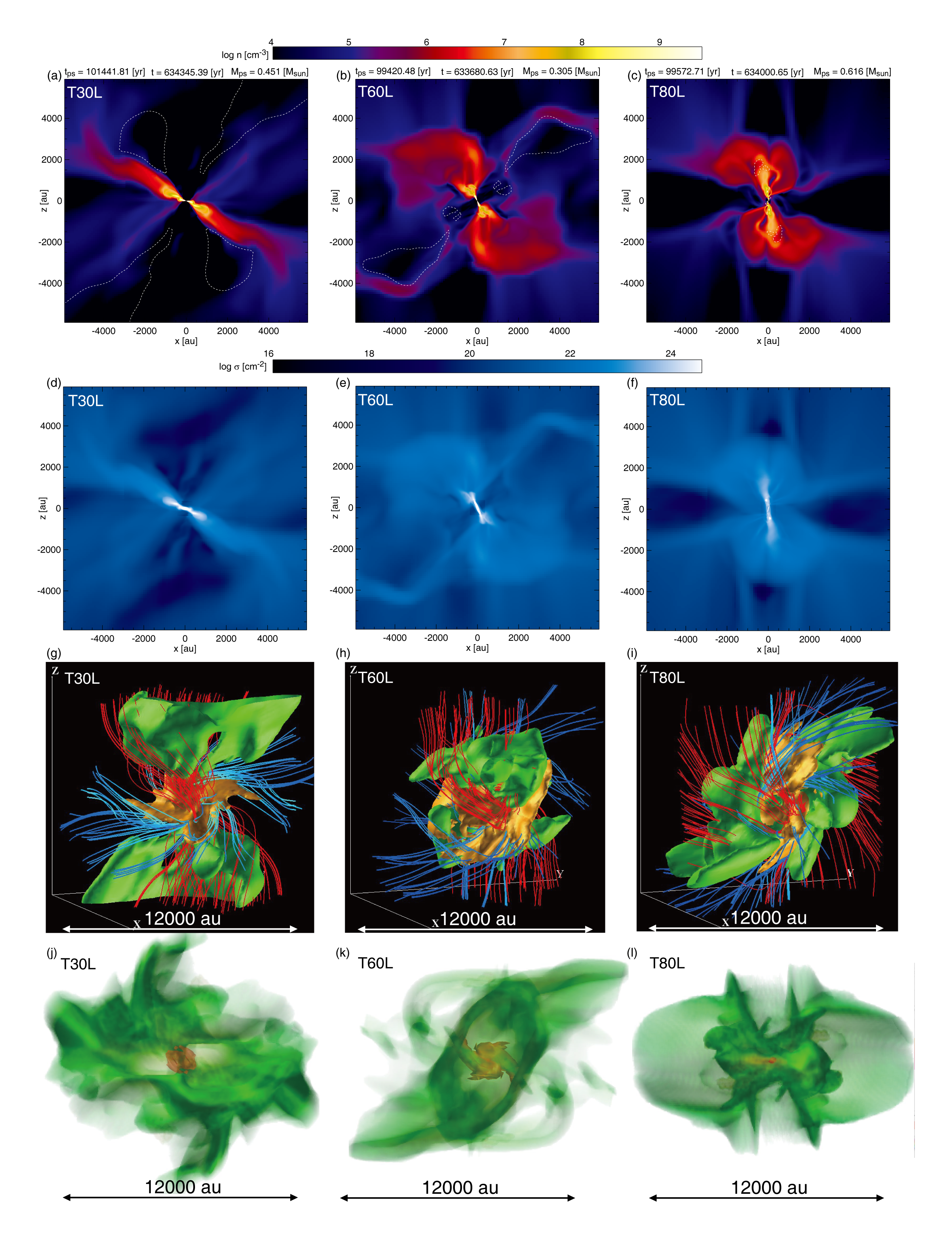}
\caption{
Large-core models T30L, T60L, and T80L: density slices on the $y=0$ plane (a--c), surface density maps (d--f), three-dimensional views (g--i), and volume renderings of the density (j--l). 
The model name is indicated in each panel. 
In panels (a--c), the white dotted line marks the boundary between outflowing and infalling gas; the region inside the white dotted line corresponds to the outflowing region. 
In panels (g--i), the high-density region (orange surface), outflow (green surface), infall streams (blue lines), and magnetic field lines (red lines) are shown. 
These panels (g--i) are constructed using the same method as Figures~\ref{fig:6} and \ref{fig:9}, but the iso-density surface (shown in orange) is defined at $n = 10^7\,\cc$.
}
 \label{fig:15}
\end{figure*}

\subsection{Scales of Asymmetric Accretion}
Observations have shown that streamers connect to the disk at small scales, whereas on larger scales some streamers extend beyond $0.1$\,pc \citep{pineda23}.  
Typical prestellar cores have radii of order $\sim 0.1$\,pc \citep{Andre2010, Arzoumanian2011, Tokuda2020}.
Our fiducial small-core models adopt $R_{\rm cl}=4,800$\,au (Table~\ref{table:1}), so any envelope asymmetry is necessarily confined within this radius.
To assess how far asymmetric accretion can persist when the parent core is larger, we perform three large-core models (T30L, T60L, and T80L) with $R_{\rm cl}=48,000$\,au (Table~\ref{table:1}), which is comparable to the upper end of observed core sizes.
This radius is modestly larger than that of typical observed prestellar cores, yet still lies well within observational constraints.

Figure~\ref{fig:15} shows that all three large-core models develop complex, asymmetric envelope structures on scales $\gtrsim 10,000$\,au.
The top and second rows show that the central circumstellar disk remains thin and shows no significant distortion, whereas the surrounding infalling envelope (pseudodisk) is strongly distorted. 
Reflecting this complexity, model T30L (Figure~\ref{fig:15}a) exhibits two bipolar outflows oriented in different directions.
The third row shows that even beyond 10,000\,au, the magnetic field lines and streamlines remain misaligned, indicating that the magnetic field is neither parallel nor perpendicular to the streams at these scales.  
The bottom row further reveals that the complexity of the infalling envelope extends well beyond 10,000\,au.

\begin{figure*}
\centering
\includegraphics[width=0.9\columnwidth]{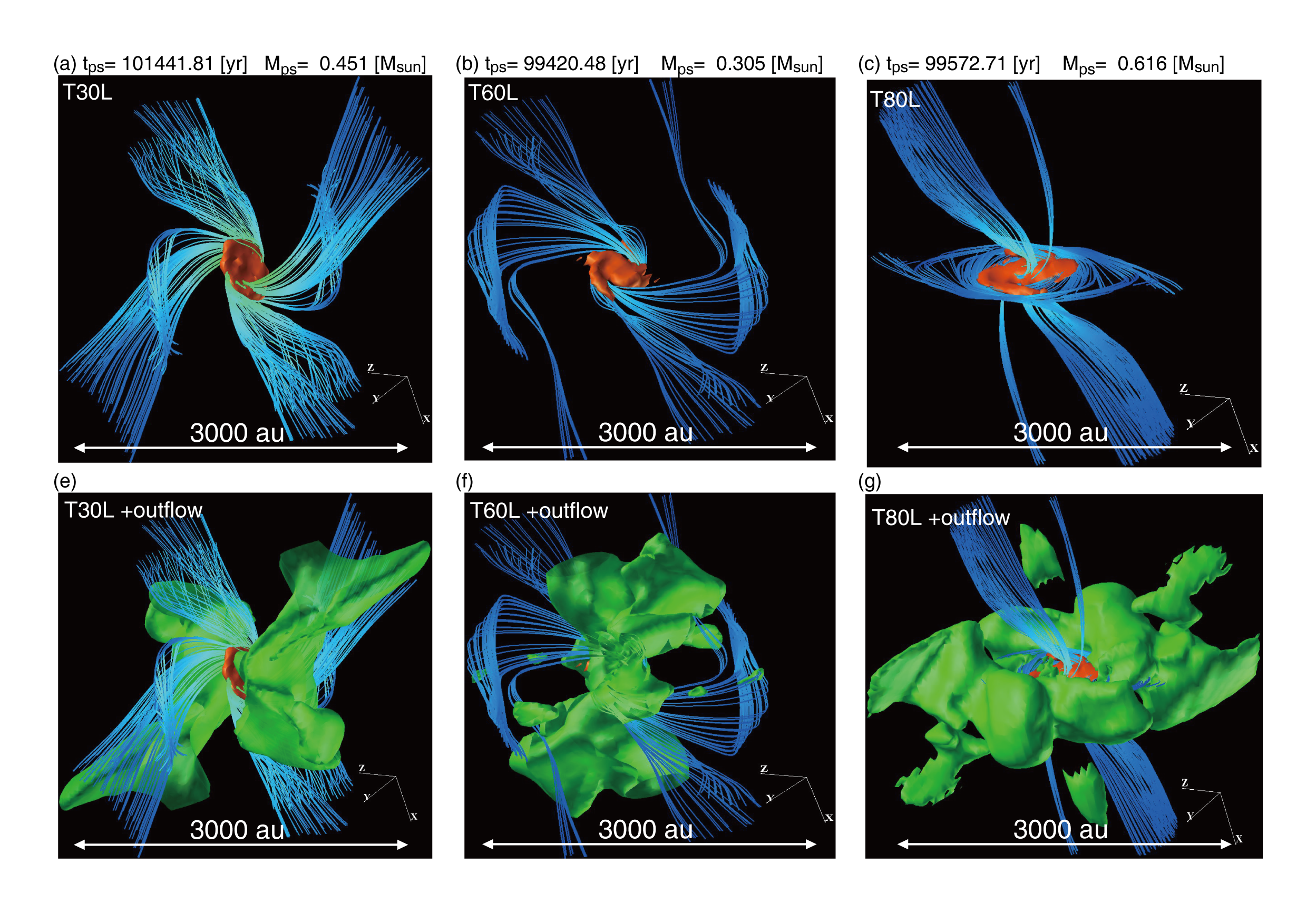}
\caption{
Three-dimensional structures for models T30L, T60L, and T80L: dense region ($10^8\,\cc$; orange surface) and inflow (blue lines) (a--c), and the corresponding outflow (green surface) (e--g).
Top panels list $t_{\rm ps}$ and $M_{\rm ps}$.
Each panel indicates the spatial scale.
}
 \label{fig:16}
\end{figure*}

Figure~\ref{fig:16} shows the regions of enhanced accretion (streams) on 3,000\,au scales, visualized with streamlines following the method of \citet{Machida2026}.  
Even on these scales, all large-core models exhibit asymmetric accretion.
For example, in model T80L, some gas wraps inward roughly along the disk plane, while other gas accretes more directly from above/below toward the inner disk regions, i.e., near the protostar (sink).
The inflow and outflow regions are strongly interleaved: streams tend to skirt the outflow cavities, and the outflows propagate outward while avoiding the dense inflow channels, as seen in observations \citep[e.g.,][]{Kido2025}.

\begin{figure*}
\centering
\includegraphics[width=1.0\columnwidth]{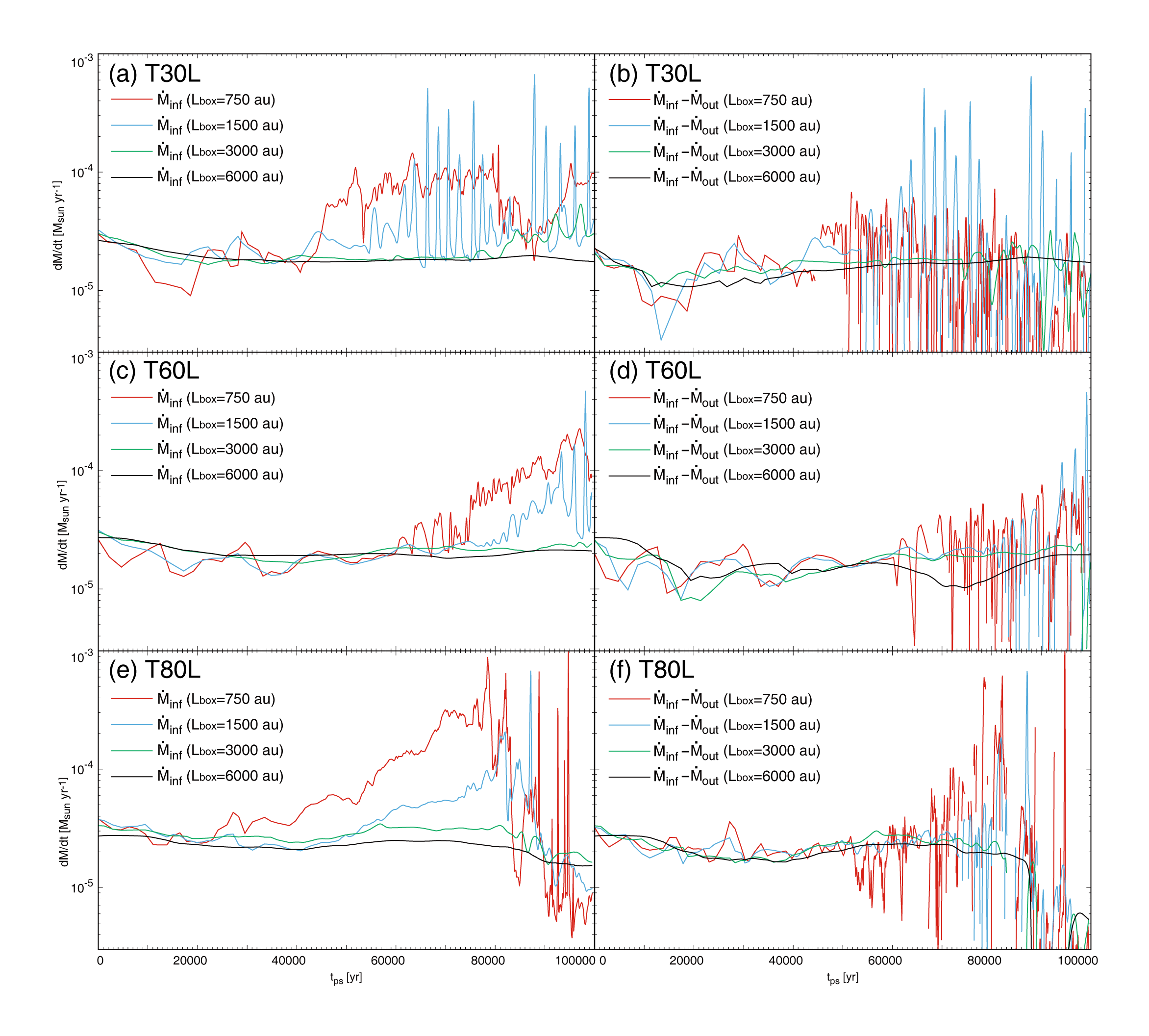}
\caption{
Mass inflow rates $\dot{M}_{\rm inf}$ (left panels) and net inflow rate $\dot{M}_{\rm inf}-\dot{M}_{\rm out}$ (right panels) at different radii for models T30L, T60L, and T80L as functions of $t_{\rm ps}$.
Truncation of the curves in panels (b), (d), and (f) indicates $\dot{M}_{\rm inf}-\dot{M}_{\rm out}<0$.
}
\label{fig:17}
\end{figure*}

Figure~\ref{fig:17} shows the inflow rates ($\dot{M}_{\rm inf}$; left) and the net inflow rate (difference between the inflow and outflow rates, $\dot{M}_{\rm inf}-\dot{M}_{\rm out}$; right) at different radii for models T30L, T60L, and T80L. 
The inflow and outflow rates at any scale or grid surface are derived in equations~(7) and (8) of \citet{machida24}.
Note that, at certain epochs, the outflow rate exceeds the inflow rate, leading to negative values $\dot{M}_{\rm inf}-\dot{M}_{\rm out}<0$.  
We omit these intervals in the right panels to focus on the net accretion onto the central region.
Within 3,000\,au, both $\dot{M}_{\rm inf}$ and $\dot{M}_{\rm inf}-\dot{M}_{\rm out}$ undergo strong temporal fluctuations during the calculation period, demonstrating highly time-dependent accretion at $t_{\rm ps}\sim 10^4 - 10^{5}$\,yr.
At 6,000\,au, fluctuations begin to appear around $t_{\rm ps} \gtrsim 10^{5}$\,yr, suggesting that longer integrations would reveal non-steady inflow on even larger scales.
Further simulations with longer evolution times are therefore required to fully characterize time variability and asymmetric accretion at $\gtrsim 10^{4}$\,au.

\subsection{Effects of Ambipolar Diffusion and the Hall Effect}
\label{sec:nonidealMHD}
In our simulations, we include only Ohmic dissipation as a non-ideal MHD effect, while ambipolar diffusion and the Hall effect are not included.
In this subsection, we discuss the possible effects of ambipolar diffusion and the Hall effect.

Ohmic dissipation becomes effective mainly in high-density regions, whereas ambipolar diffusion can play an important role at lower densities where the pseudodisk forms \citep{nakano02,Masson2016}.
Therefore, ambipolar diffusion could affect the structures discussed in this study.
The twisted pseudodisk found in our simulations originates from the misalignment between the magnetic field and the rotation axis.
Since ambipolar diffusion changes the coupling between the gas and magnetic field, it can modify the magnetic field strength and configuration during the collapse and subsequent accretion phase.
Thus, the detailed morphology of the pseudodisk and the physical properties of the disk and outflow may differ when ambipolar diffusion is included \citep{machida20b}.

The Hall effect is also neglected in our simulations.
During the collapse of a molecular cloud core, the Hall effect can become important on scales of approximately 100\,au \citep{Tsukamoto2015, Tsukamoto2017, Wurster2018, Marchand2019}.
Unlike Ohmic dissipation and ambipolar diffusion, the Hall effect depends on the relative orientation of the magnetic field and rotation.
Thus, in contrast to our Ohmic-only calculations, models with $\mathbf{B}\cdot\mathbf{\Omega}>0$ and $\mathbf{B}\cdot\mathbf{\Omega}<0$ are no longer dynamically equivalent when the Hall effect is included.
Depending on the sign of $\mathbf{B}\cdot\mathbf{\Omega}$, the Hall effect can enhance or weaken magnetic braking.
Consequently, even when all other initial conditions are identical, the disk size and mass can differ depending on the magnetic field orientation.
The Hall effect can also modify the directions of the angular momentum and magnetic field vectors and may therefore affect the circumstellar structures investigated in this study.

The distorted pseudodisk and asymmetric accretion found in our simulations originate from the misalignment between the magnetic field and rotation axis.
Although ambipolar diffusion and the Hall effect can modify the magnetic field configuration and the relative orientation between the magnetic field and rotation axis, these effects are unlikely to eliminate the misalignment completely.
As demonstrated in this study, even a small misalignment can produce a distorted pseudodisk.
Therefore, we expect that distorted pseudodisks and streamer-like accretion structures could still develop when ambipolar diffusion and the Hall effect are included.
Their inclusion may quantitatively modify the morphology of the pseudodisk and the properties of the disk, outflow, and accretion flows, while the basic mechanism identified in this study may remain applicable.
Future calculations including all three non-ideal MHD effects are required to quantitatively assess their impact.

\subsection{Possible Origins of Asymmetric Accretion Structures}
In this subsection, we discuss possible origins of asymmetric accretion based on our results and previous studies.
Previous numerical studies have shown that complex and asymmetric accretion structures can develop in turbulent and multi-scale star-forming environments.
Cloud-scale and zoom-in simulations have reported streamer-like structures and narrow accretion channels connecting circumstellar disks to the surrounding turbulent medium \citep{Kuffmeier2017,Kuffmeier2018,Kuffmeier2019,Seifried2013,Seifried2015,Lebreuilly2021,Lebreuilly2024,Lebreuilly2025,Yang2025}.
Recent zoom-in simulations by \citet{Granzow2026} have further shown that large-scale streamers can significantly alter the size, mass, and orientation of circumstellar disks.
\citet{Tu2024} also reported ``gravo-magneto sheetlets,'' which have morphological similarities to the distorted pseudodisks found in our simulations.
In these studies, complex gas motions and structures can develop in the presence of cloud-scale turbulence and interactions with the surrounding environment. 

Our simulations adopt a different and simpler setup.
We start from an isolated molecular cloud core without initial turbulence and do not consider continuous gas supply from scales beyond the initial cloud core.
Our setup is designed to investigate how a core evolves when the global magnetic field and rotation axis are initially misaligned and how this misalignment affects the resulting circumstellar structures and gas accretion.
Despite the absence of initial turbulence and continuous gas supply from larger scales, our simulations produce distorted pseudodisks and strongly asymmetric accretion.
In addition, the distorted pseudodisk is accompanied by narrow, streamer-like accretion structures, as shown in Section~\ref{sec:results}.

Our results therefore indicate that the misalignment between the magnetic field and rotation axis can provide one possible mechanism for producing streamer-like structures and asymmetric accretion, even in the absence of initial turbulence and continuous gas supply from larger scales.
This does not imply that the streamer-like structures and asymmetric accretion observed in star-forming regions originate solely from magnetic-field--rotation misalignment.
In realistic star-forming environments, cloud-scale turbulence, gas supply from larger scales, and magnetic-field--rotation misalignment may all contribute to the development of complex circumstellar structures and accretion flows.
Thus, our isolated-core calculations should be regarded as demonstrating one possible origin of asymmetric accretion rather than reproducing the full complexity of accretion from molecular-cloud scales.
Future multi-scale calculations including both turbulence and magnetic-field--rotation misalignment will be important for determining the relative roles of these effects.

\section{Summary}
We performed long-term, MHD simulations with ohmic dissipation of collapsing pre stellar cores to investigate how misalignment between the rotation axis and magnetic field drives asymmetric accretion onto circumstellar disks. 
Our suite includes an unmagnetized control run and magnetized models with a fixed mass-to-flux ratio ($\mu_0=3$), using both a fiducial small core ($M_{\rm cl}=0.81\,\msun$, $R_{\rm cl}=4,800$\,au) and a large core ($M_{\rm cl}=8.1\,\msun$, $R_{\rm cl}=48,000$\,au).
The key control parameter is the initial misalignment angle $\theta_0$, and we followed the evolution up to $\sim10^{5}$\,yr after protostar formation.

In the unmagnetized model ($B_0=0$; B00) and the aligned model ($\theta_0=0^\circ$; T00), an undistorted circumstellar disk forms. 
In the aligned model, the disk plane is perpendicular to both the magnetic field and angular momentum vectors, and the outflow propagates roughly along the initial magnetic field and rotation axis. 
Below, we summarize the main results for the misaligned models.

\begin{itemize}
\item  
All misaligned models form a rotationally supported disk that remains comparatively symmetric, while the surrounding magnetically shaped infalling envelope (pseudodisk) becomes strongly distorted and twisted after protostar formation.
Because the rotation axis of the pseudodisk is misaligned with the disk normal, the envelope becomes twisted over time. 
Gas accretion from the pseudodisk proceeds through several discrete, channel-like flows, with regions of enhanced accretion forming distinct streams. 
As a result, accretion onto the central region is neither axisymmetric nor spherically symmetric, and the infalling envelope exhibits complex three-dimensional structures, often spiral-like.

\item 
Protostellar outflows are launched even in models with highly complex envelope structures.
Inflow and outflow are strongly interleaved: inflows accrete inward while skirting outflow cavities, and outflows propagate outward while avoiding dense inflow channels.
By transporting angular momentum in specific directions, outflows reduce the angular momentum along the disk normal, 
contributing to the time evolution of the orientations of the disk and pseudodisk.
Because strong outflows are launched roughly along the disk normal, the outflow direction also changes accordingly, further enhancing the complexity of the surrounding gas distribution. 
Depending on the magnetic field configuration, secondary outflows may additionally be driven.

\item 
The outflow properties strongly depend on the misalignment angle. 
For small misalignment angles ($\theta_0 \lesssim 30^\circ$), outflows propagate roughly along the initial magnetic field direction, whereas for larger misalignments ($\theta_0 \gtrsim 30^\circ$), the outflow direction does not necessarily align with the initial field. 
Outflows are strongest in models with $\theta_0 \sim 30^\circ$--$60^\circ$, where a large fraction of gas is expelled, resulting in smaller final stellar and disk masses. 
In contrast, for $\theta_0 \lesssim 30^\circ$ and $\theta_0 \gtrsim  60^\circ$, magnetic braking efficiently removes angular momentum while allowing more mass to accrete, leading to more massive protostars. 

\item 
The mass inflow rate toward the central region and the mass outflow rate vary over long timescales as the structure of the infalling envelope evolves. 
In contrast, accretion from the disk onto the protostar shows mainly short-timescale variability (e.g., driven by disk gravitational instability) and is comparatively steady when averaged over longer periods.

\end{itemize}

We examined diagnostics relevant to observations.
First, the relative orientation between gas streams and magnetic fields is not unique.
In small-$\theta_0$ models, dense infall is often nearly perpendicular to the field in the inner few hundred au, but both parallel and perpendicular configurations occur.
In large-$\theta_0$ models, the angles are broadly distributed.
Thus, magnetic-field orientation (polarization observations) alone may not uniquely determine the flow direction.
Second, most infalling gas approaches the free-fall speed set by the protostellar mass, although a small fraction of dense gas in the pseudodisk can accrete significantly more slowly.

Finally, the spatial extent of asymmetric accretion depends on the parent core size.
In the small-core models, the gas structure remains complex even on scales larger than 2,000\,au, well beyond the disk scale, where the magnetically formed pseudodisk continues to be deformed by rotation and disturbed by outflows. 
In the large-core models, asymmetric accretion and complex envelope structures persist on scales exceeding 10,000\,au. 
Channel flows and accretion streams are also observed on scales larger than 3,000\,au.

This study does not rule out the possibility that streamer-like structures arise from initial turbulence or late-time accretion of gas and clumps from outside the core.
However, our results demonstrate that even in the absence of such external effects, the intrinsic physical properties of magnetized molecular cloud cores naturally give rise to streams and asymmetric envelope structures.

\appendix
\section{Dependence on the Disk and Outflow Definitions}
\label{sec:A1}
\begin{figure*}
\centering
\includegraphics[width=1.0\columnwidth]{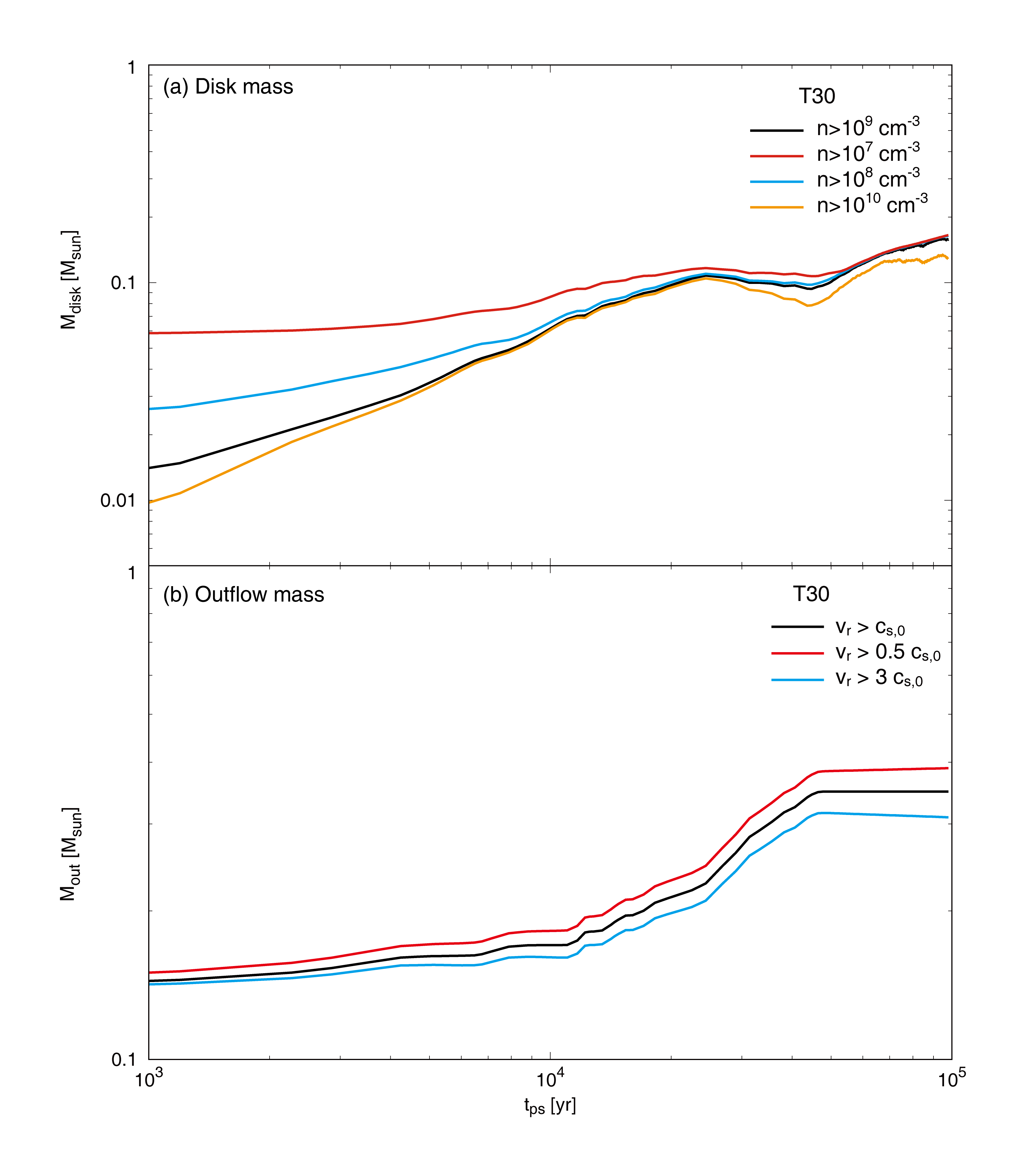}
\caption{
Dependence of the disk and outflow masses on the adopted definitions for model T30. 
The upper panel shows the disk mass calculated using density thresholds of $n>10^7$, $10^8$, $10^9$, and $10^{10}\,{\rm cm^{-3}}$. 
The lower panel shows the outflow mass calculated using velocity thresholds of $v_r>0.5\,c_{s,0}$, $v_r>c_{s,0}$, and $v_r>3\,c_{s,0}$, where $c_{s,0}$ is the initial sound speed.
}
\label{fig:18}
\end{figure*}
To examine the dependence of the disk and outflow masses on their definitions, we recalculate these quantities by varying the density and velocity thresholds. 
Figure~\ref{fig:18} shows the results for model T30. 
For the disk, we adopt density thresholds of $n=10^7$, $10^8$, $10^9$, and $10^{10}\,{\rm cm^{-3}}$. 
Although the disk mass depends on the adopted threshold at early times, the difference becomes small at later times, and the overall evolutionary trend is similar for all thresholds. 
Inspection of the two- and three-dimensional density and velocity distributions indicates that the high-density region selected by our fiducial criterion ($n>10^9\,\cc$) approximately traces the rotationally supported disk, although this density criterion does not provide a strict kinematic definition of the disk.

For the outflow, we adopt velocity thresholds of $v_r>0.5\,c_{s,0}$, $v_r>c_{s,0}$, and $v_r>3\,c_{s,0}$, where $c_{s,0}$ is the initial sound speed. 
The outflow mass shows only a moderate dependence on the velocity threshold, and its overall evolution is essentially unchanged. 
Thus, although the derived disk and outflow masses depend to some extent on the adopted thresholds, the overall evolutionary trends and our conclusions are not significantly affected by these choices.


\section{Relative Orientation between Magnetic Field and Infalling Gas Velocity}
\label{sec:A2}
\begin{figure*}
\centering
\includegraphics[width=1.0\columnwidth]{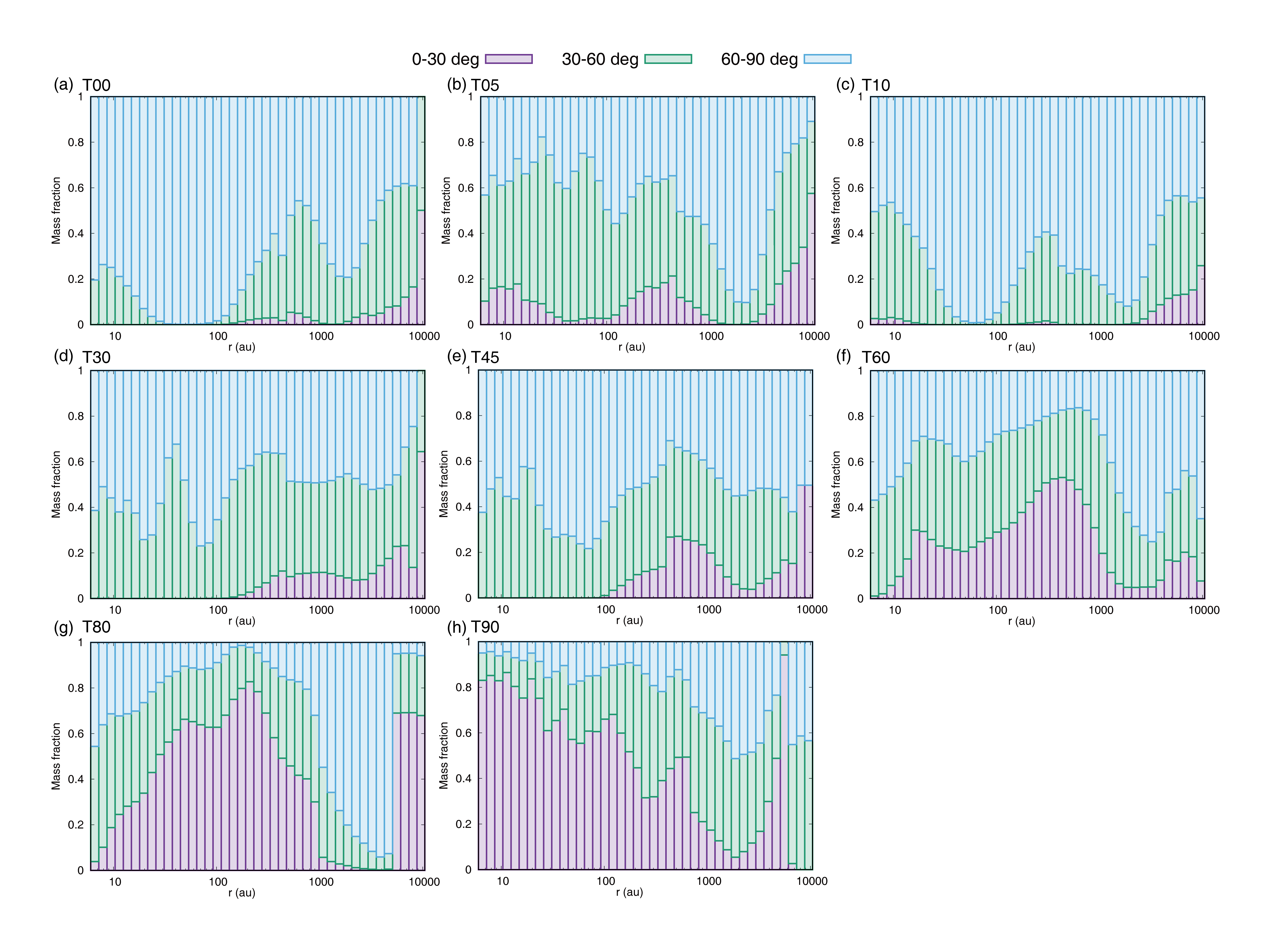}
\caption{
Radial distributions of the mass fractions of dense infalling gas classified according to the angle $\theta_{\rm VB}$ between the magnetic field and velocity vectors for models T00--T90 at the same epochs as Figure~\ref{fig:8}.
Only gas with $n>10^5\,{\rm cm^{-3}}$ and $v_r<0$ is included in the analysis.
The three components represent gas with $0^\circ \leq \theta_{\rm VB}<30^\circ$, $30^\circ \leq \theta_{\rm VB}<60^\circ$, and $60^\circ \leq \theta_{\rm VB}\leq90^\circ$.
The mass fraction of each component is calculated separately in each radial bin.
}
\label{fig:19}
\end{figure*}
To quantitatively examine the relative orientation between the magnetic field and the infalling gas velocity, we calculate the angle $\theta_{\rm VB}$ between the magnetic field and velocity vectors for each computational cell.
Here, we consider dense infalling gas with $n > 10^5\,{\rm cm^{-3}}$ and $v_r<0$.
The angle $\theta_{\rm VB}$ is defined in the range $0^\circ$--$90^\circ$, where $\theta_{\rm VB}=0^\circ$ and $90^\circ$ correspond to the magnetic field and velocity vectors being parallel and perpendicular, respectively.
We divide the gas into three angle ranges, $0^\circ \leq \theta_{\rm VB}<30^\circ$, $30^\circ \leq \theta_{\rm VB}<60^\circ$, and $60^\circ \leq \theta_{\rm VB}\leq90^\circ$.
For each radial bin, the mass fraction for a given angle range $k$ is calculated as
\begin{equation}
f_k(r)=\frac{\sum_{i\in k}\Delta M_i}{\sum_i \Delta M_i},
\end{equation}
where $\Delta M_i$ is the mass of the $i$th cell. 

Figure~\ref{fig:19} shows the radial distribution of the mass fractions for all models at the end of the simulations.
The relative orientation between the magnetic field and infalling velocity depends strongly on both radius and the initial misalignment angle.
In models with small initial misalignment angles, particularly T00 and T10, the component with $60^\circ \leq \theta_{\rm VB}\leq90^\circ$ dominates over a wide range of radii.
As the initial misalignment angle increases, the fraction of the component with $0^\circ \leq \theta_{\rm VB}<30^\circ$ generally increases.
For example, in models T60, T80, and T90, the nearly parallel component becomes significant or dominant over some radial ranges.

In many models, the dominant angle range changes with radius.
This radial variation indicates that the local direction of the infalling gas relative to the magnetic field cannot be characterized simply as either parallel or perpendicular.
Thus, although the initial misalignment angle affects the statistical distribution of $\theta_{\rm VB}$, it does not uniquely determine the direction of the infalling gas relative to the local magnetic field.

\section*{Acknowledgements}
This research used the computational resources of the HPCI system provided by the Cyber Science Center at Tohoku University and the Cybermedia Center at Osaka University (Project ID: hp230035, hp240010, and hp250007).
Simulations reported in this paper were also performed by 2024 and 2025 Koubo Kadai on Earth Simulator (NEC SX-ACE) at JAMSTEC. 
The present study was supported by JSPS KAKENHI Grant (JP25K07369: M.N.M., and JP21K13960 and JP21H01123 / JP23K20864: S.H.).
This work was supported by a NAOJ ALMA Scientific Research grant (No. 2022-22B). 
S.B. was supported by a Discovery Grant from NSERC.

\bibliography{machida}{}

@ARTICLE{aso15,
       author = {{Aso}, Yusuke and {Ohashi}, Nagayoshi and {Saigo}, Kazuya and {Koyamatsu}, Shin and {Aikawa}, Yuri and {Hayashi}, Masahiko and {Machida}, Masahiro N. and {Saito}, Masao and {Takakuwa}, Shigehisa and {Tomida}, Kengo and {Tomisaka}, Kohji and {Yen}, Hsi-Wei},
        title = "{ALMA Observations of the Transition from Infall Motion to Keplerian Rotation around the Late-phase Protostar TMC-1A}",
      journal = {\apj},
         year = 2015,
        month = oct,
       volume = {812},
       number = {1},
          eid = {27},
        pages = {27},
          doi = {10.1088/0004-637X/812/1/27},
archivePrefix = {arXiv},
       eprint = {1508.07013},
 primaryClass = {astro-ph.SR},
       adsurl = {https://ui.adsabs.harvard.edu/abs/2015ApJ...812...27A}
}

@ARTICLE{Basu24,
       author = {{Basu}, Shantanu and {Sharkawi}, Mahmoud and {Machida}, Masahiro N.},
        title = "{Outflows Driven from a Magnetic Pseudodisk}",
      journal = {\apj},
         year = 2024,
        month = apr,
       volume = {964},
       number = {2},
          eid = {116},
        pages = {116},
          doi = {10.3847/1538-4357/ad1bf3},
archivePrefix = {arXiv},
       eprint = {2401.04260},
 primaryClass = {astro-ph.SR},
       adsurl = {https://ui.adsabs.harvard.edu/abs/2024ApJ...964..116B}
}

@ARTICLE{crutcher99,
       author = {{Crutcher}, Richard M.},
        title = "{Magnetic Fields in Molecular Clouds: Observations Confront Theory}",
      journal = {\apj},
         year = 1999,
        month = aug,
       volume = {520},
       number = {2},
        pages = {706-713},
          doi = {10.1086/307483},
       adsurl = {https://ui.adsabs.harvard.edu/abs/1999ApJ...520..706C}
}

@ARTICLE{crutcher10,
       author = {{Crutcher}, Richard M. and {Wandelt}, Benjamin and {Heiles}, Carl and {Falgarone}, Edith and {Troland}, Thomas H.},
        title = "{Magnetic Fields in Interstellar Clouds from Zeeman Observations: Inference of Total Field Strengths by Bayesian Analysis}",
      journal = {\apj},
         year = 2010,
        month = dec,
       volume = {725},
       number = {1},
        pages = {466-479},
          doi = {10.1088/0004-637X/725/1/466},
       adsurl = {https://ui.adsabs.harvard.edu/abs/2010ApJ...725..466C}
}

@ARTICLE{hirano20,
       author = {{Hirano}, Shingo and {Tsukamoto}, Yusuke and {Basu}, Shantanu and {Machida}, Masahiro N.},
        title = "{The Effect of Misalignment between the Rotation Axis and Magnetic Field on the Circumstellar Disk}",
      journal = {\apj},
         year = 2020,
        month = aug,
       volume = {898},
       number = {2},
          eid = {118},
        pages = {118},
          doi = {10.3847/1538-4357/ab9f9d},
archivePrefix = {arXiv},
       eprint = {2006.13233},
 primaryClass = {astro-ph.SR},
       adsurl = {https://ui.adsabs.harvard.edu/abs/2020ApJ...898..118H}
}

@ARTICLE{Lee14,
       author = {{Lee}, Chin-Fei and {Hirano}, Naomi and {Zhang}, Qizhou and {Shang}, Hsien and {Ho}, Paul T.~P. and {Krasnopolsky}, Ruben},
        title = "{ALMA Results of the Pseudodisk, Rotating Disk, and Jet in the Continuum and HCO$^{+}$ in the Protostellar System HH 212}",
      journal = {\apj},
         year = 2014,
        month = may,
       volume = {786},
       number = {2},
          eid = {114},
        pages = {114},
          doi = {10.1088/0004-637X/786/2/114},
archivePrefix = {arXiv},
       eprint = {1403.5853},
 primaryClass = {astro-ph.GA},
       adsurl = {https://ui.adsabs.harvard.edu/abs/2014ApJ...786..114L}
}

@ARTICLE{larson69,
       author = {{Larson}, Richard B.},
        title = "{Numerical calculations of the dynamics of collapsing proto-star}",
      journal = {\mnras},
         year = 1969,
        month = jan,
       volume = {145},
        pages = {271},
          doi = {10.1093/mnras/145.3.271},
       adsurl = {https://ui.adsabs.harvard.edu/abs/1969MNRAS.145..271L}
}

@ARTICLE{machida04,
       author = {{Machida}, Masahiro N. and {Tomisaka}, Kohji and {Matsumoto}, Tomoaki},
        title = "{First MHD simulation of collapse and fragmentation of magnetized molecular cloud cores}",
      journal = {\mnras},
         year = 2004,
        month = feb,
       volume = {348},
       number = {1},
        pages = {L1-L5},
          doi = {10.1111/j.1365-2966.2004.07402.x},
archivePrefix = {arXiv},
       eprint = {astro-ph/0311339},
 primaryClass = {astro-ph},
       adsurl = {https://ui.adsabs.harvard.edu/abs/2004MNRAS.348L...1M}
}

@ARTICLE{machida05a,
       author = {{Machida}, Masahiro N. and {Matsumoto}, Tomoaki and {Tomisaka}, Kohji and {Hanawa}, Tomoyuki},
        title = "{Collapse and fragmentation of rotating magnetized clouds - I. Magnetic flux-spin relation}",
      journal = {\mnras},
         year = 2005,
        month = sep,
       volume = {362},
       number = {2},
        pages = {369-381},
          doi = {10.1111/j.1365-2966.2005.09297.x},
archivePrefix = {arXiv},
       eprint = {astro-ph/0506439},
 primaryClass = {astro-ph},
       adsurl = {https://ui.adsabs.harvard.edu/abs/2005MNRAS.362..369M}
}

@ARTICLE{Machida2006ApJ,
       author = {{Machida}, Masahiro N. and {Matsumoto}, Tomoaki and {Hanawa}, Tomoyuki and {Tomisaka}, Kohji},
        title = "{Evolution of Rotating Molecular Cloud Core with Oblique Magnetic Field}",
      journal = {\apj},
         year = 2006,
        month = jul,
       volume = {645},
       number = {2},
        pages = {1227-1245},
          doi = {10.1086/504423},
archivePrefix = {arXiv},
       eprint = {astro-ph/0602034},
 primaryClass = {astro-ph},
       adsurl = {https://ui.adsabs.harvard.edu/abs/2006ApJ...645.1227M}
}

@ARTICLE{machida06,
       author = {{Machida}, Masahiro N. and {Inutsuka}, Shu-ichiro and {Matsumoto}, Tomoaki},
        title = "{Second Core Formation and High-Speed Jets: Resistive Magnetohydrodynamic Nested Grid Simulations}",
      journal = {\apjl},
         year = 2006,
        month = aug,
       volume = {647},
       number = {2},
        pages = {L151-L154},
          doi = {10.1086/507179},
archivePrefix = {arXiv},
       eprint = {astro-ph/0603456},
 primaryClass = {astro-ph},
       adsurl = {https://ui.adsabs.harvard.edu/abs/2006ApJ...647L.151M}
}

@ARTICLE{machida10,
       author = {{Machida}, Masahiro N. and {Inutsuka}, Shu-ichiro and {Matsumoto}, Tomoaki},
        title = "{Formation Process of the Circumstellar Disk: Long-term Simulations in the Main Accretion Phase of Star Formation}",
      journal = {\apj},
         year = 2010,
        month = dec,
       volume = {724},
       number = {2},
        pages = {1006-1020},
          doi = {10.1088/0004-637X/724/2/1006},
archivePrefix = {arXiv},
       eprint = {1001.1404},
 primaryClass = {astro-ph.SR},
       adsurl = {https://ui.adsabs.harvard.edu/abs/2010ApJ...724.1006M}
}

@ARTICLE{machida11,
       author = {{Machida}, Masahiro N. and {Matsumoto}, Tomoaki},
        title = "{The origin and formation of the circumstellar disc}",
      journal = {\mnras},
         year = 2011,
        month = jun,
       volume = {413},
       number = {4},
        pages = {2767-2784},
          doi = {10.1111/j.1365-2966.2011.18349.x},
archivePrefix = {arXiv},
       eprint = {1008.0920},
 primaryClass = {astro-ph.SR},
       adsurl = {https://ui.adsabs.harvard.edu/abs/2011MNRAS.413.2767M}
}

@ARTICLE{machida12,
       author = {{Machida}, Masahiro N. and {Matsumoto}, Tomoaki},
        title = "{Impact of protostellar outflow on star formation: effects of the initial cloud mass}",
      journal = {\mnras},
         year = 2012,
        month = mar,
       volume = {421},
       number = {1},
        pages = {588-607},
          doi = {10.1111/j.1365-2966.2011.20336.x},
archivePrefix = {arXiv},
       eprint = {1108.3564},
 primaryClass = {astro-ph.SR},
       adsurl = {https://ui.adsabs.harvard.edu/abs/2012MNRAS.421..588M}
}

@ARTICLE{machida13,
       author = {{Machida}, Masahiro N. and {Hosokawa}, Takashi},
        title = "{Evolution of protostellar outflow around low-mass protostar}",
      journal = {\mnras},
         year = 2013,
        month = may,
       volume = {431},
       number = {2},
        pages = {1719-1744},
          doi = {10.1093/mnras/stt291},
archivePrefix = {arXiv},
       eprint = {1302.4176},
 primaryClass = {astro-ph.SR},
       adsurl = {https://ui.adsabs.harvard.edu/abs/2013MNRAS.431.1719M}
}

@ARTICLE{machida14,
       author = {{Machida}, Masahiro N. and {Inutsuka}, Shu-ichiro and {Matsumoto}, Tomoaki},
        title = "{Conditions for circumstellar disc formation: effects of initial cloud configuration and sink treatment}",
      journal = {\mnras},
         year = 2014,
        month = mar,
       volume = {438},
       number = {3},
        pages = {2278-2306},
          doi = {10.1093/mnras/stt2343},
archivePrefix = {arXiv},
       eprint = {1307.1747},
 primaryClass = {astro-ph.SR},
       adsurl = {https://ui.adsabs.harvard.edu/abs/2014MNRAS.438.2278M}
}

@ARTICLE{Machida2014ApJL,
       author = {{Machida}, Masahiro N.},
        title = "{Protostellar Jets Enclosed by Low-velocity Outflows}",
      journal = {\apjl},
         year = 2014,
        month = nov,
       volume = {796},
       number = {1},
          eid = {L17},
        pages = {L17},
          doi = {10.1088/2041-8205/796/1/L17},
archivePrefix = {arXiv},
       eprint = {1411.7124},
 primaryClass = {astro-ph.SR},
       adsurl = {https://ui.adsabs.harvard.edu/abs/2014ApJ...796L..17M}
}

@ARTICLE{machida20b,
       author = {{Machida}, Masahiro N. and {Basu}, Shantanu},
        title = "{Different modes of star formation - II. Gas accretion phase of initially subcritical star-forming clouds}",
      journal = {\mnras},
         year = 2020,
        month = may,
       volume = {494},
       number = {1},
        pages = {827-845},
          doi = {10.1093/mnras/staa672},
archivePrefix = {arXiv},
       eprint = {2003.03078},
 primaryClass = {astro-ph.SR},
       adsurl = {https://ui.adsabs.harvard.edu/abs/2020MNRAS.494..827M}
}

@ARTICLE{machida20,
       author = {{Machida}, Masahiro N. and {Hosokawa}, Takashi},
        title = "{Failed and delayed protostellar outflows with high-mass accretion rates}",
      journal = {\mnras},
         year = 2020,
        month = dec,
       volume = {499},
       number = {3},
        pages = {4490-4514},
          doi = {10.1093/mnras/staa3139},
archivePrefix = {arXiv},
       eprint = {2010.04866},
 primaryClass = {astro-ph.GA},
       adsurl = {https://ui.adsabs.harvard.edu/abs/2020MNRAS.499.4490M}
}

@ARTICLE{machida21,
       author = {{Machida}, Masahiro N.},
        title = "{Can high-velocity protostellar jets help to drive low-velocity outflow?}",
      journal = {\mnras},
         year = 2021,
        month = dec,
       volume = {508},
       number = {3},
        pages = {3208-3225},
          doi = {10.1093/mnras/stab2626},
archivePrefix = {arXiv},
       eprint = {2109.06367},
 primaryClass = {astro-ph.GA},
       adsurl = {https://ui.adsabs.harvard.edu/abs/2021MNRAS.508.3208M}
}

@ARTICLE{machida24,
       author = {{Machida}, Masahiro N. and {Basu}, Shantanu},
        title = "{Cloud Dissipation and Disk Wind in the Late Phase of Star Formation}",
      journal = {\apj},
         year = 2024,
        month = jul,
       volume = {970},
       number = {1},
          eid = {41},
        pages = {41},
          doi = {10.3847/1538-4357/ad4997},
archivePrefix = {arXiv},
       eprint = {2405.08271},
 primaryClass = {astro-ph.SR},
       adsurl = {https://ui.adsabs.harvard.edu/abs/2024ApJ...970...41M}
}

@ARTICLE{Machida2026,
       author = {{Machida}, Masahiro N. and {Basu}, Shantanu and {Hirano}, Shingo},
        title = "{Twisted Pseudodisk and Asymmetric Mass Accretion on the Circumstellar Disk}",
      journal = {\apj},
         year = 2026,
        month = mar,
       volume = {1000},
       number = {1},
          eid = {11},
        pages = {11},
          doi = {10.3847/1538-4357/ae41be},
archivePrefix = {arXiv},
       eprint = {2602.07741},
 primaryClass = {astro-ph.SR},
       adsurl = {https://ui.adsabs.harvard.edu/abs/2026ApJ..1000...11M}
}

@ARTICLE{masunaga00,
       author = {{Masunaga}, Hirohiko and {Inutsuka}, Shu-ichiro},
        title = "{A Radiation Hydrodynamic Model for Protostellar Collapse. II. The Second Collapse and the Birth of a Protostar}",
      journal = {\apj},
         year = 2000,
        month = mar,
       volume = {531},
       number = {1},
        pages = {350-365},
          doi = {10.1086/308439},
       adsurl = {https://ui.adsabs.harvard.edu/abs/2000ApJ...531..350M}
}

@ARTICLE{matsushita17,
       author = {{Matsushita}, Yuko and {Machida}, Masahiro N. and {Sakurai}, Yuya and {Hosokawa}, Takashi},
        title = "{Massive outflows driven by magnetic effects in star-forming clouds with high mass accretion rates}",
      journal = {\mnras},
         year = 2017,
        month = sep,
       volume = {470},
       number = {1},
        pages = {1026-1049},
          doi = {10.1093/mnras/stx893},
archivePrefix = {arXiv},
       eprint = {1704.03185},
 primaryClass = {astro-ph.SR},
       adsurl = {https://ui.adsabs.harvard.edu/abs/2017MNRAS.470.1026M}
}

@ARTICLE{ohashi14,
       author = {{Ohashi}, Nagayoshi and {Saigo}, Kazuya and {Aso}, Yusuke and {Aikawa}, Yuri and {Koyamatsu}, Shin and {Machida}, Masahiro N. and {Saito}, Masao and {Takahashi}, Sanemichi Z. and {Takakuwa}, Shigehisa and {Tomida}, Kengo and {Tomisaka}, Kohji and {Yen}, Hsi-Wei},
        title = "{Formation of a Keplerian Disk in the Infalling Envelope around L1527 IRS: Transformation from Infalling Motions to Kepler Motions}",
      journal = {\apj},
         year = 2014,
        month = dec,
       volume = {796},
       number = {2},
          eid = {131},
        pages = {131},
          doi = {10.1088/0004-637X/796/2/131},
archivePrefix = {arXiv},
       eprint = {1410.0172},
 primaryClass = {astro-ph.GA},
       adsurl = {https://ui.adsabs.harvard.edu/abs/2014ApJ...796..131O}
}

@INPROCEEDINGS{pineda23,
       author = {{Pineda}, J.~E. and {Arzoumanian}, D. and {Andre}, P. and {Friesen}, R.~K. and {Zavagno}, A. and {Clarke}, S.~D. and {Inoue}, T. and {Chen}, C. and {Lee}, Y. and {Soler}, J.~D. and {Kuffmeier}, M.},
        title = "{From Bubbles and Filaments to Cores and Disks: Gas Gathering and Growth of Structure Leading to the Formation of Stellar Systems}",
    booktitle = {Protostars and Planets VII},
         year = 2023,
       editor = {{Inutsuka}, S. and {Aikawa}, Y. and {Muto}, T. and {Tomida}, K. and {Tamura}, M.},
       series = {Astronomical Society of the Pacific Conference Series},
       volume = {534},
        month = jul,
        pages = {233},
          doi = {10.48550/arXiv.2205.03935},
archivePrefix = {arXiv},
       eprint = {2205.03935},
 primaryClass = {astro-ph.GA},
       adsurl = {https://ui.adsabs.harvard.edu/abs/2023ASPC..534..233P}
}

@ARTICLE{riaz24,
       author = {{Riaz}, B. and {Stamatellos}, D. and {Machida}, M.~N.},
        title = "{Observations of spiral and streamer on a candidate proto-brown dwarf}",
      journal = {\mnras},
         year = 2024,
        month = apr,
       volume = {529},
       number = {4},
        pages = {3601-3609},
          doi = {10.1093/mnras/stae724},
archivePrefix = {arXiv},
       eprint = {2403.07367},
 primaryClass = {astro-ph.SR},
       adsurl = {https://ui.adsabs.harvard.edu/abs/2024MNRAS.529.3601R}
}

@ARTICLE{sato23,
       author = {{Sato}, Asako and {Tokuda}, Kazuki and {Machida}, Masahiro N. and {Tachihara}, Kengo and {Harada}, Naoto and {Yamasaki}, Hayao and {Hirano}, Shingo and {Onishi}, Toshikazu and {Matsushita}, Yuko},
        title = "{Secondary Outflow Driven by the Protostar Ser-emb 15 in Serpens}",
      journal = {\apj},
         year = 2023,
        month = dec,
       volume = {958},
       number = {2},
          eid = {102},
        pages = {102},
          doi = {10.3847/1538-4357/ad0132},
archivePrefix = {arXiv},
       eprint = {2310.06055},
 primaryClass = {astro-ph.GA},
       adsurl = {https://ui.adsabs.harvard.edu/abs/2023ApJ...958..102S}
}

@ARTICLE{tokuda14,
       author = {{Tokuda}, Kazuki and {Onishi}, Toshikazu and {Saigo}, Kazuya and {Kawamura}, Akiko and {Fukui}, Yasuo and {Matsumoto}, Tomoaki and {Inutsuka}, Shu-ichiro and {Machida}, Masahiro N. and {Tomida}, Kengo and {Tachihara}, Kengo},
        title = "{ALMA Observations of a High-density Core in Taurus: Dynamical Gas Interaction at the Possible Site of a Multiple Star Formation}",
      journal = {\apjl},
         year = 2014,
        month = jul,
       volume = {789},
       number = {1},
          eid = {L4},
        pages = {L4},
          doi = {10.1088/2041-8205/789/1/L4},
archivePrefix = {arXiv},
       eprint = {1405.7574},
 primaryClass = {astro-ph.SR},
       adsurl = {https://ui.adsabs.harvard.edu/abs/2014ApJ...789L...4T}
}

@ARTICLE{tokuda16,
       author = {{Tokuda}, Kazuki and {Onishi}, Toshikazu and {Matsumoto}, Tomoaki and {Saigo}, Kazuya and {Kawamura}, Akiko and {Fukui}, Yasuo and {Inutsuka}, Shu-ichiro and {Machida}, Masahiro N. and {Tomida}, Kengo and {Tachihara}, Kengo and {Andr{\'e}}, Philippe},
        title = "{Revealing a Detailed Mass Distribution of a High-density Core MC27/L1521F in Taurus with ALMA}",
      journal = {\apj},
         year = 2016,
        month = jul,
       volume = {826},
       number = {1},
          eid = {26},
        pages = {26},
          doi = {10.3847/0004-637X/826/1/26},
archivePrefix = {arXiv},
       eprint = {1604.04954},
 primaryClass = {astro-ph.GA},
       adsurl = {https://ui.adsabs.harvard.edu/abs/2016ApJ...826...26T}
}

@ARTICLE{tokuda17,
       author = {{Tokuda}, Kazuki and {Onishi}, Toshikazu and {Saigo}, Kazuya and {Hosokawa}, Takashi and {Matsumoto}, Tomoaki and {Inutsuka}, Shu-ichiro and {Machida}, Masahiro N. and {Tomida}, Kengo and {Kunitomo}, Masanobu and {Kawamura}, Akiko and {Fukui}, Yasuo and {Tachihara}, Kengo},
        title = "{A Detached Protostellar Disk around a {\ensuremath{\sim}}0.2 M $_{{\ensuremath{\odot}}}$ Protostar in a Possible Site of a Multiple Star Formation in a Dynamical Environment in Taurus}",
      journal = {\apj},
         year = 2017,
        month = nov,
       volume = {849},
       number = {2},
          eid = {101},
        pages = {101},
          doi = {10.3847/1538-4357/aa8e9e},
archivePrefix = {arXiv},
       eprint = {1709.07113},
 primaryClass = {astro-ph.GA},
       adsurl = {https://ui.adsabs.harvard.edu/abs/2017ApJ...849..101T}
}

@ARTICLE{tokuda18,
       author = {{Tokuda}, Kazuki and {Onishi}, Toshikazu and {Saigo}, Kazuya and {Matsumoto}, Tomoaki and {Inoue}, Tsuyoshi and {Inutsuka}, Shu-ichiro and {Fukui}, Yasuo and {Machida}, Masahiro N. and {Tomida}, Kengo and {Hosokawa}, Takashi and {Kawamura}, Akiko and {Tachihara}, Kengo},
        title = "{Warm CO Gas Generated by Possible Turbulent Shocks in a Low-mass Star-forming Dense Core in Taurus}",
      journal = {\apj},
         year = 2018,
        month = jul,
       volume = {862},
       number = {1},
          eid = {8},
        pages = {8},
          doi = {10.3847/1538-4357/aac898},
archivePrefix = {arXiv},
       eprint = {1804.05603},
 primaryClass = {astro-ph.GA},
       adsurl = {https://ui.adsabs.harvard.edu/abs/2018ApJ...862....8T}
}

@ARTICLE{Tokuda2020,
       author = {{Tokuda}, Kazuki and {Fujishiro}, Kakeru and {Tachihara}, Kengo and {Takashima}, Tatsuyuki and {Fukui}, Yasuo and {Zahorecz}, Sarolta and {Saigo}, Kazuya and {Matsumoto}, Tomoaki and {Tomida}, Kengo and {Machida}, Masahiro N. and {Inutsuka}, Shu-ichiro and {Andr{\'e}}, Philippe and {Kawamura}, Akiko and {Onishi}, Toshikazu},
        title = "{FRagmentation and Evolution of Dense Cores Judged by ALMA (FREJA). I. Overview: Inner {\ensuremath{\sim}}1000 au Structures of Prestellar/Protostellar Cores in Taurus}",
      journal = {\apj},
         year = 2020,
        month = aug,
       volume = {899},
       number = {1},
          eid = {10},
        pages = {10},
          doi = {10.3847/1538-4357/ab9ca7},
archivePrefix = {arXiv},
       eprint = {2006.06361},
 primaryClass = {astro-ph.GA},
       adsurl = {https://ui.adsabs.harvard.edu/abs/2020ApJ...899...10T}
}

@ARTICLE{tokuda23,
       author = {{Tokuda}, Kazuki and {Fukaya}, Naofumi and {Tachihara}, Kengo and {Omura}, Mitsuki and {Harada}, Naoto and {Nozaki}, Shingo and {Shoshi}, Ayumu and {Machida}, Masahiro N.},
        title = "{An ALMA-resolved View of 7000 au Protostellar Gas Ring around the Class I Source CrA-IRS 2 as a Possible Sign of Magnetic Flux Advection}",
      journal = {\apjl},
         year = 2023,
        month = oct,
       volume = {956},
       number = {1},
          eid = {L16},
        pages = {L16},
          doi = {10.3847/2041-8213/acfca9},
archivePrefix = {arXiv},
       eprint = {2309.13821},
 primaryClass = {astro-ph.GA},
       adsurl = {https://ui.adsabs.harvard.edu/abs/2023ApJ...956L..16T}
}

@ARTICLE{Tokuda24, 
       author = {{Tokuda}, Kazuki and {Harada}, Naoto and {Omura}, Mitsuki and {Matsumoto}, Tomoaki and {Onishi}, Toshikazu and {Saigo}, Kazuya and {Shoshi}, Ayumu and {Nozaki}, Shingo and {Tachihara}, Kengo and {Fukaya}, Naofumi and {Fukui}, Yasuo and {Inutsuka}, Shu-ichiro and {Machida}, Masahiro N.},
        title = "{Discovery of Asymmetric Spike-like Structures of the 10 au Disk around the Very Low-luminosity Protostar Embedded in the Taurus Dense Core MC 27/L1521F with ALMA}",
      journal = {\apj},
         year = 2024,
        month = apr,
       volume = {965},
       number = {2},
          eid = {99},
        pages = {99},
          doi = {10.3847/1538-4357/ad2f9a},
archivePrefix = {arXiv},
       eprint = {2403.00305},
 primaryClass = {astro-ph.SR},
       adsurl = {https://ui.adsabs.harvard.edu/abs/2024ApJ...965...99T}
}

@INPROCEEDINGS{tsukamoto23b,
       author = {{Tsukamoto}, Y. and {Maury}, A. and {Commercon}, B. and {Alves}, F.~O. and {Cox}, E.~G. and {Sakai}, N. and {Ray}, T. and {Zhao}, B. and {Machida}, M.~N.},
        title = "{The Role of Magnetic Fields in the Formation of Protostars, Disks, and Outflows}",
    booktitle = {Protostars and Planets VII},
         year = 2023,
       editor = {{Inutsuka}, S. and {Aikawa}, Y. and {Muto}, T. and {Tomida}, K. and {Tamura}, M.},
       series = {Astronomical Society of the Pacific Conference Series},
       volume = {534},
        month = jul,
        pages = {317},
          doi = {10.48550/arXiv.2209.13765},
archivePrefix = {arXiv},
       eprint = {2209.13765},
 primaryClass = {astro-ph.SR},
       adsurl = {https://ui.adsabs.harvard.edu/abs/2023ASPC..534..317T}
}

@ARTICLE{nakano02,
       author = {{Nakano}, Takenori and {Nishi}, Ryoichi and {Umebayashi}, Toyoharu},
        title = "{Mechanism of Magnetic Flux Loss in Molecular Clouds}",
      journal = {\apj},
         year = 2002,
        month = jul,
       volume = {573},
       number = {1},
        pages = {199-214},
          doi = {10.1086/340587},
archivePrefix = {arXiv},
       eprint = {astro-ph/0203223},
 primaryClass = {astro-ph},
       adsurl = {https://ui.adsabs.harvard.edu/abs/2002ApJ...573..199N}
}

@ARTICLE{matsumoto17,
       author = {{Matsumoto}, Tomoaki and {Machida}, Masahiro N. and {Inutsuka}, Shu-ichiro},
        title = "{Circumstellar Disks and Outflows in Turbulent Molecular Cloud Cores: Possible Formation Mechanism for Misaligned Systems}",
      journal = {\apj},
         year = 2017,
        month = apr,
       volume = {839},
       number = {1},
          eid = {69},
        pages = {69},
          doi = {10.3847/1538-4357/aa6a1c},
archivePrefix = {arXiv},
       eprint = {1703.09139},
 primaryClass = {astro-ph.SR},
       adsurl = {https://ui.adsabs.harvard.edu/abs/2017ApJ...839...69M}
}

@ARTICLE{machida16,
       author = {{Machida}, Masahiro N. and {Matsumoto}, Tomoaki and {Inutsuka}, Shu-ichiro},
        title = "{Conditions for circumstellar disc formation - II. Effects of initial cloud stability and mass accretion rate}",
      journal = {\mnras},
         year = 2016,
        month = dec,
       volume = {463},
       number = {4},
        pages = {4246-4267},
          doi = {10.1093/mnras/stw2256},
       adsurl = {https://ui.adsabs.harvard.edu/abs/2016MNRAS.463.4246M}
}

@ARTICLE{Galli1993,
       author = {{Galli}, Daniele and {Shu}, Frank H.},
        title = "{Collapse of Magnetized Molecular Cloud Cores. II. Numerical Results}",
      journal = {\apj},
         year = 1993,
        month = nov,
       volume = {417},
        pages = {243},
          doi = {10.1086/173306},
       adsurl = {https://ui.adsabs.harvard.edu/abs/1993ApJ...417..243G}
}

@ARTICLE{Pineda2020,
       author = {{Pineda}, Jaime E. and {Segura-Cox}, Dominique and {Caselli}, Paola and {Cunningham}, Nichol and {Zhao}, Bo and {Schmiedeke}, Anika and {Maureira}, Mar{\'\i}a Jos{\'e} and {Neri}, Roberto},
        title = "{A protostellar system fed by a streamer of 10,500 au length}",
      journal = {Nature Astronomy},
         year = 2020,
        month = jan,
       volume = {4},
        pages = {1158-1163},
          doi = {10.1038/s41550-020-1150-z},
archivePrefix = {arXiv},
       eprint = {2007.13430},
 primaryClass = {astro-ph.GA},
       adsurl = {https://ui.adsabs.harvard.edu/abs/2020NatAs...4.1158P}
}

@ARTICLE{Yen2019,
       author = {{Yen}, Hsi-Wei and {Gu}, Pin-Gao and {Hirano}, Naomi and {Koch}, Patrick M. and {Lee}, Chin-Fei and {Liu}, Hauyu Baobab and {Takakuwa}, Shigehisa},
        title = "{HL Tau Disk in HCO$^{+}$ (3-2) and (1-0) with ALMA: Gas Density, Temperature, Gap, and One-arm Spiral}",
      journal = {\apj},
         year = 2019,
        month = aug,
       volume = {880},
       number = {2},
          eid = {69},
        pages = {69},
          doi = {10.3847/1538-4357/ab29f8},
archivePrefix = {arXiv},
       eprint = {1906.05535},
 primaryClass = {astro-ph.SR},
       adsurl = {https://ui.adsabs.harvard.edu/abs/2019ApJ...880...69Y}
}

@ARTICLE{Kido2025,
       author = {{Kido}, Miyu and {Yen}, Hsi-Wei and {Sai}, Jinshi and {Takakuwa}, Shigehisa and {Ohashi}, Nagayoshi and {Aikawa}, Yuri and {Aso}, Yusuke and {Flores}, Christian and {Han}, Ilseung and {Koch}, Patrick M. and {Kwon}, Woojin and {Lee}, Jeong-Eun and {Li}, Zhi-Yun and {Looney}, Leslie W. and {Narang}, Mayank and {Saigo}, Kazuya and {Sharma}, Rajeeb and {Thieme}, Travis J. and {Tomida}, Kengo and {Williams}, Jonathan P.},
        title = "{Early Planet Formation in Embedded Disks (eDisk). XXI. Limited Role of Streamers in Mass Supply to the Disk in the Class 0 Protostar IRAS 16544-1604}",
      journal = {\apj},
         year = 2025,
        month = jun,
       volume = {985},
       number = {2},
          eid = {166},
        pages = {166},
          doi = {10.3847/1538-4357/adc72a},
archivePrefix = {arXiv},
       eprint = {2504.00495},
 primaryClass = {astro-ph.SR},
       adsurl = {https://ui.adsabs.harvard.edu/abs/2025ApJ...985..166K}
}

@ARTICLE{Misugi2024,
       author = {{Misugi}, Yoshiaki and {Inutsuka}, Shu-ichiro and {Arzoumanian}, Doris and {Tsukamoto}, Yusuke},
        title = "{Evolution of the Angular Momentum of Molecular Cloud Cores in Magnetized Molecular Filaments}",
      journal = {\apj},
         year = 2024,
        month = mar,
       volume = {963},
       number = {2},
          eid = {106},
        pages = {106},
          doi = {10.3847/1538-4357/ad1990},
archivePrefix = {arXiv},
       eprint = {2312.16920},
 primaryClass = {astro-ph.GA},
       adsurl = {https://ui.adsabs.harvard.edu/abs/2024ApJ...963..106M}
}

@ARTICLE{Hull2013,
       author = {{Hull}, Charles L.~H. and {Plambeck}, Richard L. and {Bolatto}, Alberto D. and {Bower}, Geoffrey C. and {Carpenter}, John M. and {Crutcher}, Richard M. and {Fiege}, Jason D. and {Franzmann}, Erica and {Hakobian}, Nicholas S. and {Heiles}, Carl and {Houde}, Martin and {Hughes}, A. Meredith and {Jameson}, Katherine and {Kwon}, Woojin and {Lamb}, James W. and {Looney}, Leslie W. and {Matthews}, Brenda C. and {Mundy}, Lee and {Pillai}, Thushara and {Pound}, Marc W. and {Stephens}, Ian W. and {Tobin}, John J. and {Vaillancourt}, John E. and {Volgenau}, N.~H. and {Wright}, Melvyn C.~H.},
        title = "{Misalignment of Magnetic Fields and Outflows in Protostellar Cores}",
      journal = {\apj},
         year = 2013,
        month = may,
       volume = {768},
       number = {2},
          eid = {159},
        pages = {159},
          doi = {10.1088/0004-637X/768/2/159},
archivePrefix = {arXiv},
       eprint = {1212.0540},
 primaryClass = {astro-ph.SR},
       adsurl = {https://ui.adsabs.harvard.edu/abs/2013ApJ...768..159H}
}

@ARTICLE{Hirano2019,
       author = {{Hirano}, Shingo and {Machida}, Masahiro N.},
        title = "{Origin of misalignments: protostellar jet, outflow, circumstellar disc, and magnetic field}",
      journal = {\mnras},
         year = 2019,
        month = jun,
       volume = {485},
       number = {4},
        pages = {4667-4674},
          doi = {10.1093/mnras/stz740},
archivePrefix = {arXiv},
       eprint = {1903.05283},
 primaryClass = {astro-ph.SR},
       adsurl = {https://ui.adsabs.harvard.edu/abs/2019MNRAS.485.4667H}
}

@ARTICLE{Machida2020d,
       author = {{Machida}, Masahiro N. and {Hirano}, Shingo and {Kitta}, Hideyuki},
        title = "{Misalignment of magnetic fields, outflows, and discs in star-forming clouds}",
      journal = {\mnras},
         year = 2020,
        month = jan,
       volume = {491},
       number = {2},
        pages = {2180-2197},
          doi = {10.1093/mnras/stz3159},
archivePrefix = {arXiv},
       eprint = {1911.03851},
 primaryClass = {astro-ph.SR},
       adsurl = {https://ui.adsabs.harvard.edu/abs/2020MNRAS.491.2180M}
}

@ARTICLE{Tu2024,
       author = {{Tu}, Yisheng and {Li}, Zhi-Yun and {Lam}, Ka Ho and {Tomida}, Kengo and {Hsu}, Chun-Yen},
        title = "{Protostellar discs fed by dense collapsing gravomagneto sheetlets}",
      journal = {\mnras},
         year = 2024,
        month = feb,
       volume = {527},
       number = {4},
        pages = {10131-10150},
          doi = {10.1093/mnras/stad3843},
archivePrefix = {arXiv},
       eprint = {2307.16774},
 primaryClass = {astro-ph.EP},
       adsurl = {https://ui.adsabs.harvard.edu/abs/2024MNRAS.52710131T}
}

@ARTICLE{Matsushita2018,
       author = {{Matsushita}, Yuko and {Sakurai}, Yuya and {Hosokawa}, Takashi and {Machida}, Masahiro N.},
        title = "{Massive outflows driven by magnetic effects - II. Comparison with observations}",
      journal = {\mnras},
         year = 2018,
        month = mar,
       volume = {475},
       number = {1},
        pages = {391-403},
          doi = {10.1093/mnras/stx3070},
archivePrefix = {arXiv},
       eprint = {1711.05910},
 primaryClass = {astro-ph.SR},
       adsurl = {https://ui.adsabs.harvard.edu/abs/2018MNRAS.475..391M}
}

@ARTICLE{Hirano2025,
       author = {{Hirano}, Shingo and {Aikawa}, Yuri and {Machida}, Masahiro N.},
        title = "{Velocity Structure of Circumstellar Environment around Class 0/I Protostars: Uncertainty in the Protostellar Mass Estimation Using Circumstellar Velocities}",
      journal = {\apj},
         year = 2025,
        month = apr,
       volume = {982},
       number = {2},
          eid = {166},
        pages = {166},
          doi = {10.3847/1538-4357/adb96f},
archivePrefix = {arXiv},
       eprint = {2501.17492},
 primaryClass = {astro-ph.SR},
       adsurl = {https://ui.adsabs.harvard.edu/abs/2025ApJ...982..166H}
}

@ARTICLE{Nozaki2023,
       author = {{Nozaki}, Shingo and {Machida}, Masahiro N.},
        title = "{Environmental effects of star-forming cores on mass accretion rate}",
      journal = {\mnras},
         year = 2023,
        month = mar,
       volume = {519},
       number = {4},
        pages = {5017-5030},
          doi = {10.1093/mnras/stac3819},
archivePrefix = {arXiv},
       eprint = {2212.13024},
 primaryClass = {astro-ph.SR},
       adsurl = {https://ui.adsabs.harvard.edu/abs/2023MNRAS.519.5017N}
}

@ARTICLE{Nozaki2025,
       author = {{Nozaki}, Shingo and {Machida}, Masahiro N.},
        title = "{How the External Medium Outside Prestellar Cores Affects Protostellar Growth: Variations in Accretion Rate and Evolution of Disks and Outflows}",
      journal = {\apj},
         year = 2025,
        month = dec,
       volume = {995},
       number = {1},
          eid = {127},
        pages = {127},
          doi = {10.3847/1538-4357/ae18d4},
archivePrefix = {arXiv},
       eprint = {2510.24065},
 primaryClass = {astro-ph.GA},
       adsurl = {https://ui.adsabs.harvard.edu/abs/2025ApJ...995..127N}
}

@ARTICLE{Okoda2021,
       author = {{Okoda}, Yuki and {Oya}, Yoko and {Francis}, Logan and {Johnstone}, Doug and {Inutsuka}, Shu-ichiro and {Ceccarelli}, Cecilia and {Codella}, Claudio and {Chandler}, Claire and {Sakai}, Nami and {Aikawa}, Yuri and {Alves}, Felipe O. and {Balucani}, Nadia and {Bianchi}, Eleonora and {Bouvier}, Mathilde and {Caselli}, Paola and {Caux}, Emmanuel and {Charnley}, Steven and {Choudhury}, Spandan and {De Simone}, Marta and {Dulieu}, Francois and {Dur{\'a}n}, Aurora and {Evans}, Lucy and {Favre}, C{\'e}cile and {Fedele}, Davide and {Feng}, Siyi and {Fontani}, Francesco and {Hama}, Tetsuya and {Hanawa}, Tomoyuki and {Herbst}, Eric and {Hirota}, Tomoya and {Imai}, Muneaki and {Isella}, Andrea and {J{\'\i}menez-Serra}, Izaskun and {Kahane}, Claudine and {Lefloch}, Bertrand and {Loinard}, Laurent and {L{\'o}pez-Sepulcre}, Ana and {Maud}, Luke T. and {Maureira}, Mar{\'\i}a Jos{\'e} and {Menard}, Francois and {Mercimek}, Seyma and {Miotello}, Anna and {Moellenbrock}, George and {Mori}, Shoji and {Murillo}, Nadia M. and {Nakatani}, Riouhei and {Nomura}, Hideko and {Oba}, Yasuhiro and {O'Donoghue}, Ross and {Ohashi}, Satoshi and {Ospina-Zamudio}, Juan and {Pineda}, Jaime E. and {Podio}, Linda and {Rimola}, Albert and {Sakai}, Takeshi and {Segura-Cox}, Dominique and {Shirley}, Yancy and {Svoboda}, Brian and {Taquet}, Vianney and {Testi}, Leonardo and {Vastel}, Charlotte and {Viti}, Serena and {Watanabe}, Naoki and {Watanabe}, Yoshimasa and {Witzel}, Arezu and {Xue}, Ci and {Zhang}, Yichen and {Zhao}, Bo and {Yamamoto}, Satoshi},
        title = "{FAUST. II. Discovery of a Secondary Outflow in IRAS 15398-3359: Variability in Outflow Direction during the Earliest Stage of Star Formation?}",
      journal = {\apj},
         year = 2021,
        month = mar,
       volume = {910},
       number = {1},
          eid = {11},
        pages = {11},
          doi = {10.3847/1538-4357/abddb1},
archivePrefix = {arXiv},
       eprint = {2101.07404},
 primaryClass = {astro-ph.SR},
       adsurl = {https://ui.adsabs.harvard.edu/abs/2021ApJ...910...11O}
}

@ARTICLE{Sai2024,
       author = {{Sai}, Jinshi and {Yen}, Hsi-Wei and {Machida}, Masahiro N. and {Ohashi}, Nagayoshi and {Aso}, Yusuke and {Maury}, Ana{\"e}lle J. and {Maret}, S{\'e}bastien},
        title = "{Multiple Outflows around a Single Protostar IRAS 15398‑3359}",
      journal = {\apj},
         year = 2024,
        month = may,
       volume = {966},
       number = {2},
          eid = {192},
        pages = {192},
          doi = {10.3847/1538-4357/ad34b7},
archivePrefix = {arXiv},
       eprint = {2403.10857},
 primaryClass = {astro-ph.SR},
       adsurl = {https://ui.adsabs.harvard.edu/abs/2024ApJ...966..192S}
}

@ARTICLE{Hennebelle2009,
       author = {{Hennebelle}, P. and {Ciardi}, A.},
        title = "{Disk formation during collapse of magnetized protostellar cores}",
      journal = {\aap},
         year = 2009,
        month = nov,
       volume = {506},
       number = {2},
        pages = {L29-L32},
          doi = {10.1051/0004-6361/200913008},
archivePrefix = {arXiv},
       eprint = {0909.3190},
 primaryClass = {astro-ph.GA},
       adsurl = {https://ui.adsabs.harvard.edu/abs/2009A&A...506L..29H}
}

@ARTICLE{Takakuwa2013,
       author = {{Takakuwa}, Shigehisa and {Saito}, Masao and {Lim}, Jeremy and {Saigo}, Kazuya},
        title = "{Evidence for Infalling Gas of Low Angular Momentum toward the L1551 NE Keplerian Circumbinary Disk}",
      journal = {\apj},
         year = 2013,
        month = oct,
       volume = {776},
       number = {1},
          eid = {51},
        pages = {51},
          doi = {10.1088/0004-637X/776/1/51},
archivePrefix = {arXiv},
       eprint = {1308.2054},
 primaryClass = {astro-ph.GA},
       adsurl = {https://ui.adsabs.harvard.edu/abs/2013ApJ...776...51T}
}

@ARTICLE{Chou2014,
       author = {{Chou}, Ti-Lin and {Takakuwa}, Shigehisa and {Yen}, Hsi-Wei and {Ohashi}, Nagayoshi and {Ho}, Paul T.~P.},
        title = "{Transition from the Infalling Envelope to the Keplerian Disk around L1551 IRS 5}",
      journal = {\apj},
         year = 2014,
        month = nov,
       volume = {796},
       number = {1},
          eid = {70},
        pages = {70},
          doi = {10.1088/0004-637X/796/1/70},
archivePrefix = {arXiv},
       eprint = {1410.3927},
 primaryClass = {astro-ph.SR},
       adsurl = {https://ui.adsabs.harvard.edu/abs/2014ApJ...796...70C}
}

@ARTICLE{Takakuwa2020,
       author = {{Takakuwa}, Shigehisa and {Saigo}, Kazuya and {Matsumoto}, Tomoaki and {Saito}, Masao and {Lim}, Jeremy and {Yen}, Hsi-Wei and {Ohashi}, Nagayoshi and {Ho}, Paul T.~P. and {Looney}, Leslie W.},
        title = "{Circumbinary Disks of the Protostellar Binary Systems in the L1551 Region}",
      journal = {\apj},
         year = 2020,
        month = jul,
       volume = {898},
       number = {1},
          eid = {10},
        pages = {10},
          doi = {10.3847/1538-4357/ab9b7c},
archivePrefix = {arXiv},
       eprint = {2006.05723},
 primaryClass = {astro-ph.GA},
       adsurl = {https://ui.adsabs.harvard.edu/abs/2020ApJ...898...10T}
}

@ARTICLE{Sai2022,
       author = {{Sai}, Jinshi and {Ohashi}, Nagayoshi and {Maury}, Ana{\"e}lle J. and {Maret}, S{\'e}bastien and {Yen}, Hsi-Wei and {Aso}, Yusuke and {Gaudel}, Mathilde},
        title = "{Which Part of Dense Cores Feeds Material to Protostars? The Case of L1489 IRS}",
      journal = {\apj},
         year = 2022,
        month = jan,
       volume = {925},
       number = {1},
          eid = {12},
        pages = {12},
          doi = {10.3847/1538-4357/ac341d},
archivePrefix = {arXiv},
       eprint = {2110.14210},
 primaryClass = {astro-ph.SR},
       adsurl = {https://ui.adsabs.harvard.edu/abs/2022ApJ...925...12S}
}

@ARTICLE{Machida2011PASJ,
       author = {{Machida}, Masahiro N. and {Inutsuka}, Shu-ichiro and {Matsumoto}, Tomoaki},
        title = "{Effect of Magnetic Braking on Circumstellar Disk Formation in a Strongly Magnetized Cloud}",
      journal = {\pasj},
         year = 2011,
        month = jun,
       volume = {63},
       number = {3},
        pages = {555-573},
          doi = {10.1093/pasj/63.3.555},
archivePrefix = {arXiv},
       eprint = {1009.2140},
 primaryClass = {astro-ph.SR},
       adsurl = {https://ui.adsabs.harvard.edu/abs/2011PASJ...63..555M}
}

@ARTICLE{Arzoumanian2011,
       author = {{Arzoumanian}, D. and {Andr{\'e}}, Ph. and {Didelon}, P. and {K{\"o}nyves}, V. and {Schneider}, N. and {Men'shchikov}, A. and {Sousbie}, T. and {Zavagno}, A. and {Bontemps}, S. and {di Francesco}, J. and {Griffin}, M. and {Hennemann}, M. and {Hill}, T. and {Kirk}, J. and {Martin}, P. and {Minier}, V. and {Molinari}, S. and {Motte}, F. and {Peretto}, N. and {Pezzuto}, S. and {Spinoglio}, L. and {Ward-Thompson}, D. and {White}, G. and {Wilson}, C.~D.},
        title = "{Characterizing interstellar filaments with Herschel in IC 5146}",
      journal = {\aap},
         year = 2011,
        month = may,
       volume = {529},
          eid = {L6},
        pages = {L6},
          doi = {10.1051/0004-6361/201116596},
archivePrefix = {arXiv},
       eprint = {1103.0201},
 primaryClass = {astro-ph.GA},
       adsurl = {https://ui.adsabs.harvard.edu/abs/2011A&A...529L...6A}
}

@ARTICLE{Andre2010,
       author = {{Andr{\'e}}, Ph. and {Men'shchikov}, A. and {Bontemps}, S. and {K{\"o}nyves}, V. and {Motte}, F. and {Schneider}, N. and {Didelon}, P. and {Minier}, V. and {Saraceno}, P. and {Ward-Thompson}, D. and {di Francesco}, J. and {White}, G. and {Molinari}, S. and {Testi}, L. and {Abergel}, A. and {Griffin}, M. and {Henning}, Th. and {Royer}, P. and {Mer{\'\i}n}, B. and {Vavrek}, R. and {Attard}, M. and {Arzoumanian}, D. and {Wilson}, C.~D. and {Ade}, P. and {Aussel}, H. and {Baluteau}, J.-P. and {Benedettini}, M. and {Bernard}, J.-Ph. and {Blommaert}, J.~A.~D.~L. and {Cambr{\'e}sy}, L. and {Cox}, P. and {di Giorgio}, A. and {Hargrave}, P. and {Hennemann}, M. and {Huang}, M. and {Kirk}, J. and {Krause}, O. and {Launhardt}, R. and {Leeks}, S. and {Le Pennec}, J. and {Li}, J.~Z. and {Martin}, P.~G. and {Maury}, A. and {Olofsson}, G. and {Omont}, A. and {Peretto}, N. and {Pezzuto}, S. and {Prusti}, T. and {Roussel}, H. and {Russeil}, D. and {Sauvage}, M. and {Sibthorpe}, B. and {Sicilia-Aguilar}, A. and {Spinoglio}, L. and {Waelkens}, C. and {Woodcraft}, A. and {Zavagno}, A.},
        title = "{From filamentary clouds to prestellar cores to the stellar IMF: Initial highlights from the Herschel Gould Belt Survey}",
      journal = {\aap},
         year = 2010,
        month = jul,
       volume = {518},
          eid = {L102},
        pages = {L102},
          doi = {10.1051/0004-6361/201014666},
archivePrefix = {arXiv},
       eprint = {1005.2618},
 primaryClass = {astro-ph.GA},
       adsurl = {https://ui.adsabs.harvard.edu/abs/2010A&A...518L.102A}
}

@ARTICLE{Kido2023,
       author = {{Kido}, Miyu and {Takakuwa}, Shigehisa and {Saigo}, Kazuya and {Ohashi}, Nagayoshi and {Tobin}, John J. and {J{\o}rgensen}, Jes K. and {Aikawa}, Yuri and {Aso}, Yusuke and {Encalada}, Frankie J. and {Flores}, Christian and {Gavino}, Sacha and {de Gregorio-Monsalvo}, Itziar and {Han}, Ilseung and {Hirano}, Shingo and {Koch}, Patrick M. and {Kwon}, Woojin and {Lai}, Shih-Ping and {Lee}, Chang Won and {Lee}, Jeong-Eun and {Li}, Zhi-Yun and {Lin}, Zhe-Yu Daniel and {Looney}, Leslie W. and {Mori}, Shoji and {Narayanan}, Suchitra and {Plunkett}, Adele L. and {Phuong}, Nguyen Thi and {(Insa Choi)}, Jinshi Sai and {Santamar{\'\i}a-Miranda}, Alejandro and {Sharma}, Rajeeb and {Sheehan}, Patrick D. and {Thieme}, Travis J. and {Tomida}, Kengo and {van't Hoff}, Merel L.~R. and {Williams}, Jonathan P. and {Yamato}, Yoshihide and {Yen}, Hsi-Wei},
        title = "{Early Planet Formation in Embedded Disks (eDisk). VII. Keplerian Disk, Disk Substructure, and Accretion Streamers in the Class 0 Protostar IRAS 16544-1604 in CB 68}",
      journal = {\apj},
         year = 2023,
        month = aug,
       volume = {953},
       number = {2},
          eid = {190},
        pages = {190},
          doi = {10.3847/1538-4357/acdd7a},
archivePrefix = {arXiv},
       eprint = {2306.15443},
 primaryClass = {astro-ph.EP},
       adsurl = {https://ui.adsabs.harvard.edu/abs/2023ApJ...953..190K}
}

@ARTICLE{Flores2023,
       author = {{Flores}, Christian and {Ohashi}, Nagayoshi and {Tobin}, John J. and {J{\o}rgensen}, Jes K. and {Takakuwa}, Shigehisa and {Li}, Zhi-Yun and {Lin}, Zhe-Yu Daniel and {van't Hoff}, Merel L.~R. and {Plunkett}, Adele L. and {Yamato}, Yoshihide and {Sai (Insa Choi)}, Jinshi and {Koch}, Patrick M. and {Yen}, Hsi-Wei and {Aikawa}, Yuri and {Aso}, Yusuke and {de Gregorio-Monsalvo}, Itziar and {Kido}, Miyu and {Kwon}, Woojin and {Lee}, Jeong-Eun and {Lee}, Chang Won and {Looney}, Leslie W. and {Santamar{\'\i}a-Miranda}, Alejandro and {Sharma}, Rajeeb and {Thieme}, Travis J. and {Williams}, Jonathan P. and {Han}, Ilseung and {Narayanan}, Suchitra and {Lai}, Shih-Ping},
        title = "{Early Planet Formation in Embedded Disks (eDisk). XII. Accretion Streamers, Protoplanetary Disk, and Outflow in the Class I Source Oph IRS 63}",
      journal = {\apj},
         year = 2023,
        month = nov,
       volume = {958},
       number = {1},
          eid = {98},
        pages = {98},
          doi = {10.3847/1538-4357/acf7c1},
archivePrefix = {arXiv},
       eprint = {2310.14617},
 primaryClass = {astro-ph.SR},
       adsurl = {https://ui.adsabs.harvard.edu/abs/2023ApJ...958...98F}
}

@ARTICLE{Tanious2024,
       author = {{Tanious}, M. and {Le Gal}, R. and {Neri}, R. and {Faure}, A. and {Gupta}, A. and {Law}, C.~J. and {Huang}, J. and {Cuello}, N. and {Williams}, J.~P. and {M{\'e}nard}, F.},
        title = "{Anatomy of the Class I protostar L1489 IRS with NOEMA. I. Disk, streamers, outflow(s) and bubbles at 3 mm}",
      journal = {\aap},
         year = 2024,
        month = jul,
       volume = {687},
          eid = {A92},
        pages = {A92},
          doi = {10.1051/0004-6361/202348785},
archivePrefix = {arXiv},
       eprint = {2403.18905},
 primaryClass = {astro-ph.SR},
       adsurl = {https://ui.adsabs.harvard.edu/abs/2024A&A...687A..92T}
}

@ARTICLE{Cacciapuoti2024,
       author = {{Cacciapuoti}, L. and {Macias}, E. and {Gupta}, A. and {Testi}, L. and {Miotello}, A. and {Espaillat}, C. and {K{\"u}ffmeier}, M. and {van Terwisga}, S. and {Tobin}, J. and {Grant}, S. and {Manara}, C.~F. and {Segura-Cox}, D. and {Wendeborn}, J. and {Klessen}, R.~S. and {Maury}, A.~J. and {Lebreuilly}, U. and {Hennebelle}, P. and {Molinari}, S.},
        title = "{A dusty streamer infalling onto the disk of a class I protostar. ALMA dual-band constraints on grain properties and the mass-infall rate}",
      journal = {\aap},
         year = 2024,
        month = feb,
       volume = {682},
          eid = {A61},
        pages = {A61},
          doi = {10.1051/0004-6361/202347486},
archivePrefix = {arXiv},
       eprint = {2311.13723},
 primaryClass = {astro-ph.EP},
       adsurl = {https://ui.adsabs.harvard.edu/abs/2024A&A...682A..61C}
}

@ARTICLE{Aso2023,
       author = {{Aso}, Yusuke and {Kwon}, Woojin and {Ohashi}, Nagayoshi and {J{\o}rgensen}, Jes K. and {Tobin}, John J. and {Aikawa}, Yuri and {de Gregorio-Monsalvo}, Itziar and {Han}, Ilseung and {Kido}, Miyu and {Koch}, Patrick M. and {Lai}, Shih-Ping and {Lee}, Chang Won and {Lee}, Jeong-Eun and {Li}, Zhi-Yun and {Lin}, Zhe-Yu Daniel and {Looney}, Leslie W. and {Narayanan}, Suchitra and {Phuong}, Nguyen Thi and {Sai}, Jinshi (Insa Choi) and {Saigo}, Kazuya and {Santamar{\'\i}a-Miranda}, Alejandro and {Sharma}, Rajeeb and {Takakuwa}, Shigehisa and {Thieme}, Travis J. and {Tomida}, Kengo and {Williams}, Jonathan P. and {Yen}, Hsi-Wei},
        title = "{Early Planet Formation in Embedded Disks (eDisk). VI. Kinematic Structures around the Very-low-mass Protostar IRAS 16253-2429}",
      journal = {\apj},
         year = 2023,
        month = sep,
       volume = {954},
       number = {1},
          eid = {101},
        pages = {101},
          doi = {10.3847/1538-4357/ace624},
archivePrefix = {arXiv},
       eprint = {2309.01891},
 primaryClass = {astro-ph.EP},
       adsurl = {https://ui.adsabs.harvard.edu/abs/2023ApJ...954..101A}
}

@ARTICLE{Lee2019,
       author = {{Lee}, Chin-Fei and {Kwon}, Woojin and {Jhan}, Kai-Syun and {Hirano}, Naomi and {Hwang}, Hsiang-Chih and {Lai}, Shih-Ping and {Ching}, Tao-Chung and {Rao}, Ramprasad and {Ho}, Paul T.~P.},
        title = "{A Pseudodisk Threaded with a Toroidal and Pinched Poloidal Magnetic Field Morphology in the HH 211 Protostellar System}",
      journal = {\apj},
         year = 2019,
        month = jul,
       volume = {879},
       number = {2},
          eid = {101},
        pages = {101},
          doi = {10.3847/1538-4357/ab2458},
archivePrefix = {arXiv},
       eprint = {1905.09417},
 primaryClass = {astro-ph.GA},
       adsurl = {https://ui.adsabs.harvard.edu/abs/2019ApJ...879..101L}
}

@ARTICLE{Ahmad2026,
       author = {{Ahmad}, Adnan Ali and {Commer{\c{c}}on}, Beno{\^\i}t and {Lynch}, Elliot and {Lovascio}, Francesco and {Charnoz}, Sebastien and {Marschall}, Raphael and {Morbidelli}, Alessandro},
        title = "{Numerical insights into disk accretion, eccentricity, and kinematics in the Class 0 phase}",
      journal = {\aap},
         year = 2026,
        month = mar,
       volume = {707},
          eid = {A97},
        pages = {A97},
          doi = {10.1051/0004-6361/202558019},
archivePrefix = {arXiv},
       eprint = {2601.08916},
 primaryClass = {astro-ph.SR},
       adsurl = {https://ui.adsabs.harvard.edu/abs/2026A&A...707A..97A}
}

@ARTICLE{Hirano2026,
       author = {{Hirano}, Shingo and {Aikawa}, Yuri and {Machida}, Masahiro N.},
        title = "{Multiple Protostellar Outflows from a Single Protostar with a Misaligned Disk}",
      journal = {\apj},
         year = 2026,
        month = may,
       volume = {1003},
       number = {1},
          eid = {99},
        pages = {99},
          doi = {10.3847/1538-4357/ae6326},
archivePrefix = {arXiv},
       eprint = {2602.20691},
 primaryClass = {astro-ph.SR},
       adsurl = {https://ui.adsabs.harvard.edu/abs/2026ApJ..1003...99H}
}

@ARTICLE{Kuffmeier2017,
       author = {{Kuffmeier}, Michael and {Haugb{\o}lle}, Troels and {Nordlund}, {\r{A}}ke},
        title = "{Zoom-in Simulations of Protoplanetary Disks Starting from GMC Scales}",
      journal = {\apj},
         year = 2017,
        month = sep,
       volume = {846},
       number = {1},
          eid = {7},
        pages = {7},
          doi = {10.3847/1538-4357/aa7c64},
archivePrefix = {arXiv},
       eprint = {1611.10360},
 primaryClass = {astro-ph.SR},
       adsurl = {https://ui.adsabs.harvard.edu/abs/2017ApJ...846....7K}
}

@ARTICLE{Kuffmeier2018,
       author = {{Kuffmeier}, Michael and {Frimann}, S{\o}ren and {Jensen}, Sigurd S. and {Haugb{\o}lle}, Troels},
        title = "{Episodic accretion: the interplay of infall and disc instabilities}",
      journal = {\mnras},
         year = 2018,
        month = apr,
       volume = {475},
       number = {2},
        pages = {2642-2658},
          doi = {10.1093/mnras/sty024},
archivePrefix = {arXiv},
       eprint = {1710.00931},
 primaryClass = {astro-ph.SR},
       adsurl = {https://ui.adsabs.harvard.edu/abs/2018MNRAS.475.2642K}
}

@ARTICLE{Kuffmeier2019,
       author = {{Kuffmeier}, M. and {Calcutt}, H. and {Kristensen}, L.~E.},
        title = "{The bridge: a transient phenomenon of forming stellar multiples. Sequential formation of stellar companions in filaments around young protostars}",
      journal = {\aap},
         year = 2019,
        month = aug,
       volume = {628},
          eid = {A112},
        pages = {A112},
          doi = {10.1051/0004-6361/201935504},
archivePrefix = {arXiv},
       eprint = {1907.02083},
 primaryClass = {astro-ph.SR},
       adsurl = {https://ui.adsabs.harvard.edu/abs/2019A&A...628A.112K}
}

@ARTICLE{Seifried2013,
       author = {{Seifried}, D. and {Banerjee}, R. and {Pudritz}, R.~E. and {Klessen}, R.~S.},
        title = "{Turbulence-induced disc formation in strongly magnetized cloud cores}",
      journal = {\mnras},
         year = 2013,
        month = jul,
       volume = {432},
       number = {4},
        pages = {3320-3331},
          doi = {10.1093/mnras/stt682},
archivePrefix = {arXiv},
       eprint = {1302.4901},
 primaryClass = {astro-ph.GA},
       adsurl = {https://ui.adsabs.harvard.edu/abs/2013MNRAS.432.3320S}
}

@ARTICLE{Seifried2015,
       author = {{Seifried}, D. and {Banerjee}, R. and {Klessen}, R.~S.},
        title = "{Turbulence and its effect on protostellar disk formation}",
      journal = {arXiv e-prints},
         year = 2015,
        month = sep,
          eid = {arXiv:1509.04467},
        pages = {arXiv:1509.04467},
          doi = {10.48550/arXiv.1509.04467},
archivePrefix = {arXiv},
       eprint = {1509.04467},
 primaryClass = {astro-ph.SR},
       adsurl = {https://ui.adsabs.harvard.edu/abs/2015arXiv150904467S}
}

@ARTICLE{Yang2025,
       author = {{Yang}, Trey Qingyun and {Federrath}, Christoph},
        title = "{Protostellar disc structure and dynamics during star formation from cloud-scale initial conditions}",
      journal = {\mnras},
         year = 2025,
        month = aug,
       volume = {541},
       number = {2},
        pages = {1969-1987},
          doi = {10.1093/mnras/staf1088},
archivePrefix = {arXiv},
       eprint = {2501.07626},
 primaryClass = {astro-ph.SR},
       adsurl = {https://ui.adsabs.harvard.edu/abs/2025MNRAS.541.1969Y}
}

@ARTICLE{Lebreuilly2021,
       author = {{Lebreuilly}, Ugo and {Hennebelle}, Patrick and {Colman}, Tine and {Commer{\c{c}}on}, Beno{\^\i}t and {Klessen}, Ralf and {Maury}, Ana{\"e}lle and {Molinari}, Sergio and {Testi}, Leonardo},
        title = "{Protoplanetary Disk Birth in Massive Star-forming Clumps: The Essential Role of the Magnetic Field}",
      journal = {\apjl},
         year = 2021,
        month = aug,
       volume = {917},
       number = {1},
          eid = {L10},
        pages = {L10},
          doi = {10.3847/2041-8213/ac158c},
archivePrefix = {arXiv},
       eprint = {2107.08491},
 primaryClass = {astro-ph.SR},
       adsurl = {https://ui.adsabs.harvard.edu/abs/2021ApJ...917L..10L}
}

@ARTICLE{Lebreuilly2024,
       author = {{Lebreuilly}, U. and {Hennebelle}, P. and {Maury}, A. and {Gonz{\'a}lez}, M. and {Traficante}, A. and {Klessen}, R. and {Testi}, L. and {Molinari}, S.},
        title = "{Influence of protostellar outflows on star and protoplanetary disk formation in a massive star-forming clump}",
      journal = {\aap},
         year = 2024,
        month = mar,
       volume = {683},
          eid = {A13},
        pages = {A13},
          doi = {10.1051/0004-6361/202347913},
archivePrefix = {arXiv},
       eprint = {2309.05397},
 primaryClass = {astro-ph.SR},
       adsurl = {https://ui.adsabs.harvard.edu/abs/2024A&A...683A..13L}
}

@ARTICLE{Lebreuilly2025,
       author = {{Lebreuilly}, Ugo and {Traficante}, Alessio and {Nucara}, Alice and {Tung}, Ngo-Duy and {Hennebelle}, Patrick and {Molinari}, Sergio and {Klessen}, Ralf S. and {Testi}, Leonardo and {Pelkonen}, Veli-Matti and {Benedettini}, Milena and {Coletta}, Alessandro and {Elia}, Davide and {Mininni}, Chiara and {Pezzuto}, Stefania and {Soler}, Juan D. and {Suin}, Paolo and {Toci}, Claudia},
        title = "{The Rosetta Stone Project: I. A suite of radiative magnetohydrodynamics simulations of high-mass star-forming clumps}",
      journal = {\aap},
         year = 2025,
        month = sep,
       volume = {701},
          eid = {A217},
        pages = {A217},
          doi = {10.1051/0004-6361/202554774},
archivePrefix = {arXiv},
       eprint = {2507.08436},
 primaryClass = {astro-ph.SR},
       adsurl = {https://ui.adsabs.harvard.edu/abs/2025A&A...701A.217L}
}

@ARTICLE{Matsumoto2004,
       author = {{Matsumoto}, Tomoaki and {Tomisaka}, Kohji},
        title = "{Directions of Outflows, Disks, Magnetic Fields, and Rotation of Young Stellar Objects in Collapsing Molecular Cloud Cores}",
      journal = {\apj},
         year = 2004,
        month = nov,
       volume = {616},
       number = {1},
        pages = {266-282},
          doi = {10.1086/424897},
archivePrefix = {arXiv},
       eprint = {astro-ph/0408086},
 primaryClass = {astro-ph},
       adsurl = {https://ui.adsabs.harvard.edu/abs/2004ApJ...616..266M}
}

@ARTICLE{Joos2012,
       author = {{Joos}, M. and {Hennebelle}, P. and {Ciardi}, A.},
        title = "{Protostellar disk formation and transport of angular momentum during magnetized core collapse}",
      journal = {\aap},
         year = 2012,
        month = jul,
       volume = {543},
          eid = {A128},
        pages = {A128},
          doi = {10.1051/0004-6361/201118730},
archivePrefix = {arXiv},
       eprint = {1203.1193},
 primaryClass = {astro-ph.SR},
       adsurl = {https://ui.adsabs.harvard.edu/abs/2012A&A...543A.128J}
}

@ARTICLE{Li2013,
       author = {{Li}, Zhi-Yun and {Krasnopolsky}, Ruben and {Shang}, Hsien},
        title = "{Does Magnetic-field-Rotation Misalignment Solve the Magnetic Braking Catastrophe in Protostellar Disk Formation?}",
      journal = {\apj},
         year = 2013,
        month = sep,
       volume = {774},
       number = {1},
          eid = {82},
        pages = {82},
          doi = {10.1088/0004-637X/774/1/82},
archivePrefix = {arXiv},
       eprint = {1301.6545},
 primaryClass = {astro-ph.SR},
       adsurl = {https://ui.adsabs.harvard.edu/abs/2013ApJ...774...82L}
}

@ARTICLE{Tsukamoto2018,
       author = {{Tsukamoto}, Y. and {Okuzumi}, S. and {Iwasaki}, K. and {Machida}, M.~N. and {Inutsuka}, S.},
        title = "{Does Misalignment between Magnetic Field and Angular Momentum Enhance or Suppress Circumstellar Disk Formation?}",
      journal = {\apj},
         year = 2018,
        month = nov,
       volume = {868},
       number = {1},
          eid = {22},
        pages = {22},
          doi = {10.3847/1538-4357/aae4dc},
archivePrefix = {arXiv},
       eprint = {1811.06675},
 primaryClass = {astro-ph.SR},
       adsurl = {https://ui.adsabs.harvard.edu/abs/2018ApJ...868...22T}
}

@ARTICLE{Bate2018,
       author = {{Bate}, Matthew R.},
        title = "{On the diversity and statistical properties of protostellar discs}",
      journal = {\mnras},
         year = 2018,
        month = apr,
       volume = {475},
       number = {4},
        pages = {5618-5658},
          doi = {10.1093/mnras/sty169},
archivePrefix = {arXiv},
       eprint = {1801.07721},
 primaryClass = {astro-ph.SR},
       adsurl = {https://ui.adsabs.harvard.edu/abs/2018MNRAS.475.5618B}
}

@ARTICLE{Tsukamoto2015,
       author = {{Tsukamoto}, Y. and {Iwasaki}, K. and {Okuzumi}, S. and {Machida}, M.~N. and {Inutsuka}, S.},
        title = "{Bimodality of Circumstellar Disk Evolution Induced by the Hall Current}",
      journal = {\apjl},
         year = 2015,
        month = sep,
       volume = {810},
       number = {2},
          eid = {L26},
        pages = {L26},
          doi = {10.1088/2041-8205/810/2/L26},
archivePrefix = {arXiv},
       eprint = {1506.07242},
 primaryClass = {astro-ph.SR},
       adsurl = {https://ui.adsabs.harvard.edu/abs/2015ApJ...810L..26T}
}

@ARTICLE{Tsukamoto2017,
       author = {{Tsukamoto}, Yusuke and {Okuzumi}, Satoshi and {Iwasaki}, Kazunari and {Machida}, Masahiro N. and {Inutsuka}, Shu-ichiro},
        title = "{The impact of the Hall effect during cloud core collapse: Implications for circumstellar disk evolution}",
      journal = {\pasj},
         year = 2017,
        month = dec,
       volume = {69},
       number = {6},
          eid = {95},
        pages = {95},
          doi = {10.1093/pasj/psx113},
archivePrefix = {arXiv},
       eprint = {1706.04363},
 primaryClass = {astro-ph.SR},
       adsurl = {https://ui.adsabs.harvard.edu/abs/2017PASJ...69...95T}
}

@ARTICLE{Marchand2019,
       author = {{Marchand}, P. and {Tomida}, K. and {Commer{\c{c}}on}, B. and {Chabrier}, G.},
        title = "{Impact of the Hall effect in star formation, improving the angular momentum conservation}",
      journal = {\aap},
         year = 2019,
        month = nov,
       volume = {631},
          eid = {A66},
        pages = {A66},
          doi = {10.1051/0004-6361/201936215},
archivePrefix = {arXiv},
       eprint = {1909.09025},
 primaryClass = {astro-ph.SR},
       adsurl = {https://ui.adsabs.harvard.edu/abs/2019A&A...631A..66M}
}

@ARTICLE{Masson2016,
       author = {{Masson}, J. and {Chabrier}, G. and {Hennebelle}, P. and {Vaytet}, N. and {Commer{\c{c}}on}, B.},
        title = "{Ambipolar diffusion in low-mass star formation. I. General comparison with the ideal magnetohydrodynamic case}",
      journal = {\aap},
         year = 2016,
        month = mar,
       volume = {587},
          eid = {A32},
        pages = {A32},
          doi = {10.1051/0004-6361/201526371},
archivePrefix = {arXiv},
       eprint = {1509.05630},
 primaryClass = {astro-ph.SR},
       adsurl = {https://ui.adsabs.harvard.edu/abs/2016A&A...587A..32M}
}

@ARTICLE{Wurster2018,
       author = {{Wurster}, James and {Bate}, Matthew R. and {Price}, Daniel J.},
        title = "{Hall effect-driven formation of gravitationally unstable discs in magnetized molecular cloud cores}",
      journal = {\mnras},
         year = 2018,
        month = nov,
       volume = {480},
       number = {4},
        pages = {4434-4442},
          doi = {10.1093/mnras/sty2212},
archivePrefix = {arXiv},
       eprint = {1808.04376},
 primaryClass = {astro-ph.SR},
       adsurl = {https://ui.adsabs.harvard.edu/abs/2018MNRAS.480.4434W}
}

@ARTICLE{Granzow2026,
       author = {{Granzow Holm}, Christian and {Lambrechts}, Michiel and {Kuffmeier}, Michael and {Johansen}, Anders and {Haugb{\o}lle}, Troels and {Nordlund}, {\r{A}}ke},
        title = "{Accretion across scales: streamers, surface-layer transport, and rapid replenishment in young protoplanetary discs}",
      journal = {arXiv e-prints},
         year = 2026,
        month = aug,
          eid = {arXiv:2608.02204},
        pages = {arXiv:2608.02204},
          doi = {10.48550/arXiv.2608.02204},
archivePrefix = {arXiv},
       eprint = {2608.02204},
 primaryClass = {astro-ph.EP},
       adsurl = {https://ui.adsabs.harvard.edu/abs/2026arXiv260802204G}
}
\bibliographystyle{aasjournal}
\end{document}